\documentclass[a4paper,10pt]{article}
\usepackage[numbers]{natbib}
\usepackage{fullpage}
\usepackage{amsmath}
\usepackage{amssymb}
\usepackage{amsthm}
\usepackage[english]{babel}

\usepackage{url}
\usepackage{multirow}
\usepackage{bigstrut}    		
\usepackage{rotating}
\usepackage{acronym}
\usepackage{listings}
\usepackage{verbatim}
\usepackage{subcaption}
\usepackage{colortbl}
\usepackage{xfrac}
\usepackage[inline]{enumitem}
\usepackage{placeins}
\usepackage{todonotes}

\usepackage{tkz-graph}
\usepackage{tikz}
\usetikzlibrary{backgrounds}
\usetikzlibrary{trees,positioning,arrows}
\usetikzlibrary{calc}

\definecolor{Gray}{gray}{0.80}

\newcommand{\quotes}[1]{``.1"}

\newcommand{\overbar}[1]{\mkern 1.5mu\overline{\mkern-1.5mu#1\mkern-1.5mu}\mkern 1.5mu}
\newtheorem{example}{Example}

\begin{document}

\title{Towards Meaningful Statements in IR Evaluation\\
Mapping Evaluation Measures to Interval Scales}


\author{Marco Ferrante\thanks{University of Padua, Department of Mathematics, ``Tullio Levi-Civita'', Italy, ferrante@math.unipd.it} 
\and Nicola Ferro\thanks{University of Padua,
Department of Information Engineering, Italy, ferro@dei.unipd.it} 
\and Norbert Fuhr\thanks{University of Duisburg-Essen, Germany, norbert.fuhr@uni-due.de}}


%
%
%

\maketitle
\begin{abstract}
Recently, it was shown that most popular IR measures are not interval-scaled, implying that decades of experimental IR research used potentially improper methods, which may have produced questionable results. However, it was unclear if and to what extent these findings apply to actual evaluations and this opened a debate in the community with researchers standing on opposite positions about whether this should be considered an issue (or not) and to what extent.

In this paper, we first give an introduction to the representational measurement theory explaining why certain operations and significance tests are permissible only with scales of a certain level. For that, we introduce the notion of meaningfulness specifying the conditions under which
the truth (or falsity) of a statement is invariant under permissible transformations of a scale. 
Furthermore, we show how  the recall base and the length of the run may make comparison and aggregation across topics problematic.
Then we propose a straightforward and powerful approach for turning an evaluation measure into an interval scale, and describe an experimental evaluation of the differences between using the original measures and the interval-scaled ones.
For all the regarded measures – namely Precision, Recall, Average Precision, (Normalized) Discounted Cumulative Gain, Rank-Biased Precision and Reciprocal Rank - we observe substantial effects, both on the order of average values and on the outcome of significance tests. For the latter, previously significant differences turn out to be insignificant, while insignificant ones become significant. The effect varies remarkably between the tests considered but overall, on average, we observed a $25\%$ change in the decision about which systems are significantly different and which are not.
\end{abstract}

\section{Introduction}

By virtue or by necessity, \acf{IR} has always been deeply rooted in experimentation and evaluation has been a formidable driver of innovation and advancement in the field, as also witnessed by the success of the major evaluation initiatives -- \ac{TREC}\footnote{\url{https://trec.nist.gov/}} in the United States~\cite{zz-HarmanVoorhees2005-editor}, \ac{CLEF}\footnote{\url{http://www.clef-initiative.eu/}} in Europe~\cite{zz-FerroPeters2019-editor}, \ac{NTCIR}\footnote{\url{http://research.nii.ac.jp/ntcir/}} in Japan and Asia~\cite{zz-SakaiEtAl2020-editor}, and \ac{FIRE}\footnote{\url{http://fire.irsi.res.in/}} in India -- not only from the scientific and technological point of view but also from the economic impact one~\cite{TRECImpact2010}.

Central to experimentation and evaluation is how to measure the performance of an \ac{IR} system and there is a rich set of IR literature discussing existing evaluation measures or introducing new ones as well as proposing frameworks to model them~\cite{Carterette2011,MoffatEtAl2013}. The major goal is to quantify users' experience of retrieval quality for certain types of search behavior, like e.g. users stopping at the first relevant document, or after the first ten results. Most of the measures proposed are based on plausible arguments and often accompanied by experimental studies, also investigating how close they are to end-user experience and satisfaction~\cite{JiangAllan2016,ZhangEtAl2017b,ZhangEtAl2020}. However, little attention has been given to a proper theoretic basis of the evaluation measures, leading to possibly flawed measures and affecting the \emph{validity} of the scientific results based on them, especially their \emph{internal validity}, i.e. ``the ability to draw conclusions about causal relationships from the results of a study''~\cite[p.~157]{CozbyBates2018}

A few years ago, \citet{Robertson2006} raised the question of which scales are used by \ac{IR} evaluation measures, since they determine which operations make sense on the values of a measure, as originally proposed by~\citet{Stevens1946}. Scales have increasing  properties: a \emph{nominal scale} allows for determination of equality and for the computation of the mode; an \emph{ordinal scale} allows only for determination of greater or less and for the computation of medians and percentiles; an \emph{interval  scale} allows also for determination of equality of intervals or differences and for the computation of mean, standard deviation, rank-order correlation; finally, a \emph{ratio scale} allows also for the determination of equality of ratios and for the computation of coefficient of variation. Recently, \citet{FerranteEtAl2017,FerranteEtAl2018b} have theoretically shown that some of the most known and used IR measures, like \ac{AP} or \ac{DCG}, are not interval-scales. As a consequence, we should neither compute means, standard deviations and confidence intervals, nor perform significance tests that require an interval scale.  Over the decades there has been much debate about \citeauthor{Stevens1946}'s prescriptions~\cite{Lord1953,VellemanWilkinson1993,Hand1996,Michell1986} and this debate has also spawn to the \ac{IR} field with \citet{Fuhr2017} suggesting strict adherence to \citeauthor{Stevens1946}'s prescriptions and \citet{Sakai2020} arguing for a more lenient approach.

Our vision is that it is now time for the \ac{IR} field to accurately investigate and understand the scale properties of its evaluation measures and their implications on the validity of our experimental findings. As a matter of fact, we are not aware of any experimental \ac{IR} paper that regarded evaluation measures as ordinal scales, thus refraining from computing (and comparing) means; also, most papers using evaluation measures apply parametric tests, which should be used only from interval scales onwards. This means that improper methods have been potentially applied. Independently from your stance in the above long-standing debate, the key question about \ac{IR} experimental findings is: are we on the safe side or are we at risk? Are we in a situation like using a rubber band to measure and compare lengths? Are we facing a state of the affairs where decades of \ac{IR} research may have produced questionable results?

We do not have the answer to these questions but our intention with this paper is to lay the foundations and set all the pieces needed to have the means and instruments to answer these questions and to let the \ac{IR} community discuss these issues on a common ground in order to reach shared conclusions.

Therefore, the contributions of the paper are as follows:
\begin{enumerate}
	\item introduction to the \emph{representational measurement theory}~\cite{KrantzEtAl1971,LuceEtAl1990,SuppesEtAl1989}, clearly explaining why (or why not) certain operations and significance tests should be permissible on a given scale and presenting the different stances on this long-standing debate;
	\item introduction to the notion of \emph{meaningfulness}~\cite{FalmagneNarens1983,Narens2002,Roberts1985}, i.e. the conditions under which the truth (or falsity) of a statement is invariant under permissible transformations of a scale. To the best of our knowledge, this concept has never investigated or applied in \ac{IR} but it is fundamental to the validity of the inferences we draw;
	\item discussion and demonstration of further measurement issues, specific to \ac{IR} and beyond the debate on permissible operations. In particular, we show how the recall base and the length of the run may make averaging across topics (or other forms of aggregate statistics) problematic, at best;
	\item proposal of a straightforward and powerful approach for turning an evaluation measure into an interval scale, by transforming its values into their rank position. In this way, we provide a means for improving the meaningfulness and validity of our inferences, still preserving the different user models embedded by the various evaluation measures;
	\item experimental evaluation of the differences between using the original measures and the interval-scaled ones, by relying on several \ac{TREC} collections. For all the regarded measures -- namely Precision, Recall, \ac{AP}, \ac{DCG}, \ac{nDCG}, \ac{RBP}, and \ac{RR} -- we observe substantial effects, both on the order of average values and on the outcome of significance tests. For the latter, previously significant differences turn out to be insignificant, while insignificant ones become significant. The effect varies remarkably between the tests considered but overall, on average, we observed a $25\%$ change in decisions about what is significant and what is not. 
\end{enumerate}

The paper is organized as follows: Section~\ref{sec:measurement} provides an overview of the representational theory of measurement, of the different types of scale, and the notion of meaningfulness. Section~\ref{sec:measurement-ir} deeply discusses measurement and meaningfulness issues specific to \ac{IR}. Section~\ref{sec:related} briefly summarizes related works. Section~\ref{sec:methodology} explains our methodology for transforming evaluation measures into interval scales. Section~\ref{sec:setup} introduces the experimental setup while Section~\ref{sec:experiments} discusses the results of the experiments. Finally, Section~\ref{sec:conclusions} draws some conclusions and outlooks for future works.

\section{Measurement}
\label{sec:measurement}

\subsection{Overview}
\label{subsec:measurement-overview}

The \emph{representational theory of measurement}~\cite{KrantzEtAl1971,SuppesEtAl1989,LuceEtAl1990} is one of the most developed approaches to measurement, suitable for many areas of science ranging to physics and engineering to psychology. The basic idea is that real world \emph{objects} have \emph{attributes} which constitute their relevant features and induce a set of relationship among them; the set of objects $E$ together with the relationships $R^E_1, R^E_2, \ldots$ among them comprise the so-called \acf{ERS} $\mathbf{E} = \big\langle E, R^E_1, R^E_2, \ldots \big\rangle$. Then, we look for a mapping between the real word objects $E$ and numbers $N$ in such a way that the relationships $R^E_1, R^E_2, \ldots$ among the objects match with relationships $R^N_1, R^N_2, \ldots$ among numbers; the set of numbers $N$ together with the relationships $R^N_1, R^N_2, \ldots$ constitutes the so-called \acf{NRS} $\mathbf{N} = \big\langle N, R^N_1, R^N_2, \ldots \big\rangle$. 

More precisely, the representational theory of measurement seeks for an \emph{homomorphism} $\phi$ which maps $E$ onto $N$ in such a way that $\forall R^E_i, \, \forall e_1, e_2, \ldots, e_k \in E \mid (e_1, e_2, \ldots, e_k) \in R^E_i$ it holds that  $\exists \, n_1=\phi(e_1), n_2=\phi(e_2), \ldots n_k=\phi(e_k) \in N \mid (n_1, n_2, \ldots, n_k) \in R^N_i$. The homomorphism $\phi$ is called a \emph{scale of measurement}. Note that, in general, we seek for an homomorphism and not an isomorphism because two different real word objects might be mapped into the same number.

The most typical example is length. Suppose the \ac{ERS} $\mathbf{E} = \big\langle E, \succsim, \circ \big\rangle$ is a set of rods with an order relationship $\succsim$ among rods and a concatenation operation $\circ$ among them. If the attribute under examination is the length of a rod, we can map the \ac{ERS} to the \ac{NRS} $\mathbf{N} = \big\langle \mathbb{R}_0^+, \geq, + \big\rangle$ such that $\forall e_1, e_2, e_3 \in E$ it holds $e_1 \succsim e_2 \Leftrightarrow \phi(e_1) \geq \phi(e_2)$ and $e_1 \circ e_2\sim e_3 \Leftrightarrow \phi(e_1) + \phi(e_2) = \phi(e_3)$, that is if a rod is longer than another one the number assigned to the first one is bigger than the number  assigned to the second one and the concatenation of two rods corresponds to the sum of the two numbers assigned to them.

The core of the representational theory of measurements is to seek for a \emph{representation} theorem and a \emph{uniqueness} theorem for the scale of measurement in order to fully define it.

The \emph{representation theorem} ensures that if the \ac{ERS} satisfies given properties, it is possible to construct an homomorphism to a certain \ac{NRS}. In the previous example, the representation theorem defines which properties the order relation $\succsim$ and the concatenation $\circ$ have to satisfy in order to construct a real-valued function $\phi$ which is order preserving and additive. It is important to underline that the representational theory of measurement seeks for ``operations'' among real word objects -- e.g. we can put two rods side by side to order them or we can lay two rods end by end to concatenate them -- and if these ``operations'' satisfy given properties they can be reflected into corresponding operations among numbers, where numbers are just a proxy of what happens among real world objects but are much more convenient to manipulate.

In general, given an \ac{ERS} and an \ac{NRS}, it is possible to create more than one homomorphism between them. For example, it is possible to express length by using meters or yards and both of them are legitimate scales for length.  The \emph{uniqueness theorem} is concerned with determining which are the \emph{permissible transformations} $\phi \rightarrow \phi^\prime$ such that $\phi$ and $\phi^\prime$ are \emph{both} homomorphisms of the given \ac{ERS} into the \emph{same} \ac{NRS}. In our example, any transformation $\phi^\prime = \alpha \phi, \alpha > 0$ is permissible for length. Therefore, the uniqueness theorem guarantees that the ``structure'' of a scale of measurement is invariant to changes in the numerical assignment, which preserve the relationships.

\subsection{Classification of the Scales of Measurement}
\label{subsec:measurement-scales}

\citet{Stevens1946} introduced a classification of scales based on their permissible transformations, described below.

\subsubsection{Nominal scale} 

It is used when entities of the real world can be placed into different classes or categories on the basis of their attribute under examination. The \ac{ERS} consists only of different classes without any notion of ordering among them and any distinct numeric representation of the classes is an acceptable measure but there is no notion of magnitude associated with numbers. Therefore, any arithmetic operation on the numeric representation has no meaning.
    
    The class of permissible transformations is the set of all \emph{one-to-one mappings}, i.e. bijective functions: $\phi^\prime = \mathrm{f}(\phi)$, since they preserve the distinction among classes.

\begin{example}[Nominal Scale]
    Consider a classification of people by their country, e.g. France, Germany, Greece, Italy, Spain, and so on. We could define the two following measurements:
	\begin{equation*}
		\phi = 
			\begin{cases}
				5 & \text{if France} \\
				4 & \text{if Germany} \\
				3 & \text{if Greece} \\
				2 & \text{if Italy} \\
				1 & \text{if Spain} \\
				\cdots & \text{if } \cdots \\
			\end{cases}
		\hspace*{3em}
		\phi^\prime = 
			\begin{cases}
				41 & \text{if France} \\
				13 & \text{if Germany} \\
				-10 & \text{if Greece} \\
				23 & \text{if Italy} \\
				17 & \text{if Spain} \\
				\cdots & \text{if } \cdots \\
			\end{cases}
	\end{equation*}
	both $\phi$ and $\phi^\prime$ are valid measures, which can be related with a one-to-one mapping. Note that even if  $\phi$ looks like being ordered, there is actually no meaning in the associated magnitudes and so it should not be confused with an ordinal scale. Moreover, even if it is alway possible to operate with numbers, using $\phi$ and performing $4 - 3 = 1$, which would correspond to  $\text{Germany} - \text{Greece} \stackrel{?}{=} \text{Spain}$, has no specific meaning, as well as using $\phi^\prime$ and performing $13 - (-10) = 23$, which would correspond to  $\text{Germany} - \text{Greece} \stackrel{?}{=} \text{Italy}$, even in disagreement with the previous case.
\label{ex:nominal-scale}
\end{example}

\subsubsection{Ordinal scale} 

It can be considered as a nominal scale where, in addition, there is a notion of ordering among the different classes or categories. The \ac{ERS} consists of classes that are ordered with respect to the attribute under examination and any distinct numeric representation which preserves the ordering is acceptable. Therefore, the magnitude of the numbers is used just to represent the ranking among classes. As a consequence, addition, subtraction or other mathematical operations have no meaning.
    
    The class of permissible transformations is the set of all the \emph{monotonic increasing functions}, since they preserve the ordering: $\phi^\prime = \mathrm{f}(\phi)$.

\begin{example}[Ordinal Scale]
	 The European Commission Regulation 607/2009~\cite{EC-L193-60} and the follow-up regulation 2019/33~\cite{EC-L9-2} set the following increasing scale to classify sparkling wines on the basis of their sugar content:
	\begin{itemize}
		\item \emph{pas dos\'{e}} (brut nature): sugar content is less than 3 grams per litre; let us call this range $s_0 = [0, 3]$;
		\item \emph{extra brut}: sugar content is between 0 and 6 grams per litre; let us call this range $s_1 = [0, 6]$; 
		\item \emph{brut} : sugar content is less than 12 grams per litre; let us call this range $s_2 = [0, 12]$; 
		\item \emph{extra dry}: sugar content is between 12 and 17 grams per litre; let us call this range $s_3 = (12, 17]$;
		\item \emph{sec} (dry):  sugar content is between 17 and 32 grams per litre; let us call this range $s_4 = (17, 32]$;
		\item \emph{demi-sec} (medium dry):  sugar content is between 32 and 50 grams per litre; let us call this range $s_5 = (32, 50]$;
		\item \emph{doux} (sweet): sugar content is greater than 50 grams per litre; let us call this range $s_6 = (50, 2000]$, where 2000 grams per litre is roughly the saturation of sugar in water, which is much higher than those of sugar in alcohol.
	\end{itemize}	 
	 	  
	We can introduce two alternative ordinal scales $\phi$ and $\phi^\prime$ of the above wine scale where $\phi$ is given by the maximum of a range while $\phi^\prime$ is given by a monotonic transformation $\phi^\prime =\phi^2$:
	\begin{equation*}
		\phi = 
			\begin{cases}
				3		& \text{if pas dos\'{e}} \\
				6   		& \text{if extra brut} \\
				12 		& \text{if brut} \\
				17  		& \text{if extra dry} \\
				32  		& \text{if sec} \\
				50 		& \text{if demi-sec} \\
				2000		& \text{if doux} \\
			\end{cases}
		\hspace*{3em}
		\phi^\prime = 
			\begin{cases}
				9		& \text{if pas dos\'{e}} \\
				36		& \text{if extra brut} \\
				144		& \text{if brut} \\
				289  		& \text{if extra dry} \\
				1024 	& \text{if sec} \\
				2500 	& \text{if demi-sec} \\
				4000000	& \text{if doux} \\
			\end{cases}
	\end{equation*}

	As in the case of the previous Example~\ref{ex:nominal-scale}, mathematical operations have no specific meaning, even if, especially in the case of $\phi$, we may be tempted to perform operations like $\frac{\text{brut}}{\text{extra brut}} = \frac{12}{6} = 2$ to express statements like ``brut may be twice as sweet as extra brut''. However, such statement cannot be expressed on the $\phi$ or $\phi^\prime$ scale and it actually comes from implicitly changing scale to the \emph{mass concentration} scale of the solution, which is a ratio scale (see below) where the division operation would make sense. Also addition and subtraction have no meaning, so  $\text{brut} - \text{extra brut} = 12 - 6 = 6$ is not a way to express statements like ``brut may have $6 \, \sfrac{g}{l}$ of sugar more than extra brut'', for the same reasons above. We could perform operations such as $\mathrm{sgn}(\phi(e_1) - \phi(e_2))$ or $\mathrm{sgn}(\phi^\prime(e_1) - \phi^\prime(e_2))$ but this would be just a more involute way of expressing the order among categories, which is the only property guaranteed by ordinal scales.
	\label{ex:ordinal-scale}
\end{example}

\subsubsection{Interval scale}

Besides relying on ordered classes, it also captures information about the size of the intervals that separate the classes.  The \ac{ERS} consists of classes that are ordered with respect to the attribute under examination and where the \emph{size of the ``gap''} among two classes is somehow understood; more precisely,  fundamental to the definition of an interval scale is that \emph{intervals must be equi-spaced}. An interval scale preserves order, as an ordinal one, and differences among classes have meaning -- but not their ratio. Therefore, addition and subtraction are acceptable operations but not multiplication and division.
    
    The class of permissible transformations is the set of all \emph{affine transformations}: $\phi^\prime = \alpha\phi + \beta, \, \alpha > 0$.
    
    Note that while ratios of classes $\frac{\phi(e_1)}{\phi(e_2)}$ have no meaning on an interval scale, the ratio of differences among classes, i.e. the ratio of intervals, is allowed and invariant $\frac{\phi^\prime(a) - \phi^\prime(b)}{\phi^\prime(c) - \phi^\prime(d)} = \frac{[\alpha\phi(a) + \beta] - [\alpha\phi(b) + \beta]}{[\alpha\phi(c) + \beta] - [\alpha\phi(d) + \beta]} = \frac{\phi(a) - \phi(b)}{\phi(c) - \phi(d)}$.

\begin{example}[Interval Scale]
A typical example of interval scale is temperature, which can be expressed on either the Fahrenheit or the Celsius scale, where the affine transformation $F = \frac{9}{5}C + 32$ allows us to pass from one to the other. When talking about temperature it does not make sense to say that $20 \, ^\circ\text{C}$ is twice as hot as $10 \, ^\circ\text{C}$, i.e. multiplication and division are not allowed; you can also note that the division operation is not invariant to the transformation, since $\frac{20 \, ^\circ\text{C}}{10 \, ^\circ\text{C}} = 2$ but $\frac{68 \, ^\circ\text{F}}{50 \, ^\circ\text{F}} = 1.36$. However, it makes sense to say that the increase between $10 \, ^\circ\text{C}$ and $20 \, ^\circ\text{C}$ is the same as the increase  between $20 \, ^\circ\text{C}$ and $30 \, ^\circ\text{C}$, i.e. addition and subtractions are allowed; you can also note that the subtraction operation is invariant to the transformation since $30 \, ^\circ\text{C}  - 20 \, ^\circ\text{C} = 20 \, ^\circ\text{C} - 10 \, ^\circ\text{C} = 10 \, ^\circ\text{C}$ and $86 \, ^\circ\text{F}  - 68 \, ^\circ\text{F} = 68 \, ^\circ\text{F} - 50 \, ^\circ\text{F} = 18 \, ^\circ\text{F}$. Moreover, the ratio of intervals $\frac{20 \, ^\circ\text{C} \; - \; 10 \, ^\circ\text{C}}{30 \, ^\circ\text{C} \; - \; 20 \, ^\circ\text{C}} = 1$ is invariant to the transformation $\frac{68 \, ^\circ\text{F} \; - \; 50 \, ^\circ\text{F}}{86 \, ^\circ\text{F} \; - \; 68 \, ^\circ\text{F}} = 1$.

Central to the notion of temperature is the fact that the size of the ``gap'' has the same meaning all over the scale; indeed, 1 degree represents the same amount of thermal energy all over the scale. i.e. the gaps are \emph{equi-spaced}.
	\label{ex:interval-scale}
\end{example}    
    
\subsubsection{Ratio scale}

It allows us to compute ratios among the different classes.  The \ac{ERS} consists of classes that are ordered, where there is a notion of ``gap'' among two classes and where the ``proportion'' among two classes is somehow understood. It preserves order and differences as well as ratios. Therefore, all the arithmetic operations are allowed.
    
    The class of permissible transformations is the set of all \emph{linear transformations}: $\phi^\prime = \alpha\phi, \, \alpha > 0$.

\begin{example}[Ratio Scale]
	A typical example of ratio scale is length which can be expressed on different scales, e.g. meters or yards, which can all be mapped one into another via a similarity transformation. For example, to pass from kilometers ($\phi$) to miles ($\phi^\prime$), we have the following transformation $\phi^\prime = 0.62 \phi$. 
	
	Another example of ratio scale is the absolute temperature on the Kelvin scale where there is a zero element, which represents the absence of any thermal motion. 	

	\label{ex:ratio-scale}
\end{example}

\subsection{Admissible Statistical Operations}
\label{subsec:admissible-operations}

\citeauthor{Stevens1946} moved a step forward and linked the notion of scale with that of admissible statistical operations which can be carried out with that scale:
\begin{itemize}
    \item \emph{Nominal scale}: the only allowable operation is counting number of items in each class, that is, in statistical terms, mode and frequency.
    \item \emph{Ordinal scale}: besides the operations already allowed for nominal scales, median, quantiles, and percentiles are appropriate, since there is a notion of ordering.
    \item \emph{Interval scale} besides the operations already allowed for ordinal scales, mean and standard deviation are allowable since they depend just on sum and subtraction\footnote{Note that when we talk about admissible operations, we mean operations between items of the scale. So, for example, a mean involves summing items of the scale, e.g. temperature, and this is possible on an interval scale. The fact that a mean also requires a division by the number $N$ of items added together is not in contrast with saying that only addition and subtraction are allowed, since $N$ is not an item of the scale.}.
    \item \emph{Ratio scale}: besides the operations already allowed for interval scales, geometric and harmonic mean, as well as coefficient of variation, are allowable since they depend on multiplication and division. 
\end{itemize}

These prescriptions originated several debates over the decades. \citet[p.~751]{Lord1953} argued that ``since the numbers don't remember where they come from, they always behave the same way, regardless'' and so any operation should be allowed even on ``football numbers'', i.e. a nominal scale; \citet{Gaito1980} reinforced this argument by distinguishing between the realm of the measurement theory, where \citeauthor{Stevens1946}'s restrictions should apply, and the realm of the statistical theory, where these restrictions should not be applied, since other assumptions, such as normal distribution of the data, are those actually needed. \citet{TownsendAshby1984} replied back showing cases where performing operations inadmissible for a given scale of measurement may mislead the conclusions drawn by statistical tests. \citet{OBrien1985} discussed the type of errors introduced when using ordinal data for representing an underlying continuous variable, classifying them into pure transformation errors, pure categorization errors, pure grouping errors, and random measurement errors. \citet{VellemanWilkinson1993} summarized the previous debate and argumented that once you are in the numerical realm every operation is admissible among numbers. Recently, \citet{ScholtenBorsboom2009} made a case of flaws in the original \citeauthor{Lord1953}'s argument and, as a striking consequence, \citeauthor{Lord1953}'s experiment would not be a counterargument to \citeauthor{Stevens1946}'s restrictions but it would rather comply with them. In a very recent textbook, \citet{SauroLewis2016} firmly supported \citeauthor{Lord1953}'s view, at least in the case of ordinal scales, but with the caveat to not make claims on the outcomes of a statistical test that violate the underlying scale. So, for example, if you are on ordinal scale and you detected a significant effect using a test which would require a ratio scale, you should not claim that that effect is twice as big as another effect but just that it is significant.

\subsection{Meaningfulness}
\label{subsec:meaningfulness}

The above observation brings the debate back to the core issue of what we should pay attention to. Indeed, both \citet{Hand1996} and \citet{Michell1986,Michell1990} argued that the problem is not what operations you can perform with numbers but what kind of \emph{inference} you wish to make from those operations and how much such inference has to be indicative of what actually happens among real world objects. Already \citet[pp.~99-100]{AdamsEtAl1965} explicitly stated that 
\begin{quote}
Statistical operations on measurements of a given scale are not appropriate or inappropriate \emph{per se} but only relative to the kinds of statements made about them. The criterion of \emph{appropriateness} for a statement about a statistical operation is that the statement be \emph{empirically meaningful} in the sense that its truth or falsity must be invariant under permissible transformations of the underlying scale
\end{quote}

These statements opened the way to the development of a full (formal) theory of \emph{meaningfulness}~\cite{FalmagneNarens1983,Narens2002,Roberts1985}, which is a central concept to clearly shape and define the questions discussed above: according to the adopted measurement scales, what processing, manipulation, and analyses can be conducted and what can we tell about the conclusions drawn from such processing?

Note that the statement ``A mouse weights more than an elephant'' is meaningful even if it is clearly false; indeed, its truth value, i.e. false, does not change whatever weight scale you use (kilograms, pounds, and so on). Therefore, as anticipated above, meaningfulness is a distinct concept from the one of truth of a statement and it is somehow close to the notion of invariance in geometry, since the truth value of a statement stays the same independently from the permissible scales used to express it.

\begin{example}[Meaningfulness for a Nominal Scale -- Example~\ref{ex:nominal-scale} continued]
	Suppose  that we observe a set of 10 people, where 5 people are Spanish, 3 German, 1 Greek, and 1 Italian. According to $\phi$ we would have $P = [1\; 1\; 1\; 1\; 1\; 4\; 4\; 4\; 3\; 2]$ while according to $\phi^\prime$ we would have $P^\prime = [17\; 17\; 17\; 17\; 17\; 13\; 13\; 13\; {-}10\; 23]$. In both cases, the statement ``Most people come from Spain'' is meaningful since, if we compute the mode of the values, it is $1$ in the case of $\phi$ and $17$ in the case of $\phi^\prime$ which both correspond to Spain. On the other hand, the statement ``The lowest quartile consist of Spanish people'' is not meaningful, since it is true with $1$ corresponding to Spain in the case of $\phi$ but is is false with $13$ corresponding to Germany in the case of $\phi^\prime$. Indeed, the first statement about the mode involves just counting, which is an allowable operation for a nominal scale, while the second statement about the lowest quartile requires a notion of ordering not present in a nominal scale.
	\label{ex:nominal-scale-meaningfulness}
\end{example}

\begin{example}[Meaningfulness for an Ordinal Scale -- Example~\ref{ex:ordinal-scale} continued]
	Suppose that we have two wineries $X$ and $Y$. The first winery $W_X$ produced five bottles as follows: \emph{extra brut}, \emph{extra brut}, \emph{brut}, \emph{extra dry}, and \emph{sec}; the second one $W_Y$ produced five bottles as follows: \emph{pas dos\'{e}}, \emph{pas dos\'{e}},  \emph{pas dos\'{e}}, \emph{brut}, and \emph{demi-sec}. Therefore, according to the scale $\phi$, we have  $\phi(W_X) = [6\; 6\; 12\; 17\; 32]$ and $\phi(W_Y) = [3\; 3\; 3\; 12\; 50]$; while according to the scale $\mathrm{\phi^\prime}$, we have $\phi^\prime(W_X) = [36\; 36\; 144\; 289\; 1024]$ and $\phi^\prime(W_Y) = [9\; 9\; 9\; 144\; 2500]$. The statement ``The median of the first winery is greater than the one of the second winery'' is meaningful since $12 > 3$ according to $\phi$ is true as well as $144 > 9$ according to $\phi^\prime$; so we could safely say that the first winery produces a little more \emph{brut}-like wines than the second one, focusing on a more standard product. On the other hand, the statement "The average of the first winery is greater than the one of the second winery'' is not meaningful since $14.6 > 14.2$ according to $\phi$ is true but $305.8 > 534.2$ according to $\phi^\prime$ is false, which would lead us to draw basically opposite conclusions based on the scale we use. Indeed, the first statement about the median involves just the notion of ordering which is allowable on an ordinal scale, while the second statement about the average requires to sum values, which is not an allowable operation.
	\label{ex:ordinal-scale-meaningfulness}
\end{example}

\begin{example}[Meaningfulness for an Interval Scale -- Example~\ref{ex:interval-scale} continued]
	The statement `Today the difference in temperature between Rome and Oslo is twice as high as it was one month ago'' is meaningful. Indeed,  if, on the Celsius scale, the temperature today in Rome is $20$ $^\circ$C and in Oslo is $10$ $^\circ$C while one month ago it was $12$ $^\circ$C and $7$ $^\circ$C, leading to $20 - 10 = 10$ which is twice as  $12 - 7 = 5$, on the Fahrenheit scale we would have $68 - 50 = 18$ which is twice as $53.6 - 44.6 = 9$.

	Suppose now that we have recorded two sets of temperatures from Paris and Rome: $T_P^C = [2\; 2\; 4\; 8\; 36]$ and $T_R^C = [1\; 2\; 4\; 15\; 34]$ in Celsius degrees and, the same, $T_P^F = [35.6\; 35.6\; 39.2\; 46.4\; 96.8]$ and $T_R^F = [33.8\; 35.6\; 39.2\; 59.0\; 93.2]$ in Fahrenheit degrees. 
	
	The statement ``The median temperature in Paris is the same as in Rome'' is meaningful, since $4 = 4$ in Celsius degrees and $39.2 = 39.2$ in Fahrenheit degrees; this is due to the fact that interval scales are also ordinal and quantiles are an allowable operation on ordinal scales. 
	
	The statement ``The mean temperature in Paris is less than in Rome'' is meaningful as well, since $10.4 < 11.2$ in Celsius degrees and $50.72 < 52.16$ in Fahrenheit degrees; this is due to the fact that addition and subtraction are allowable operations on an interval scale and, as a consequence, mean is invariant to affine transformations. Indeed, let $X = \{x_1, x_2, \ldots, x_n\}$ and $Y = \{y_1, y_2, \ldots, y_n\}$ be two set of values on an interval scale; it holds that  
	\begin{equation*}
		\begin{gathered}
			\frac{1}{n}\sum_{i=1}^n \phi^\prime(x_i) > \frac{1}{n}\sum_{i=1}^n \phi^\prime(y_i) \; \Leftrightarrow \;
			\frac{1}{n}\sum_{i=1}^n \big[\alpha\phi(x_i) + \beta\big] > \frac{1}{n}\sum_{i=1}^n \big[\alpha\phi(y_i) + \beta\big] \; \Leftrightarrow \\
			\alpha\Bigg(\frac{1}{n}\sum_{i=1}^n \phi(x_i)\Bigg) + \beta > \alpha\Bigg(\frac{1}{n}\sum_{i=1}^n \phi(y_i)\Bigg) + \beta \; \Leftrightarrow \;
			\frac{1}{n}\sum_{i=1}^n \phi(x_i) > \frac{1}{n}\sum_{i=1}^n \phi(y_i)
		\end{gathered}
	\end{equation*}
	Therefore, the statement ``The mean of $X$ is greater than the mean of $Y$'' is always meaningful.

	Finally, the statement ``The geometric mean of temperature in Paris is greater than in Rome'' is not meaningful, since $5.40 > 5.27$ in Celsius degrees and $46.74 < 48.17$ in Fahrenheit degrees; this is due to the fact that the geometric mean involves the multiplication and division of values, which is not a permitted operation on an interval scale. 

	Also note that we may be tempted to compare the results of the arithmetic mean with those of the geometric mean to gain ``more insights''. For example, we might observe that the arithmetic mean in Paris is less than in Rome -- $10.4 < 11.2$ in Celsius degrees --  but the opposite is true when we consider the geometric mean -- $5.40 > 5.27$ in Celsius degrees. We might thus highlight that this due to the fact that the first (and lowest) value $2$ in Paris is double than $1$ in Rome and that the geometric mean rewards gains at lowest values; on the other hand, the arithmetic mean rewards gains at higher values and thus $8$ in Paris is (almost) half than $15$ in Rome and it contributes less. However, while the explanation why the geometric mean may differ from the arithmetic one is surely credible, the issue here is that the geometric mean could not be relied upon, as well as conclusions drawn from it, since it is based on operations not allowed on an interval scale; indeed, if we consider exactly the same temperatures just on the Fahrenheit scale, we would reach opposite conclusions.
	\label{ex:interval-scale-meaningfulness}
\end{example}

\begin{example}[Meaningfulness for a Ratio Scale -- Example~\ref{ex:ratio-scale} continued]
	If the air distance between Rome and Padua is (about) $400$ kilometers and the air distance among Rome and Oslo is (about) $2{,}000$ kilometers, the statement ``Rome and Oslo are five times as distant as Rome and Padua'' is meaningful, even expressed in miles, since $248.54 \backsim 5 \cdot 1{,}242.74$.

	On the Kelvin scale for temperature, it does make sense to say that a thing is twice as hot as another thing if, for example, the first one is $273$ K (almost $0$ $^\circ$C, $32$ $^\circ$F) and the second one is $546$ K (almost $273$ $^\circ$C, $523.4$ $^\circ$F); you can note, however, how this statement does not hold if we consider Celsius and Fahrenheit degrees, since $\frac{0}{32} = 0$ while $\frac{273}{523.4} = 0.52$ (and none of them is exactly twice).

	Finally, let us show that the statement ``The geometric mean of $X$ is greater than the geometric mean of $Y$'' is always meaningful. Indeed, let $X = \{x_1, x_2, \ldots, x_n\}$ and $Y = \{y_1, y_2, \ldots, y_n\}$ be two set of values on a ratio scale; it holds that  
	\begin{equation*}
		\begin{gathered}
			\sqrt[n]{\prod_{i=1}^n \phi^\prime(x_i)} > \sqrt[n]{\prod_{i=1}^n \phi^\prime(y_i)} \; \Leftrightarrow \;
			\sqrt[n]{\prod_{i=1}^n \alpha\phi(x_i)} > \sqrt[n]{\prod_{i=1}^n \alpha\phi(y_i)} \; \Leftrightarrow \\
			\alpha\sqrt[n]{\prod_{i=1}^n \phi(x_i)} > \alpha\sqrt[n]{\prod_{i=1}^n \phi(y_i)}  \; \Leftrightarrow \;
			\sqrt[n]{\prod_{i=1}^n \phi(x_i)} > \sqrt[n]{\prod_{i=1}^n \phi(y_i)}
		\end{gathered}
	\end{equation*}	
	\label{ex:ratio-scale-meaningfulness}
\end{example}

\subsection{Statistical Significance Testing}
\label{subsec:significance-test}

\citet{Siegel1956} and \citet{Senders1958} have discussed the implications of \citeauthor{Stevens1946}' classification and permissible operations in the case of statistical inference and parametric and nonparametric statistical significance tests. We consider the following tests:
\begin{itemize}
	\item \textbf{Sign Test}~\cite{GibbonsChakraborti2011} is a non parametric test which looks at the signs of the differences among two paired samples $x_i$ and $y_i$; the null hypothesis is that the median of the differences is zero. 
	
	The sign test requires samples to be on an \emph{ordinal scale}, since it needs to determine the sign of their difference or, equivalently, which one is greater. Note that the sign test discards the tied samples, i.e. when $x_i = y_i$.
	
	\item \textbf{Wilcoxon Rank Sum Test} (or Mann-Whitney U Test)~\cite{Wilcoxon1945,GibbonsChakraborti2011} is a non parametric test which looks at the ranks of two paired samples $x_i$ and $y_i$; the null hypothesis is that the two samples have the same median. 
	
	The Wilcoxon rank sum test requires samples to be on an \emph{ordinal scale}, since it needs to order them for determining their rank.
	
	\item \textbf{Wilcoxon Signed Rank Test}~\cite{Wilcoxon1945,GibbonsChakraborti2011} is a non parametric test which looks at the signs and ranks of the differences among two paired samples $x_i$ and $y_i$; the null hypothesis is that the median of the differences is zero. 
	
	The Wilcoxon signed rank test requires samples to be on an \emph{interval scale}, since it regards the ranks of the differences, for which intervals must be equi-spaced.
Note that the Wilcoxon signed rank test  discards the tied samples, i.e. when $x_i = y_i$.

	\item \textbf{Student's t Test}~\cite{Student1908} is a parametric test for the null hypothesis that two paired samples $x_i$ and $y_i$ come from a normal distribution with same mean and unknown variance. 
	
	The Student's t test requires samples to be on an \emph{interval scale}, since it needs to compute means and variances.
	
	\item \textbf{\ac{ANOVA}}~\cite{Fisher1925,KutnerEtAl2005} is a parametric test for the null hypothesis that $q$ samples come from a normal distribution with same mean and unknown variance.
		
	\ac{ANOVA} requires samples to be on an \emph{interval scale}, since it needs to compute means and variances.

	\item \textbf{Kruskal-Wallis Test}~\cite{KruskalWallis1952,GibbonsChakraborti2011} is a nonparametric version of the one-way \ac{ANOVA} for the null hypothesis that $q$ samples come from a distribution with same median. It is based on the ranks of the different samples and it can be considered as an extension of the Wilcoxon rank sum test to the comparison of multiple systems at the same time.

	The Kruskal-Wallis test requires samples to be on an \emph{ordinal scale}, since it needs to order them for determining their rank.
				
	\item \textbf{Friedman Test}~\cite{Friedman1937,Friedman1939,GibbonsChakraborti2011} is a nonparametric version of the two-way \ac{ANOVA} for the null hypothesis that the effects of the $q$ samples are the same. It is based on the ranks of the different samples.
	
	The Friedman test requires samples to be on an \emph{ordinal scale}, since it needs to order them for determining their rank.
\end{itemize}

As in the case of \citeauthor{Stevens1946}' permissible operations, defining which statistical significance tests should be permitted on the basis of the scale properties of the investigated variables raised a lot of discussion and controversy. \citet{Anderson1961}, along the line of reasoning of \citeauthor{Lord1953}, argued that statistical significance tests should be used regardless of scale limitations. \citet{Gardner1975} summarizes much of the discussion up to that point, leaning towards not worrying too much about scale assumptions, and suggests that, if and when lack of compliance to measurement scale requirements biases the outcomes of significance tests, transformations can be applied to turn ordinal scales into more interval-like ones such as, for example, averaging the ranks of each score, as proposed by~\citet{Gaito1959}, or using a more complex set of rules, as developed by~\citet{AbelsonTukey1959}. \citet{WareBenson1975} replied to \citeauthor{Gardner1975}'s positions by further revising the pro and con arguments and concluding that parametric significance tests should be used only when dealing with interval and ratio scales while, in the case of ordinal scales, nonparametric significance tests should be adopted. \citet{TownsendAshby1984} further investigated the issue, highlighting some serious pitfalls you may fall in, when ignoring the scale assumptions. 

We can summarise the discussion with the conclusions of \citet[p.~391]{Marcus-Roberts1987}:
\begin{quote}
	The appropriateness of a statistical test of a hypothesis is just a matter of whether the population and sampling procedure satisfy the appropriate statistical model, and is not influenced by the properties of the measurement scale used. However, if we want to draw conclusions about a population which say something basic about the population, rather than something which is an accident of the particular scale of measurement used, then we should only test meaningful hypotheses, and meaningfulness is determined by the properties of the measurement scale in connection with the distribution of the population. 
\end{quote}
and~\citet[p.~471]{Hand1996}
\begin{quote}
	Restrictions on statistical operations arising from scale type are more important in model fitting and hypothesis testing contexts than in model generation or hypothesis generation contexts.
\end{quote}

\section{Measurement Issues in Information Retrieval}
\label{sec:measurement-ir}

\subsection{Why does Studying the Scale Properties of \ac{IR} Evaluation Measures Matter?}
\label{subsec:why-interval-IR}

Let us start our discussion by considering a not-exhaustive list of core \ac{IR} areas where scales may matter.

The most common and basic operation we perform to understand whether a system $A$ is better than a system $B$ is to average their performance over a set of topics and compare these aggregate scores. According to the discussion so far this leads to \emph{meaningful statements} only if \ac{IR} evaluation measures are, at least, interval scales.

Topic difficulty~\cite{CarmelYomTov2010} is another central theme in \ac{IR} because of its importance for adapting the behaviour of a system to the topic at hand. \citet{Voorhees2004,Voorhees2005b}, in the TREC Robust tracks, explored how to evaluate and compare systems designed to deal with difficult topics and proposed to use the geometric mean, instead of the arithmetic one, for \acf{AP}~\cite{BuckleyVoorhees2005}. However, the use of a geometric mean further raises the requirements for the evaluation measures, even calling for a ratio scale.

Statistical significance testing has a long story of adoption and investigation in \ac{IR}, from the early uses of t-test reported by~\citet{SaltonLesk1968}, to the discussion on the compliance with the distribution assumptions of significance tests by~\citet{vanRijsbergen1979}, to advocating for a more wide-spread adoption of different types of significance tests by~\citet{Hull1993,Savoy1997,Carterette2012,Sakai2014b}, to surveys on the current state of adoption of significance tests by~\citet{Sakai2016d}. Again, drawing \emph{meaningful inference} depends on the appropriate use of parametric or nonparametric tests in accordance with the scale properties of the adopted \ac{IR} evaluation measures.

Several authors have proposed the use of score transformation and standardisation techniques, such as z-score by~\citet{WebberEtAl2008b} and other types of linear (and non-linear) transformations by~\citet{Sakai2016c,UrbanoEtAl2019b}, in order to compare performance across collections and to reduce the impact of few topics skewing the performance distribution. However, in order to ensure \emph{meaningful conclusions} from these transformation, at least an interval scale would be required.

Despite so many aspects of \ac{IR} evaluation which can be affected by the scale properties of evaluation measures and despite the deep scrutiny that the above techniques have received over the years, there has been much less attention to the implications of the scale assumptions on them.

\citet{Robertson2006} was the first to discuss the admissibility of the use of the geometric mean from the \citeauthor{Stevens1946}'s perspective in the context of the \acs{TREC} Robust track. In particular, \citeauthor{Robertson2006} focused on the fact that \acf{MAP} and \acf{GMAP} may lead to different conclusions -- e.g. blind feedback is beneficial according to \ac{MAP} but detrimental according to \ac{GMAP} -- and which of them may hold more \emph{(intrinsic) validity}. In this respect, \citet[p.~80]{Robertson2006} observed that
\begin{quote}
If the interval assumption is not valid for the original measure nor for any specific transformation of it, then \emph{any} monotonic transformation of the measure is \emph{just as good a measure} as the untransformed version. If we believe that the interval assumption is good for the original measure, that would give the arithmetic mean some validity over and above the means of transformed versions. If, however, we believe that the interval assumption might be good for one of the transformed versions, we should perhaps favour the transformed version over the original. But if there is no particular reason to believe the interval assumption for any version, then all versions are equally valid. If they differ, it is because they measure different things.
\end{quote}
Since both \ac{AP} and the log-transformation of \ac{AP} (implied by the geometric mean) are not interval scales, \citeauthor{Robertson2006} concluded that no preference could be granted to \ac{MAP} or \ac{GMAP} in terms of (intrinsic) validity of their findings. In this way \citeauthor{Robertson2006} takes a neutral stance with respect to the debate on whether certain operations should be permitted or not on the basis of the scale properties. 

Note that \citeauthor{Robertson2006} somehow implicitly  indicates transformations as a possible means to turn a not-interval scale into an interval one, as also supported by~\citet{Gaito1959,AbelsonTukey1959}.

As a final remark, even if \citeauthor{Robertson2006} did not mention it explicitly, his reasoning seems to be loosely related the concept of \emph{meaningfulness} when he says [p.~80]
\begin{quote}
Good robustness would be indicated if the conclusions looked the same whatever transformation we used; if we found it easy to find transformations which would substantially change the conclusions, then we might infer that our conclusions are sensitive to the interval assumption, and that the different transformations measure different things in ways that may be important to us
\end{quote}
still keeping a neutral stance about what should or should not be done.

\citet{Fuhr2017} took a firm position and argued that \acf{MRR}~\cite{SinghalEtAl1997} should not be computed because:
\begin{enumerate*}
	\item in general, \acs{RR} is just an ordinal scale and, according to \citeauthor{Stevens1946} means cannot be computed for a ordinal scales;
	\item in particular, \acs{RR} has some counter-intuitive behaviour. 
\end{enumerate*}
On the other hand, \citet{Sakai2020} has recently disagreed with \citeauthor{Fuhr2017}:
\begin{enumerate*}
	\item in general, on the fact that  means should not be computed for an ordinal scale, using arguments similar to those discussed in Section~\ref{subsec:admissible-operations};
	\item in particular, on the use of \ac{RR} which \citeauthor{Sakai2020} finds quite useful from a practical point of view.
\end{enumerate*}

Whatever stance you wish to take about whether (or not) operations should be constrained by scale properties, from the discussion so far, it clearly emerges that \ac{IR} needs further and systematic investigation about the implications and impact of derogating from compliance with  scale properties. Moreover, most of the above discussion is just about averaging values and does not tackles the implications for statistical significance testing. Finally, and more importantly, we completely lack a thorough discussion on and any adoption of the notion of \emph{meaningfulness} in \ac{IR} and this is quite striking for a discipline so strongly rooted in experimentation and so much based on inference.

\subsection{A Formal Theory of Scale Properties for IR Evaluation Measures}
\label{subsec:formal-theory}

\citet{FerranteEtAl2015b,FerranteEtAl2017,FerranteEtAl2018b} leveraged the representational theory of measurement for developing a formal theory of \ac{IR} evaluation measures which allows us to determine the scale properties of an evaluation measure. In particular, they defined an \ac{ERS} for system runs and used two basic operations -- \emph{swap}, i.e. swapping a relevant with a not-relevant document in a ranking, and \emph{replacement}, i.e. substituting a relevant document with a not-relevant one -- to study how runs are ordered. In this way, they demonstrated that there  exists a \emph{partial order of runs} where, when runs are comparable, all the measures agree on the same way of ordering them; however, when runs are not comparable, measures may disagree on how to order them. By using properties of the partial orders and theorems from the representational theory of measurement, they were able to define an interval scale measure $\phi$ and to check whether there is any linear transformation between such measure $\phi$ and \ac{IR} evaluation measures, in order to determine if the latter are interval scales too.

In short,  \citeauthor{FerranteEtAl2018b} found that, for a single topic: 
\begin{itemize}
	\item set-based evaluation measures:
		\begin{itemize}
			\item binary relevance: precision, recall, F-measure are interval scales;
			\item multi-graded relevance: \acf{gP} and \acf{gR} are interval scales only if the relevance degrees are on a ratio scale;
		\end{itemize}
	\item rank-based evaluation measures:
		\begin{itemize}
			\item binary relevance: \acf{RBP}~\cite{MoffatZobel2008} is an interval scale only for $p = 1/2$; \acf{AP} is not an interval scale;
			\item multi-graded relevance: \acf{gRBP} is an interval scale only for $p = G/(G+1)$, where $G$ is the normalized smallest gap between the gain of two consecutive relevance degrees, and the relevance degrees themselves are on a ratio scale; \acf{DCG}~\cite{JarvelinKekalainen2002} and \acf{ERR}~\cite{ChapelleEtAl2009} are not interval scales.
		\end{itemize}
\end{itemize}

\citet{FerranteEtAl2018b} also studied what they called the \emph{induced total order}, i.e. pretending that runs in the \ac{ERS} are ordered by the actual values of a measure. Also in this case which is the most ``favourable'' to each measure, \citeauthor{FerranteEtAl2018b} have shown that \ac{AP}, \ac{RBP} with $p \neq 1/2$ (and its multi-graded version), \ac{DCG}, and \ac{ERR} are not interval scales, because their values are not equi-spaced.

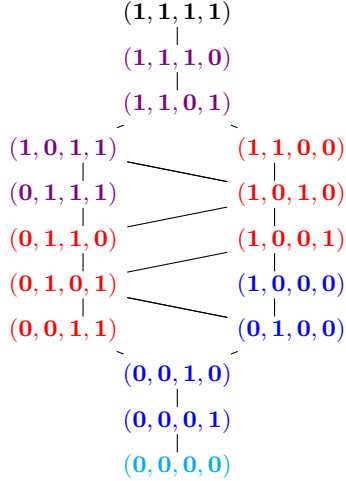
\begin{figure}[tb]
\centering
\begin{tikzpicture}[scale=0.4,font=\small,color= white,
  level 1/.style={sibling distance=5cm},
  level 2/.style={sibling distance=7.5cm},baseline=(current bounding box.center)]
  \node (root) {${\color{black}\mathbf{(1,1,1,1)}}$} 
    child{ node {$\mathbf{{\color{violet}(1,1,1,0)}}$}
       child{node{$\mathbf{{\color{violet}(1,1,0,1)}}$}
     	  child { node { $\mathbf{{\color{violet}(1,0,1,1)}}$} 
     	  	  child{ node{$\mathbf{{\color{violet}(0,1,1,1)}}$}
     	  	 	child{node{$\mathbf{{\color{red}(0,1,1,0)}}$}
     	  	 		child{node{$\mathbf{{\color{red}(0,1,0,1)}}$}
     	  	 			child{ node{$\mathbf{{\color{red}(0,0,1,1)}}$}
     	  	 				child{edge from parent[draw=none]}
     	  	 				child{ node{$\mathbf{{\color{blue}(0,0,1,0)}}$}
     	  	 					child{ node{$\mathbf{{\color{blue}(0,0,0,1)}}$}
     	  	 						child{node{$\mathbf{{\color{cyan}{\bf(0,0,0,0)}}}$}}}}}}}}}  
      	  child {node {$\mathbf{{\color{red}(1,1,0,0)}}$}
      	  	child{node{$\mathbf{{\color{red}(1,0,1,0)}}$}
      	  		child{node{$\mathbf{{\color{red}(1,0,0,1)}}$}
      	  			child{node{$\mathbf{{\color{blue}(1,0,0,0)}}$}
      	  				child{node{$\mathbf{{\color{blue}(0,1,0,0)}}$}
      	  				}}}}}}}

    ;
     \draw[-,black] ($(root)+(0cm,-4.5mm)$) -- ($(root-1)+(0cm,+4.5mm)$);
     \draw[-,black] ($(root-1)+(0cm,-4.5mm)$) -- ($(root-1-1)+(0cm,+4.5mm)$);
     \draw[-,black] (root-1-1) -- (root-1-1-2);
     \draw[-,black] (root-1-1) -- (root-1-1-1);
     \draw[-,black] ($(root-1-1-2)+(-0.55cm,-4.5mm)$) -- ($(root-1-1-2-1)+(-0.55cm,+4.5mm)$);
     \draw[-,black] ($(root-1-1-1-1-1-1)+(+0.65cm,-4.5mm)$) -- ($(root-1-1-1-1-1-1-1)+(+0.65cm,+4.5mm)$);
     \draw[-,black] (root-1-1-1-1-1-1-1) -- (root-1-1-1-1-1-1-1-2);
     \draw[-,black] ($(root-1-1-1-1-1-1-1-2)+(0cm,-4.5mm)$) -- ($(root-1-1-1-1-1-1-1-2-1)+(0cm,+4.5mm)$);
     \draw[-,black] ($(root-1-1-1-1-1-1-1-2-1)+(0cm,-4.5mm)$) -- ($(root-1-1-1-1-1-1-1-2-1-1)+(0cm,+4.5mm)$);
  \draw[-,black] (root-1-1-1) -- (root-1-1-2-1);
  \draw[-,black] (root-1-1-2-1) -- (root-1-1-1-1-1);
    \draw[-,black] (root-1-1-1-1-1-1) -- (root-1-1-2-1-1-1-1);
      \draw[-,black] (root-1-1-2-1-1) -- (root-1-1-1-1-1-1);
  \draw[-,black] (root-1-1-2-1-1-1-1) -- (root-1-1-1-1-1-1-1-2);
  \draw[-,black]
  ($(root-1-1-2-1-1)+(-0.55cm,+4.5mm)$) -- ($(root-1-1-2-1)+(-0.55cm,-4.5mm)$);
     \draw[-,black]
  ($(root-1-1-2-1-1-1)+(-0.55cm,+4.5mm)$) -- ($(root-1-1-2-1-1)+(-0.55cm,-4.5mm)$);
     \draw[-,black]
 ($ (root-1-1-2-1-1-1-1)+(-0.55cm,+4.5mm)$) -- ($(root-1-1-2-1-1-1)+(-0.55cm,-4.5mm)$);
     \draw[-,black]
  (root-1-1-2-1) -- (root-1-1-1);
       \draw[-,black]
  (root-1-1-2-1-1-1-1) -- (root-1-1-1-1-1-1);
         \draw[-,black]
  ($(root-1-1-1-1-1-1)+(+0.65cm,+4.5mm)$) -- ($(root-1-1-1-1-1)+(+0.65cm,-4.5mm)$);
         \draw[-,black]
  ($( root-1-1-1-1-1)+(+0.65cm,4.5mm)$) --  ($(root-1-1-1-1)+(+0.65cm,-4.5mm)$);
         \draw[-,black]
  ($(root-1-1-1-1)+(+0.65cm,+4.5mm)$) -- ($(root-1-1-1)+(+0.65cm,-4.5mm)$);
\end{tikzpicture}
	\caption{Hasse diagram showing the partial order of all the possible runs of length $4$. The different colours of the runs correspond to different total numbers of relevant retrieved documents.}
	\label{fig:hasse}
\end{figure}

Figure~\ref{fig:hasse} shows the \emph{Hasse diagram}~\cite{DaveyPriestley2002} which represents the partial order among all the runs of length  $N=4$. In the figure, vertices are runs while edges represent the direct predecessor relation that is, if $r\prec s$, i.e. $r$ and  $s$ are comparable, then $r$ is below $s$ in the diagram. Note that if $r$  and $s$ lie on the same horizontal level of the diagram, then they are incomparable; furthermore, elements on different levels may be incomparable as well. In the example $(1,1,0,1) \preceq (1,1,1,0)$, $(1,1,0,0) \preceq (1,1,1,0)$, and $(1,0,1,1) \preceq (1,1,1,0)$ are  all comparable; therefore, all IR measures agree on these runs and order them in the same way. On the other hand, $(1,1,0,0)$ and $(1,0,1,1)$ are not comparable, as well as $(1,1,0,0)$ and $ (0,1,1,1)$, and IR measures disagree on how to order them; as a consequence, measures will order these runs differently, producing different \ac{RoS}.

The difference in the \ac{RoS} produced by evaluation measures is what is studied when performing a correlation analysis among measures, e.g. by using Kendall's $\tau$~\cite{Kendall1948}; practical wisdom says that measures should be neither too much correlated -- otherwise it practically makes no difference using one or the other -- nor too few correlated -- otherwise it may be an indicator of some ``pathological'' behaviour of a measure. Indeed, each evaluation measure embodies a different user model~\cite{Carterette2011}, i.e. a different way in which the user interacts with the ranked result list and derives gain from the retrieved documents, and the differences between the \ac{RoS} produced by different evaluation measures, and as a consequence their Kendall's $\tau$, may be considered as the tangible manifestation of such different user models. Note that the work by~\citeauthor{FerranteEtAl2018b} provides a formal explanation of what originates differences in Kendall's $\tau$: for all the runs which are comparable in the Hasse diagram,  Kendall's $\tau$ between different measures is $1$, since all of them order these runs in the same way; for runs which are not comparable in the Hasse diagram,  Kendall's $\tau$ between different measures is less than $1$, since all of them order these runs differently; therefore, these not comparable runs are where user models differentiate themselves and can take a different stance.

\begin{figure}[tb]
	\centering
	\begin{subfigure}[t]{0.35\linewidth}
		\centering
		\includegraphics[height=5cm]{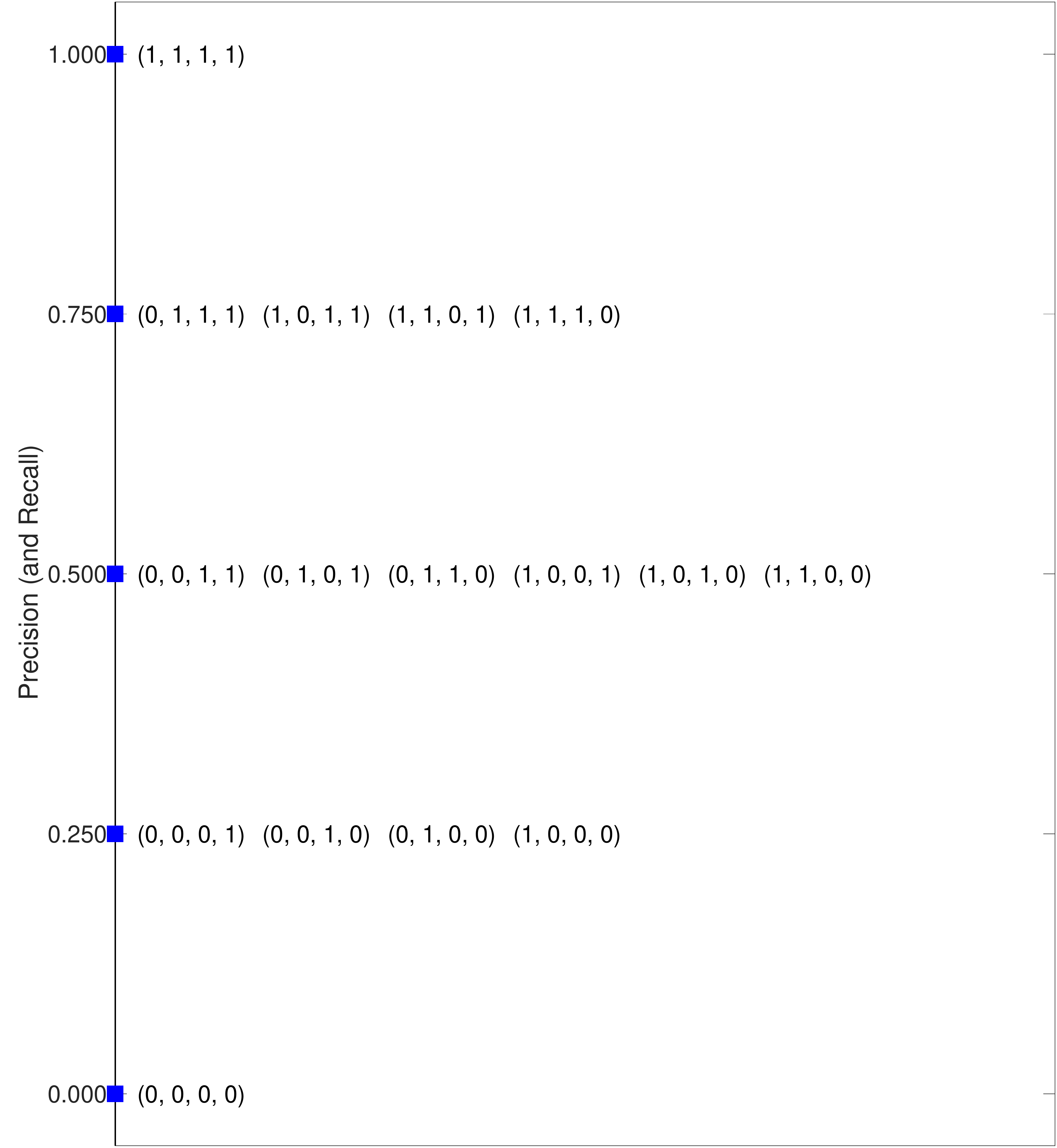}
		\caption{Precision (and Recall).}
		\label{fig:p_steps}
	\end{subfigure}\hfill	
	\begin{subfigure}[t]{0.15\linewidth}
		\centering
		\includegraphics[height=5cm]{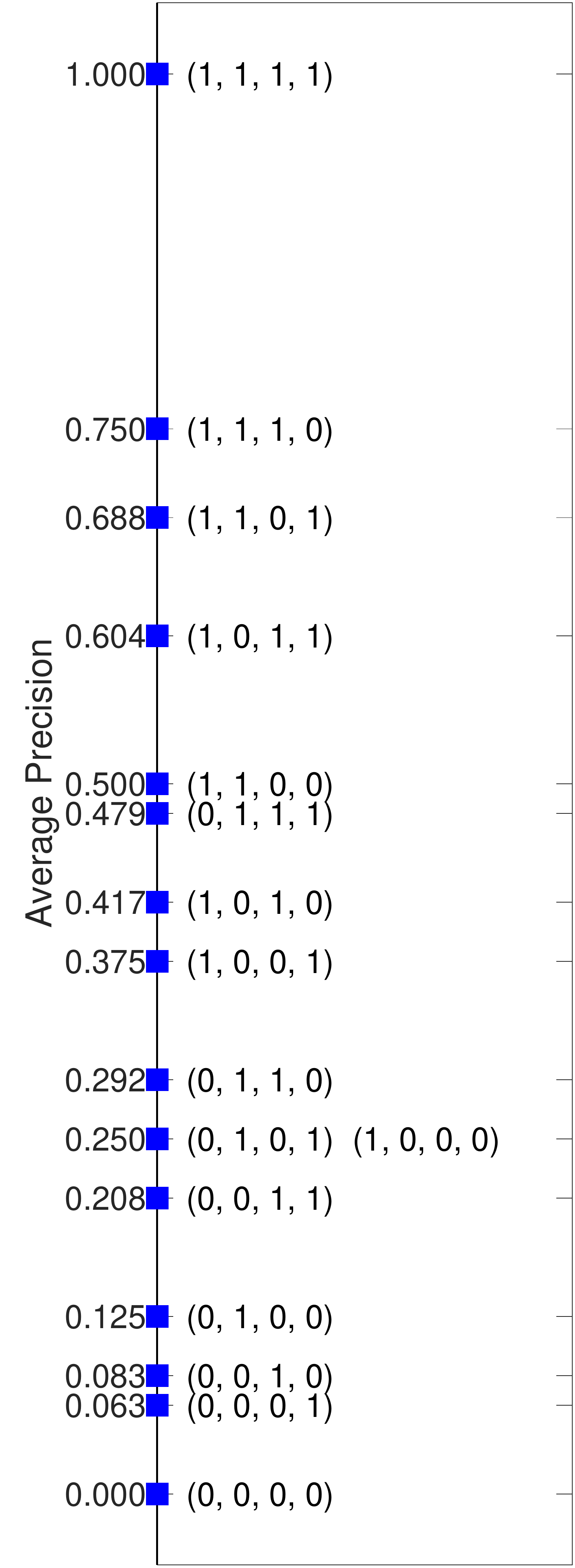}
		\caption{AP.}
		\label{fig:ap_steps}
	\end{subfigure}\hfill	
	\begin{subfigure}[t]{0.5\linewidth}
		\centering
		\includegraphics[height=5cm]{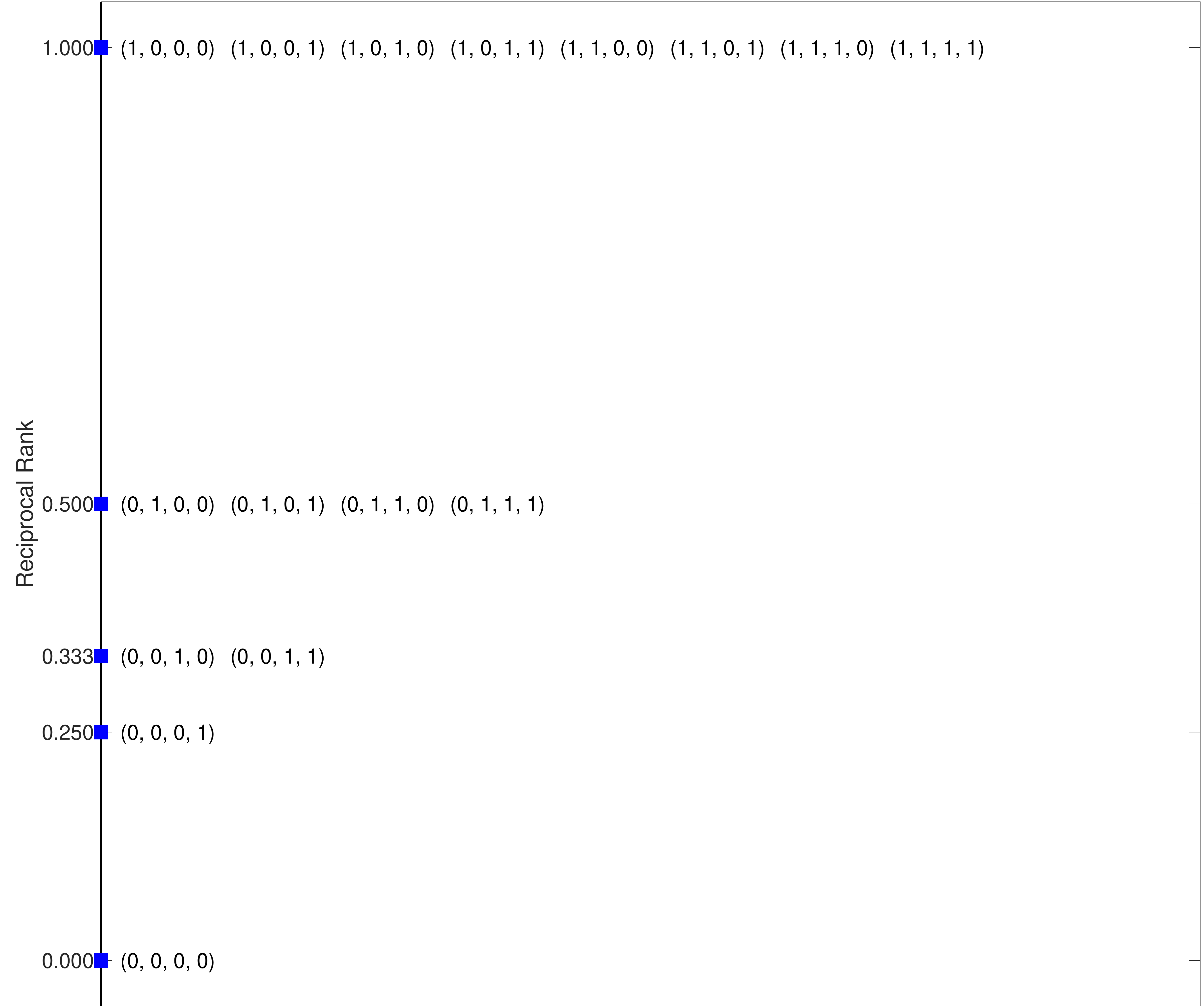}
		\caption{RR.}
		\label{fig:rr_steps}
	\end{subfigure} \\
	\begin{subfigure}[t]{0.20\linewidth}
		\centering
		\includegraphics[height=5cm]{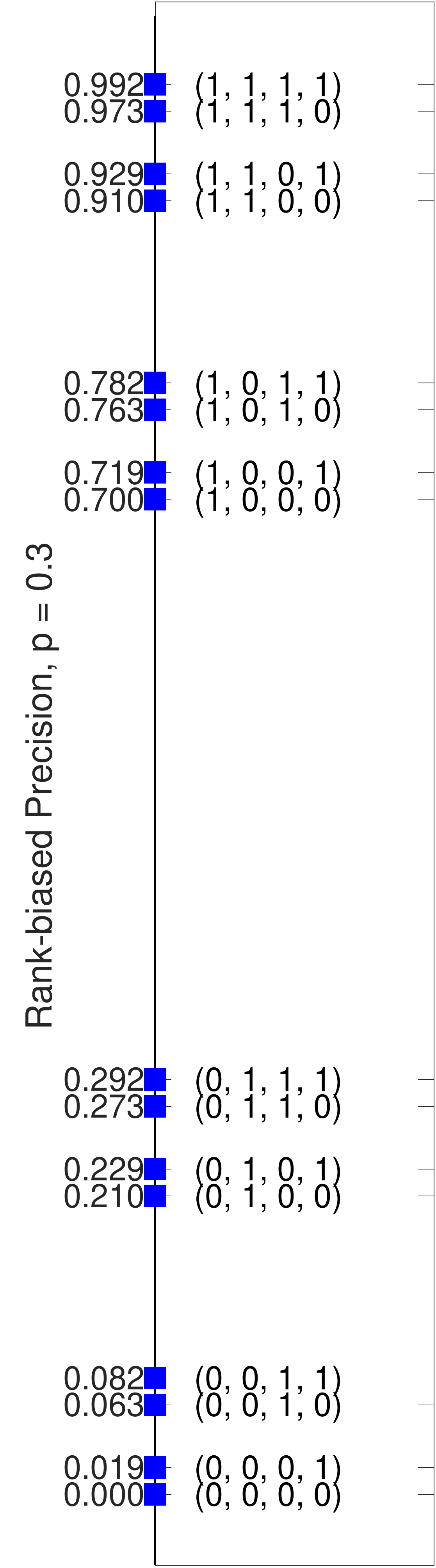}
		\caption{RBP, $p=0.3$.}
		\label{fig:rbp02_steps}
	\end{subfigure}\hfill	
	\begin{subfigure}[t]{0.20\linewidth}
		\centering
		\includegraphics[height=5cm]{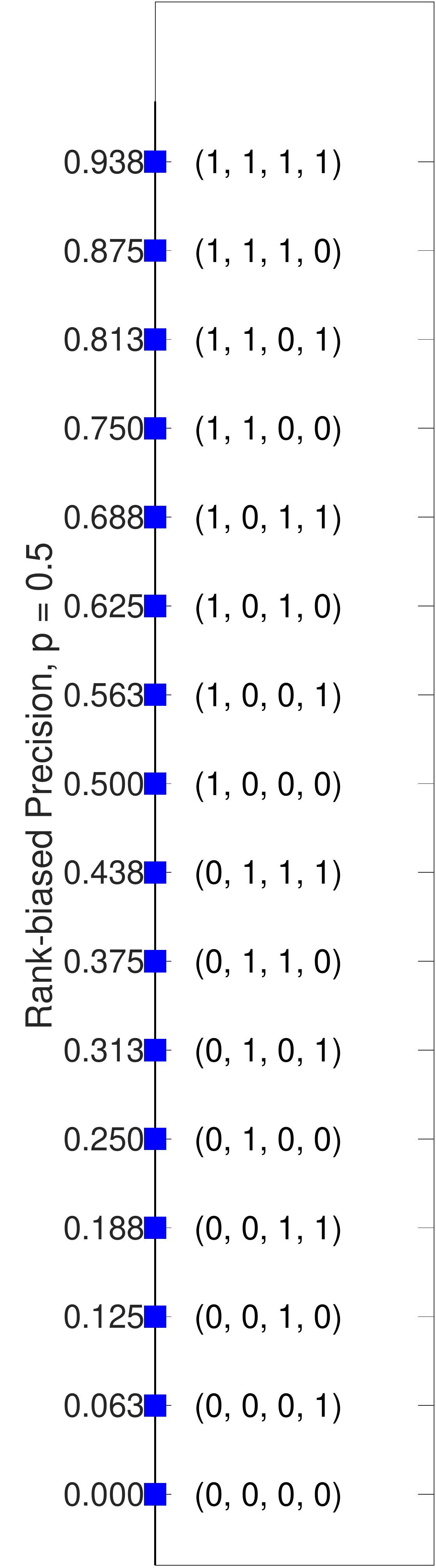}
		\caption{RBP, $p=0.5$.}
		\label{fig:rbp05_steps}
	\end{subfigure}\hfill	
	\begin{subfigure}[t]{0.20\linewidth}
		\centering
		\includegraphics[height=5cm]{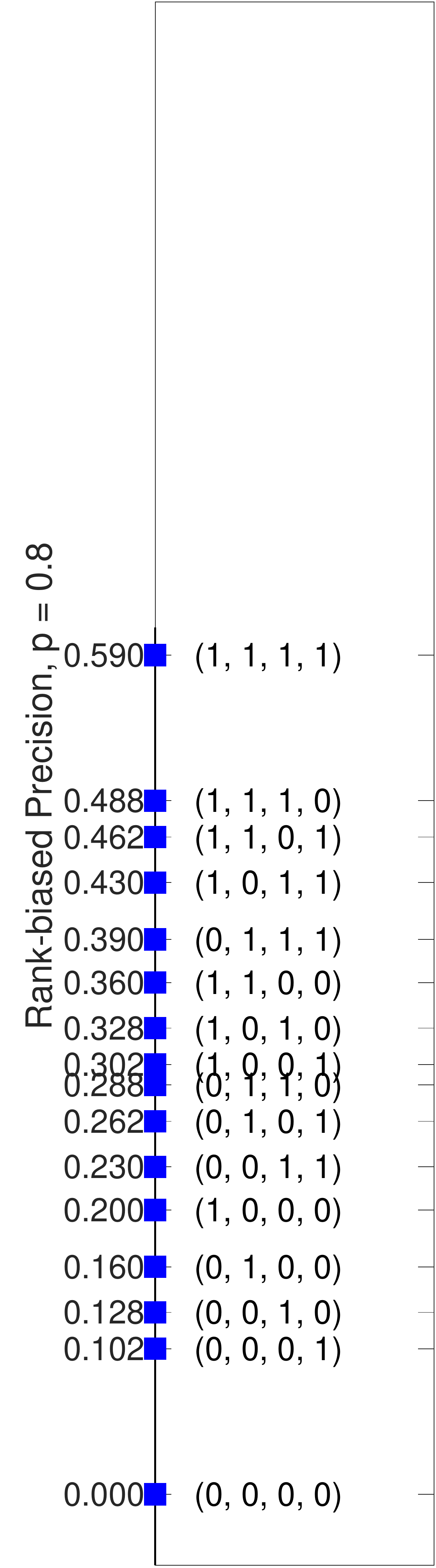}
		\caption{RBP, $p=0.8$.}
		\label{fig:rbp08_steps}
	\end{subfigure}\hfill	
	\begin{subfigure}[t]{0.20\linewidth}
		\centering
		\includegraphics[height=5cm]{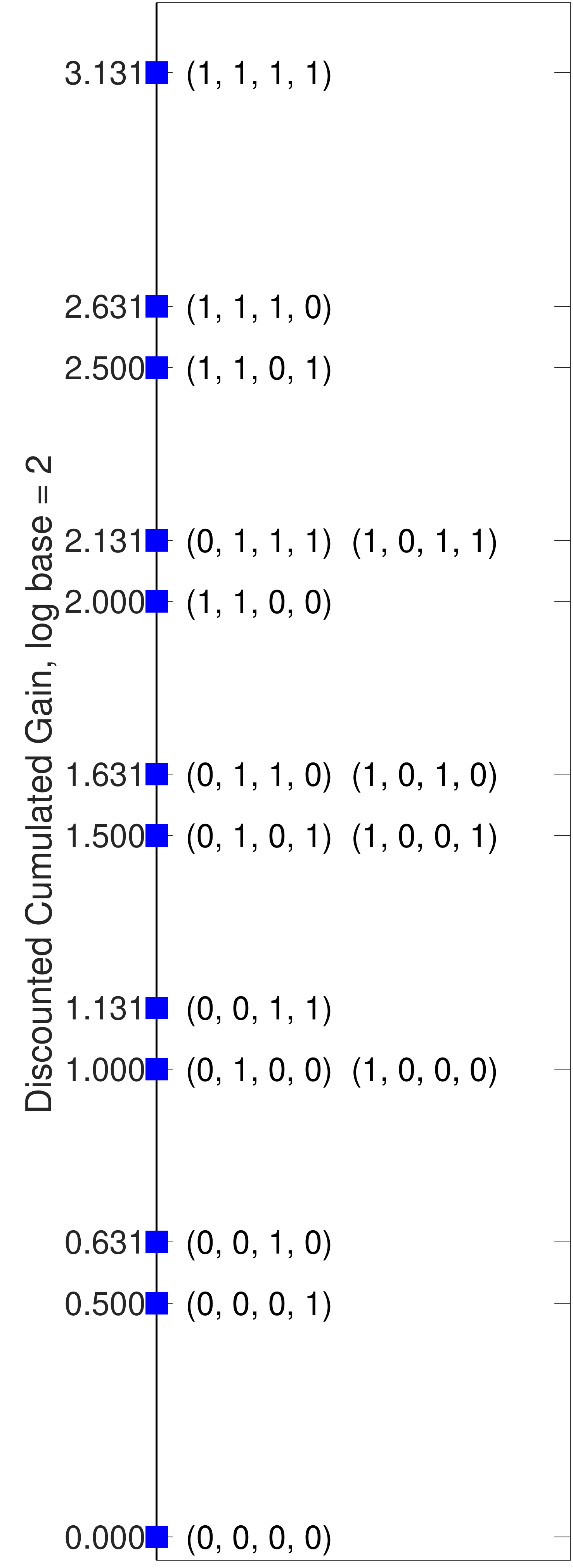}
		\caption{DCG, log base $2$.}
		\label{fig:dcg_steps}
	\end{subfigure}
	\caption{Ordering and spacing of the runs of Figure~\ref{fig:hasse} by different evaluation measures. Each blue square corresponds to a score of a given measure. On the right of the square, the run corresponding to that score is reported; in case of tied runs, i.e. runs for which the measures produces the same score, they are all listed on the right of the square.}
	\label{fig:measure_steps}
\end{figure}

However, these differences in the \ac{RoS} are not causing \ac{IR} evaluation measures to not be interval scales; they would just mean that \ac{IR} evaluation measures are different scales. The real problem with \ac{IR} evaluation measures is that their scores are not \emph{equi-spaced} and thus they cannot be interval scales, as explained in Section~\ref{subsec:measurement-scales}. This issue is depicted in Figure~\ref{fig:measure_steps} which shows how different measures -- namely, Precision (and Recall\footnote{Note that in this specific case, since the length of the run $N=4$ and the recall base $RB=4$ are the same, Precision and Recall yield to the same scores.}), \ac{AP}, \ac{RR}, \ac{RBP} with $p \in \{0.3, 0.5, 0.8\}$, and \ac{DCG} with log base $2$ -- order and space the runs shown in the Hasse diagram of Figure~\ref{fig:hasse}. 

We can observe that only Precision (Recall) and \ac{RBP} with $p=0.5$ produce equi-spaced values, while all the other measures violate this assumption, required to obtain an interval scale; in other terms, Figure~\ref{fig:measure_steps} visually represents the issue found by \citet{FerranteEtAl2018b} even when using the induced total order. We can also note that all the measures agree only on the common comparable runs -- i.e. $(0, 0, 0, 0) \preceq (0, 0, 0, 1) \preceq (0, 0, 1, 0)$ and  $(1, 1, 0, 1) \preceq (1, 1, 1, 0) \preceq (1, 1, 1, 1)$ -- but, as soon as incomparable runs come into play, they start to disagree on how to order them. Finally, looking at Figure~\ref{fig:measure_steps} we can notice how \ac{IR} measures behave differently in violating the equi-spacing assumption. \ac{RBP} with $p \in \{0.3, 0.8\}$ and \ac{DCG} follows a somehow regular pattern, where scores are not equi-spaced but they are in some way evenly clustered and they are symmetric if you fold the figure along its middle horizontal axis; on the other hand, \ac{AP} and \ac{RR} follow a much more irregular and not symmetric pattern. 

We can also note how these measures spread values in their range differently. Precision (and Recall) and \ac{DCG} spread their values all over the possible range while this is not always the case with \ac{RBP}. Indeed, \ac{RBP} assumes runs of infinite length and normalizes by the $\frac{1}{1-p}$ factor. However, we deal with runs of limited length and the $\frac{1}{1-p}$ factor is an overestimation, the bigger the overestimation the bigger is the value of $p$ and the smaller is the length of the run -- this is more clearly visible in the case of \ac{RBP} with $p=0.8$ in Figure~\ref{fig:rbp08_steps}. Finally, \ac{AP}, \ac{RBP} with $p=0.3$, and \ac{RR}, i.e. those measures farther from being interval scales, leave large portions of their possible range completely unused. In particular, \ac{AP} leaves one quarter of its range unused, in the top part roughly corresponding to the first quartile of the possible values; \ac{RR} leaves one half of its range unused, in the top part roughly corresponding to the first and second quartiles of the possible values; and, finally, \ac{RBP} with $p=0.3$ leaves half of its range empty, in the middle part roughly corresponding to the second and third quartile of the possible values.

Why does it matter how much equi-spaced the values are and how they are spread over their range?  Consider a random variable $X$ that takes values in the set $\{0,1,2,4,13\}$. Even if all these five values can be obtained with equal probability, i.e. the random variable is uniform, the mean and the median of the variable differ, being the mean equal to $4$ and the median to $2$. This shows how the lack of equi-spacing causes some sort of ``imbalance'' even in the case of a uniform variable, which may be an undesirable situation from the measurement point of view, at least if not explicitly considered and accounted for.  Furthermore, when we compute $\mathbb{P}[X\in(x-\varepsilon,x+\varepsilon)]$, i.e. the probability that the value of $X$ is equal to $x$ with an error of at most $2 \varepsilon$, this function is not constant all over the range but it is assumer greater for values around  $\{0,1,2\}$ than for those around $\{4,13\}$. As a consequence, a similar accuracy in approximating the value of $X$ produces a different precision in the measurement depending on the value $x$ that we are considering. Note that in the present toy model it happens for $\varepsilon \ge 1$, but a suitable modification of the
present model can produce the same behaviour for any $\varepsilon>0$ set in advance.

As a further example, let us consider a measure with a limited range of equi-spaced values. If we draw a set of random values taken from this range and consider its arithmetic mean, by the law of large numbers, we have that this mean converges to the middle point of the range interval. This property is independent from the distance among the subsequent values, i.e. the unit of measurement chosen. So we can use such a procedure -- the convergence of the mean towards the middle of the range -- in order to ``calibrate'' the measuring instrument, independently from the specific unit of measurement chosen. This is no more possible if we have values which are not equi-spaced.

\begin{example}[Effect of \ac{RR} not being equi-spaced]
	Let us assume that we have two queries and two systems. System A returns the first relevant document at ranks 1 and 4, respectively, while system B finds the relevant answers in both cases at rank 2. Computing the \ac{MRR} of the two systems, i.e. the average value of the \ac{RR}, we get MRR(A)=$\frac{1}{2}(1/1+1/4)=0.625$, while MRR(B)=$0.5$, telling us that system A is better than B. However, if instead of reciprocal rank, we regard the ranks themselves, we have equi-spaced values forming an interval scale (actually, even a ratio scale). In our example, system A finds the first relevant item on average at rank 2.5, which is worse than the average rank 2 of system B -- so we would get the opposite finding when we use a scale still based on the rank of relevant documents but properly equi-spaced.
\label{ex:rr-not-equispaced}
\end{example}


\begin{table}[h]
    \centering
    \begin{tabular}{c|c|c|l||c|c|c|l}
    System & $Q_1$ & $Q_2$ & \multicolumn{1}{c||}{AP} &  System & $Q_1$ & $Q_2$ & \multicolumn{1}{c}{AP}  \\ \hline
A & $(1, 1, 1, 0)$ & $(1, 0, 0, 1)$ & 0.5625 & C & $(0, 0, 1, 0)$ & $(0, 1, 0, 1)$ & 0.1665\\
B & $(1, 1, 0, 1)$ & $(1, 0, 1, 0)$ & 0.5575 & D & $(0, 0, 0, 1)$ &$(0, 1, 1, 0)$ & 0.1770
    \end{tabular}
    \caption{Example for \ac{AP} not being equi-spaced}
    \label{tab:APexample}
\end{table}

\begin{example}[Effect of \ac{AP} not being equi-spaced]
Table~\ref{tab:APexample} shows an example of two system pairs (A,B) and (C,D) and two queries, for which we compute AP values. In the first case, \ac{AP} will say that A performs better than B, while in the second case, C is worse than D. Why is this effect related to AP not being on an interval scale? Because in both the examples, the runs retrieved by the two systems for a given topic have the same relevance degrees in the first two positions and just a swap of a relevant with a not-relevant document in the last two positions. So, to the same loss of relevance for the swap in the last two positions, still keeping the same relevance in the first two positions, AP ``reacts'' in one case telling us that system A is better than system B, and in the second case that D is better than C and this is also due to the not-equispaced values of  AP, e. g. runs ranked 13 and 14 are much closer than runs ranked 10 (on the left branch of Figure~\ref{fig:hasse}) and 9, as shown in Figure~\ref{fig:ap_steps}. Note that here we are neither questioning the top-heaviness of a measure nor its capability of reflecting user preferences but rather we point out how the lack of equi-spaced values affects the assessment supported by a measure.
\label{ex:ap-not-equispaced}
\end{example}

The fact that \ac{IR} evaluation measures, apart from Precision, Recall, and \ac{RBP} with $p=0.5$, are not interval scales leads to the general issues with computing means, statistical tests, and meaningfulness discussed in Sections from~\ref{subsec:measurement-scales} to~\ref{subsec:significance-test} and shown in Examples~\ref{ex:interval-scale} and~\ref{ex:interval-scale-meaningfulness}. In addition, Examples~\ref{ex:rr-not-equispaced} and~\ref{ex:ap-not-equispaced} above show how the lack of equi-spacing may also lead to statements like ``system A is better than B'' (or viceversa) which are not always intuitive all over the scale.

\subsection{Averaging across Topics and Correlation Analysis Revisited}
\label{subsec:averaging-across-topics}

The fact that Precision and Recall are interval scales makes addition and subtraction permissible operations and, as a consequence, computing arithmetic means permissible too. Therefore, it is safe to average performance of IR systems across topics when we use Precision and Recall. But is that really true?

As said, \citet{FerranteEtAl2018b} have found an interval scale $\phi$, called \ac{SBTO}, and have shown that both Precision and Recall are an affine transformation of this interval scale and thus also an affine transformation of each other. \citet{FerranteEtAl2019c} have raised this question: if Precision and Recall are transformations of the same interval scale, they are ordinal scales too and they should rank systems in the same way. Therefore, if they produce the same \ac{RoS},  Kendall's $\tau$ correlation between them should be 1. So, why their Kendall's $\tau$ correlation is $0.8588$, using the TREC 8 Ad-hoc data?	

Let us consider how correlation analysis between evaluation measures works. Given two rankings $X$ and $Y$, their Kendall's $\tau$ correlation is given by $\tau \big( X, Y \big) = \frac{P-Q}{\sqrt{\big(P + Q + T\big)\big(P + Q + U\big) }}$, where $P$ is the total number of concordant pairs (pairs that are ranked in the same order in both vectors), $Q$ the total number of discordant pairs (pairs that are ranked in opposite order in the two vectors), $T$ and $U$ are the number of ties, respectively, in the first and in the second ranking. $\tau \in [-1, 1]$ where $\tau = 1$ indicates two perfectly concordant rankings, i.e. in the same order, $\tau = -1$ indicates two fully discordant rankings, i.e. in opposite order, and $\tau = 0$ means that 50\% of the pairs are concordant and 50\% discordant.

The typical way of performing correlation analysis is as follows: let $\phi_1$ and $\phi_2$ be  two evaluation measures; in our case, $\phi_1$ is Precision and $\phi_2$ is Recall. Let $\Phi_1$ and $\Phi_2$ be two $T \times S$ matrices where each cell contains the performance on topic $i$ of system $j$ according to measures $\phi_1$ and $\phi_2$, respectively. Therefore, $\Phi_1$ and $\Phi_2$ represent the performance of $S$ different systems (columns) over $T$ topics (rows). Let $\overbar{\Phi}_1$ and $\overbar{\Phi}_2$ be the column-wise averages of the two matrices, i.e. the average of the performance of each system across the topics. If you sort systems by their score in $\overbar{\Phi}_1$ and $\overbar{\Phi}_2$, you obtain two \ac{RoS} corresponding to $\mathrm{\phi}_1$ and $\mathrm{\phi}_2$, respectively, and you can compute  Kendall's $\tau$ correlation between these two \ac{RoS}. This is the traditional way for computing the correlation between two evaluation measures and  \citeauthor{FerranteEtAl2019c} call it \emph{overall correlation}, since it first computes the average performance across the topics and then it computes the correlation between evaluation measures. This approach leads to  a Kendall's $\tau$ correlation of $0.8588$ between Precision and Recall.

\citeauthor{FerranteEtAl2019c} proposed a different way of computing the correlation, called \emph{topic-by-topic correlation}, where, for each topic $i$, they consider the \ac{RoS} on that topic corresponding to $\phi_1$ and the one corresponding to $\phi_2$, i.e. they consider the $i$-th rows of $\Phi_1$ and $\Phi_2$, respectively; they then compute  Kendall's $\tau$ correlation among the two \ac{RoS} on that topic. Therefore, they end-up with a set of $T$ correlation values, one for each topic. Using, this way of computing correlation, \citeauthor{FerranteEtAl2019c} found that  Kendall's $\tau$ correlation between Precision and Recall is always $1$ for all the topics and this was the result expected for two interval scales which order systems in the same way.

Therefore, if you consider each topic alone, Precision and Recall are just a transformation of the same interval scale, as Celsius and Fahrenheit are, and their Kendall's $\tau$ correlation is $1$. However, if you first average across topics, which should be a permitted operation for interval scales, and then you compute  Kendall's $\tau$ correlation, it stops to be $1$. This was somehow surprising and unexpected. Indeed, as an example from another domain, if you take a matrix of scores in Celsius degrees and another one with the corresponding Fahrenheit degrees, their Kendall's $\tau$ correlation is always $1$, either if you compute it row-by-row (i.e. our topic-by-topic correlation) or if you first average across rows and then compute it (i.e. our overall correlation).

\citet[p.~305]{FerranteEtAl2019c} explained this behaviour as due to the recall base:
\begin{quote}
	Recall heavily depends on the recall base which changes for each topic and it is used to normalize the score for each topic; therefore, in a sense, recall on each topic changes the way it orders systems
\end{quote}

We further investigate this issue in Section~\ref{subsec:scale-change} below, where we provide details and demonstrations, but here we summarise the sense of our findings. The difference between overall and topic-by-topic correlation is basically due to the fact that we are using different interval scales for each topic. These scales are indeed transformations of one in the other for each topic and this is why topic-by-topic correlation is $1$; however, since we are changing scale from one topic to another, when average across topics we are mixing different scales and this is why the overall correlation is different from $1$.

\begin{example}[Recall corresponds to different scales on different topics]
	Let us consider Recall and let us assume that we have three queries $q_1, q_2, q_3$, with one, two and three relevant documents, respectively. Then, the possible values of Recall are as follows: for $q_1$ we have $0$ and $1$; for $q_2$ we have $0$, $\frac{1}{2}$  and $1$; and for $q_3$ we have $0$, $\frac{1}{3}$, $\frac{2}{3}$, and $1$.   Obviously, we have three different interval scales here -- although they are in the same range $[0, 1]$, their possible values are different. So we have to map the values onto a single scale, before we can do any statistics. There are two possibilities for doing this: 
\begin{enumerate}
    \item We take the union of the possible values. This would yield the set $\{0, \frac{1}{3}, \frac{1}{2}, \frac{2}{3}, 1\}$. However, these values are no longer equidistant, so it is not an interval scale.
	\item  We extend the union scale from above by additional values such that we have equidistant values, based on the least common denominator. Then we would have the set $ \{0, \frac{1}{6}, \frac{1}{3}, \frac{1}{2}, \frac{2}{3}, \frac{5}{6}, 1\} $ in our example. However, in this scale, the values $\frac{1}{6}$ and $\frac{5}{6}$ are not possible for our three example topics, and impossible values are not considered in the definition of the equidistance property of interval scales. Only if we had a fourth query with six relevant documents, this scale would be ok. In most cases, however, no such scale exists, and so the aggregated scale is not an interval one.
\end{enumerate} 
\label{ex:recall-different-scales}
\end{example}

The fact that we may be changing scale from topic to topic has very severe consequences. All the debate originated by \citeauthor{Stevens1946}'s permissible operations and the possibility of averaging only from interval scales onwards has always been based on the obvious assumption that the averaged values were all drawn from the same scale; no one has ever doubted that it is not possible to average values coming from different scales because this would be like mixing apples with oranges. So, what is the meaningfulness of typical statements like ``System A is (on average) better than system B'' when we are not only violating the interval scale assumptions but, even more seriously, we are mixing different scales? What about the meaningfulness of typical statements like ``System A is significantly better than system B''? The debate between using parametric or nonparametric tests concerns how much you wish to comply with the interval scale assumptions but, undoubtedly, all the significance tests, when aggregating across values, expect them to be drawn from the same scale.

If we wish to make an analogy, it is like the difference between using mass and weight, being Precision similar to mass and Recall to weight. It would be somehow safe to average the mass of bodies coming from different planets but it would not to average their weight, due to the different gravity on the different planets. The recall base is what changes the gravity from planet/topic to planet/topic in the case of Recall. 

However, even Precision is not completely ``safe'' because, when the length of the run changes, its scale changes as well. As a consequence we may end up using different scales from one run to another and this can happen not only across topics, as in the case of Recall, but also within topics, if we have two or more runs retrieving a different number of documents for that topic. This statements affects the evaluation of classical Boolean retrieval, where Precision and Recall are computed for the set of retrieved documents for each query, followed by averaging over all queries. So we have to conclude that this procedure is seriously flawed.
Luckily, in most of today's evaluations,  the length of the run has a much smaller effect because, in typical \ac{TREC} settings, almost all the runs retrieve 1,000 documents for each topic and just few of them retrieve less documents; this effect would also (practically) disappear when you consider Precision at lower cut-offs, like P@10, when it is almost guaranteed that all the runs retrieve 10 documents.

Summing up, independently from an evaluation measures being an interval scale or not, the recall base (greatly) and the length of the run (less) cause the scale to change from topic to topic and/or from run to run. This makes averaging across topics, as well as other forms of aggregation used in significance tests, problematic at best. We show how and why this happens in the case of Precision (Section~\ref{subsubsec:precision-change}) and Recall (Section~\ref{subsubsec:recall-change}), which are  the simplest measures you can think of, since they change either the length of the run or the recall base alone. We also consider the more complex case of the F-measure (Section~\ref{subsubsec:f-change}), which changes both the run length and the recall base at the same time. Therefore, we hypothesise that these issues may be even more severe in the case of more complex evaluation measures, like \ac{AP} and others, which are not even interval scales and mix recall base and run length with rank position and various forms of utility accumulation and stopping behaviours.

Finally, also the way in which we interpret the results of correlation analysis may be impacted. Indeed, we typically attribute differences in correlation values to the different user models embedded by evaluation measures. The rule-of-thumb by~\citet{Voorhees1998,Voorhees2000} is that an overall correlation above $0.9$ means that two evaluation measures are practically equivalent, an overall correlation between $0.8$ and $0.9$ means that two measures are similar, while dropping below $0.8$ indicates that measures are departing more and more. Therefore a correlation of $0.8588$ would suggest that Precision and Recall share some commonalities but they differ enough due to their user models, still not being pathologically different. However, we (now) know that they are just the transformation of the same interval scale and that this correlation value is just an artifact of mixing different scales across topics rather than an intrinsic difference in the user models of Precision and Recall.

\subsection{Why May Scales Change from Topic to Topic or from Run Length to Run Length?}
\label{subsec:scale-change}

As discussed above, \citet{FerranteEtAl2018b} have demonstrated that Precision, Recall, and F-measure are interval scales when you fix the length of the run $N$ and the recall base $RB$, i.e. they are an homomorphism with respect to the same ordering of runs in the \ac{ERS}. However, 
if we mix together runs with different bounded lengths and/or different  bounded recall bases, Precision, Recall and F-measure are no more interval  scales, they are no more an affine transformation of each other and they even order the runs in  different ways. Clearly, this is a severe issue when you need to average (or compute any other aggregate) across different topics or runs with different lengths.

Let us consider the universe set $S[N,K]$ which contains all the runs of any possible length $n$, less than or equal to $N$, and with respect to all the possible recall bases $RB$, less than or equal to $K$. To avoid trivial cases, we consider always $N$ and $K$ greater than or equal to $1$. A run in $S[N,K]$ is represented by a triple $[r, n, RB]$, where $r$ indicates the number of relevant documents retrieved by the run, $n$ is the length of the run, and $RB$ is the recall base, i.e. the total number of relevant documents for a topic. Note that, for each run in $S[N,K]$ 
it holds $n\le N$ and $RB\le K$ by construction, but we also have $r\le (n \wedge RB)$, where $x \wedge y = \min\{x,y\}$, i.e. there is a (implicit) dependence on the recall base when it comes to the number of relevant retrieved documents.

We define $S_{n,RB}$ as the set which contains all the runs with the same length $n$ and with respect to the same recall base $RB$. Therefore, we can express the universe set $S[N,K]$ as the union of such sets, namely
\begin{equation*}
S[N,K]=\bigcup_{\substack{1 \leq n \leq N \\ 1 \leq RB \leq K}} 
S_{n, RB}
\end{equation*}

$S_{n,RB}$ models the typical case of runs all with the same length for a given topic (or for a set of topics which have the same recall base). This is exactly the case for which~\citet{FerranteEtAl2018b} have demonstrated that Precision, Recall and F-Measure are interval scales and an affine transformation of each other. 
However, this holds for each $S_{n,RB}$ separately while the issue we discuss in this section is what happens when you mix different $S_{n,RB}$, i.e. when you go towards $S[N,K]$.

\subsubsection{Precision}
\label{subsubsec:precision-change}
Precision is equal to the fraction of the retrieved documents that are relevant. Therefore, for a run represented by triple $[r, n, RB]$, Precision is given by
\begin{equation*}
    Prec[r,n,RB]=\frac{r}{n} 
\end{equation*}

Let us start from $S_{n,RB}$: $Prec$ maps this set into the set $\big\{ 0, 1/n, 2/n, \ldots, (n \wedge RB)/n \big\}$ and it has been proven by~\citeauthor{FerranteEtAl2018b} that $Prec$ is an interval scale in this case.  However, already in this simpler case, there is a (implicit) dependency on the recall base, when it comes to the possible values of Precision. Therefore, even when we consider runs with the same length but for topics with different recall bases, i.e. $S_{n,RB_1}$, $S_{n,RB_2}$, $S_{n,RB_3}$, ... we are dealing with different  scales, all embedded in the single interval scale whose image is $\big\{ 0, 1/n, 2/n, \ldots, (n \wedge \max\{RB_i\})/n \big\}$.

To understand the problems arising mixing different lengths and recall bases, let us consider the general scenario of Precision defined on $S[N,K]$. This is the case where we consider the Precision measure defined on the set of the runs of any possible bounded length and recall base and we find that it is an interval scale only in the almost trivial cases of $N\le 2$. Indeed, $Prec$ maps $S[1,K]$, for any $K\ge 1$, into the set $\{0, 1\}$ and it is an interval scale since these values are equispaced. When $N=2$, $Prec$ maps $S[2,K]$, for any $K\ge 1$, into the set $\{0, 1/2, 1\}$;  Since the values are equispaced, Prec is still an interval scale. To compare the order induced on these sets by $Prec$ (and the other measures), let us consider in more detail $S[2,2]$. This set is 
\begin{equation*}
    \big\{
        [0,1,1], 
        [0,1,2], 
        [0,2,1], 
        [0,2,2], 
        [1,2,1], 
        [1,2,2], 
        [1,1,1],
        [1,1,2], 
        [2,2,2] 
    \big\} 
\end{equation*}
and $Prec[1,2,1]=Prec[1,2,2]=1/2$, while $Prec[1,1,2]=1$.

Continuing with a similar construction for $N=3$, we obtain that $Prec$ assumes the four possible values $\{0,1/3, 1/2, 1\}$, when $K=1$,  and the five possible values $\{0,1/3, 1/2, 2/3, 1\}$ when $K\ge 2$. Indeed, for runs of length at most $3$, these are all the possible values of the fraction $\frac{r}{n}$  for $1\le n \le 3$ and $0\le r \le \min\{3,K\}$. Since these  values are not equispaced, it is sufficient to state that $Prec$ is not an interval scale on $S[3,K]$.

To prove that $Prec$ is not interval on $S[N,K]$ for any finite $N>3$, let us prove again that the values on the image are not equispaced.  The three smallest values of $Prec[S[N,K]]$ 
are $0$, $1/N$ and $1/(N-1)$. Indeed, the only other possible candidate to be the third smallest value, when $K\ge 2$,  would be $2/N$, but $1/(N-1)<2/N$ when $N>2$. These three values are not equispaced since $1/N-0=1/N \neq 1/N(N-1)=1/(N-1)-1/N$, when $N> 2$, and therefore $Prec$ on $S[N,K]$ cannot be an interval scale when $N\ge 3$.

\subsubsection{Recall}
\label{subsubsec:recall-change}

The Recall measure depends explicitly on the recall base $RB$,  i.e. the total number of relevant documents available for a given topic
\begin{equation*}
Recall[r,n,RB]=\frac{r}{RB} 
\end{equation*}
Note that for any admissible run $[r,n,RB]$, i.e. for which $r\le n\wedge RB$, its recall value (implicitly) depends on $n$, creating a specular situation with respect to the one of Precision.

$Recall$ is an interval scale on $S_{n,RB}$, since it is an affine transformation of $Prec$, as demonstrated by ~\citeauthor{FerranteEtAl2018b}. However, due to the (implicit) dependency on $n$, even when we consider topics with the same recall base but runs with different length, i.e. $S_{n_1 ,RB}$, $S_{n_2, RB}$, $S_{n_3, RB}$, ... we are dealing with different scales, as discussed below.

$Recall$ is an interval scale on the sets $S[N,1]$ for any maximum length $N\ge 1$, since the image is the equispaced set $\{0, 1\}$. Applied to the sets $S[N,2]$, for any $N$, $Recall$ takes the values $\{0, 1/2, 1\}$ and, therefore, it is  an interval scale as well.
However, Precision and  Recall induce, for example on $S[2,2]$, two different orderings of the runs and  so they stop to be an affine transformation of each other, i.e. they become two different interval scales. Indeed, consider the runs $[1,2,1]$ and $[1,1,2]$: we have seen that $Prec[1,2,1]=1/2<1=Prec[1,1,2]$, while it holds that $Recall[1,2,1]=1>1/2=Recall[1,1,2]$.

When we define $Recall$ on $S[N,K]$, for $K> 2$, we have that this measure is no more interval thanks to an argument  similar to that used for Precision. Indeed, the two smallest non zero values of Recall on $S[N,K]$, are as $1/K$ and $1/(K-1)$, here obtained for a run with a unique relevant document with respect to a topic with $RB=K$ and $RB=K-1$, respectively.

Furthermore, it is immediate to see that Recall and Precision induce for any $N\ge 2$ and $K\ge 2$ two different orderings on $S[N,K]$, i.e. they become two different scales. Indeed, for any two runs $[r_1,n_1,RB_1]$ and $[r_2,n_2,RB_2]$, we have that $Prec[r_1,n_1,RB_1]<Prec[r_2,n_2,RB_2]$ if and only if $r_1/n_1<r_2/n_2$, while $Recall[r_1,n_1,RB_1]>Recall[r_2,n_2,RB_2]$ if and only if $r_1/RB_1>r_2/RB_2$. Both these condition are satisfied when
\begin{equation*}
n_2/n_1<r_2/r_1<RB_2/RB_1
\end{equation*}
For example, if we take $r_1=r_2$, $n_1=2 n_2$ and $RB_2 = 2 RB_1$, the previous condition is satisfied and the two runs are ordered in a different way by the two measures.

\subsubsection{$F_1$ Measure}
\label{subsubsec:f-change}

The $F_1$ measure is the harmonic mean of precision and recall
\begin{equation*}
F_1[r,n,RB]=\frac{2}{(Prec[r,n,RB])^{-1}+(Recall[r,n,RB])^{-1}}
\end{equation*}
Some small algebra gives us that $F_1$ is also equal to
\begin{equation*}
    \label{F1}
F_1[r,n,RB]=\frac{2 r}{n+RB} 
\end{equation*}

As before (see \citeauthor{FerranteEtAl2018b}),  we have that on $S_{n,RB}$, 
$F_1$ is an interval scale, being an affine transformation of $Prec$ and $Recall$.

On the contrary, if we  consider $F_1$ defined on $S[N,K]$, it is no more an interval scale, except for the almost trivial case $S[1,1]$, whose image is the equispaced set $\{0,1\}$.
Let us first consider $S[2,1]$ and $S[1,2]$: in both these cases the values in the image of $F_1$ are no more equispaced, since $F_1$ takes the values $\{0, 2/3, 1\}$. If we consider $F_1$ on $S[2,2]$, it takes the vales $\{0, 1/2, 2/3, 1\}$ and is still not an interval scale. Moreover, we have that $F_1$ induces yet another ordering on $S[2,2]$, since $F_1[1,2,1]=2/3=F_1[1,1,2]$, while it holds that $Prec[1,2,1]=1/2<1=Prec[1,1,2]$ and $Recall[1,2,1]=1>1/2=Recall[1,1,2]$.

When we define $F_1$ on $S[N,K]$, for $N\ge 3$ or $K\ge 3$,  we have that this measure is no more an interval scale, as can be easily 
seen since the three smallest values of the image are $0, 2/(N+K)$ and $2/(N+K-1)$, 
which are not equispaced.  At the same time, using an example similar to the one used for $Recall$, we obtain that the ordering induced on $S[N,K]$ by $F_1$ in these latter cases  differs from both the orderings induced by $Prec $ and $Recall$.

\subsubsection{Summary and Discussion}

We have demonstrated that, when we consider runs with a fixed length $n$ and with respect to a fixed recall base $RB$, i.e. we consider $S_{n, RB}$ and runs of the same length for the same topic (or, more generally, topics with the same recall base), Precision, Recall, and F-measure are interval scales and they are an affine transformation of each other. As a consequence, they order runs in the same way and their Kendall's $\tau$ is $1$.

However, when we start mixing runs with different length and/or with respect to different recall bases, the situation quickly gets more complicated. Only in the trivial (and not very useful in practice) case $N = 2$ and $K=2$, i.e. $S[2,2]$ 
where we have runs of length $1$ or $2$ and topics with $1$ or $2$ relevant documents,
Precision and Recall are still both interval scales but they stop to be an affine transformation of each other. As a consequence, they order runs in different ways and their Kendall's $\tau$ is less than $1$. F-measure already stops to be an interval scale and orders runs in yet another way than Precision and Recall, leading to a Kendall's $\tau$ less than $1$. 
For $N > 2$ and $K>2$ all of them (Precision, Recall, and F-measure) stop to be interval scales, departing from the interval assumption more and more, and they order runs in three completely different ways, again leading to a Kendall's $\tau$ less than $1$.
In the special case where we fix the length, Precision is still 
an interval scale, while if we fix a single recall base, Recall
is still an interval scale, but in both cases the other measure is no 
more interval and also order the runs in a different way.

We may be tempted to consider as positive the fact that sooner than later Precision, Recall, and F-measure start ordering runs in a different way and that their Kendall's $\tau$ is less than $1$. Indeed, this is what we expect from evaluation measures, to embed different user models and to reflect different user preferences in ordering runs. This is also one of the main motivations why there is debate and we would accept to derogate from requiring them to be interval scales: reflecting user preferences could be more important than complying with rigid assumptions.

However, we should carefully consider how this is happening. They initially are the ``same'' scale (except for an affine transformation), when we use them to measure objects with some shared characteristics, i.e. same run length $n$, and with respect to a similar context, i.e. same recall base $RB$. However, as soon as we measure objects with more mixed characteristics and contexts, they cease to be the ``same'' interval scale and only at that point they begin to order runs differently. This is more or less like saying that kilograms and pounds are the ``same'' interval scale only when we weigh people with the same height and from the same city but, as soon as we weigh people with different heights and/or coming from flatland or mountains, they become two different scales and they also possibly stop to be interval scales. This would sound odd and quite different from saying that weight and temperature are different (interval) scales because they measure different attributes/properties of an object or, in our terms, they would reflect different user preferences.

Why does this happen? Because run length $n$ and recall base $RB$ change. This is very clear and somehow more extreme in F-measure, where both $n$ and $RB$ explicitly appear in the equation of the measure. 

We hypothesize that this could be even more severe and extreme in the case of rank-based measures since not only they combine, implicitly or explicitly, the two factors $n$ and $RB$ but they also mix them with the rank of a document and various discounting and accumulation mechanisms. Figure~\ref{fig:measure_steps} gives a taste of this much more complex situation: it shows the simple (and somehow safe) case of $S_{4, 4}$ and it already emerges how different are the behaviours and patterns in violating or complying with the interval scale assumption.

Why does this matter? As already said, because we need to aggregate scores across topics and runs and to compute significance tests. We do not only have the problem of how much evaluation measures violate the interval scale assumptions, required to compute aggregates, but also the issue of not mixing apples and oranges, i.e. scores from different scales, required to make aggregates sensible. In this respect, run length is a less severe issue which can be easily mitigated in practice, either by forcing a given length or because we are interested in lower cut-offs, e.g. 5, 10, 20, 30. The effect of the recall base can be mitigated by adopting measures that do not explicitly depend on it, even if the implicit dependency due to the capping of the image values will remain.
\section{Related Works}
\label{sec:related}

\citet{vanRijsbergen1974} was the first to tackle the issue of the foundations of measurement for \ac{IR} by exploiting the representational theory of measurement. He observed that~\cite[pp.~365--366]{vanRijsbergen1974}
\begin{quote}
	The problems of measurement in information retrieval differ from those encountered in the physical sciences in one important respect. In the physical sciences there is usually an empirical ordering of the quantities we wish to measure. For example, we can establish empirically by means of a scale in which masses are equal, and which are greater or lesser than others. Such a situation docs not hold in information retrieval. In the case of the measurement of effectiveness by precision and recall, there is no \emph{absolute} sense in which one can say that one particular pair of precision/recall values is better or worse than some other pair, or, for that matter that they are comparable at all
\end{quote}
Later on, \citet[p.~33]{vanRijsbergen1981} further stressed this issue: ``There is no empirical ordering of retrieval effectiveness and therefore any measure of retrieval effectiveness will be by necessity artificial''.

\citeauthor{vanRijsbergen1974} addressed this issue by exploiting the \emph{additive conjoint measurement}~\cite{KrantzEtAl1971,LuceTukey1964}. Additive conjoint measurement was a new part of the measurement theory developed as a reaction to the views of~\citet{Campbell1920,Campbell1928} and the conclusions of Ferguson Committee of British Association for the Advancement of Science~\cite{FergusonEtAl1940}, where \citeauthor{Campbell1920} was an influential member, which considered the \emph{additive property}, i.e. the concatenation operation mentioned in Section~\ref{subsec:measurement-overview}, as fundamental to science and proper measurement; as a consequence, measurement of psychological attributes, which is lacking such additive property, was not possible in a proper scientific way. As explained by~\citet[p.~67]{Michell1990} 
\begin{quote}
	Conjoint measurement involves a situation in which two variables ($A$ and $B$) are noninteractively [e.g. non additively] related to a third ($X$). It is not required that any of the variables be already quantified, although it is necessary that the values of $X$ be orderable, and that values of $A$ and $B$ be independently identifiable (at least at a classificatory level). Then, via the order on $P$, ordinal and additive relations on $A$, $B$, and $X$ may be derived
\end{quote}

Typical examples from physics are the momentum of an object, which is affected by its mass and velocity, or the density, which is affected by its mass and volume~\cite{KrantzEtAl1971}.

\citeauthor{vanRijsbergen1974} considered retrieval effectiveness as the ``orderable $X$'' mentioned above and took precision $P$ and recall $R$ as the two variables $A$ and $B$. In particular, he demostrated that on the relational structure $(R \times P, \succsim)$ it was possible to define an additive conjoint measurement and to derive actual measures of retrieval effectiveness from it. Note that, in this way, he avoided the need to explicitly define what an ordering by retrieval effectiveness is and he considered that precision and recall contribute independently to retrieval effectiveness. The problem of how to order runs in the \ac{ERS} has been addressed some years later by~\citet{FerranteEtAl2015b,FerranteEtAl2017,FerranteEtAl2018b}. More subtly, \citeauthor{vanRijsbergen1974} treats precision and recall as two attributes which can be jointly exploited to order retrieval effectiveness but, each of them, is already a measure and quantification of retrieval effectiveness and, thus, this brings some circularity in the reasoning.
Finally, \citeauthor{vanRijsbergen1974} did not address the problem of which are the scale properties of precision and recall (or other evaluation measures), which has been later addressed by~\citeauthor{FerranteEtAl2018b}.

\citet{BollmannCherniavsky1980,BollmannCherniavsky1981} built on the conjoint measurement work by~\citeauthor{vanRijsbergen1974} and applied it to further study under which conditions the  \emph{MZ-metric}~\cite{Heine1973}. In particular, \citeauthor{BollmannCherniavsky1980} leveraged what they called \emph{transformational viewpoints}, i.e. elementary transformations of the runs which closely resemble the idea of swap and replacement used by~\citeauthor{FerranteEtAl2018b} much later on.

\citet{Bollmann1984} studied set-based measures by showing that measures complying with a monotonicity and an Archimedean axiom are a linear combination of the number of relevant retrieved documents and the number of not relevant not retrieved documents and how this could be related to collections and sub-collections. He thus addressed a problem somehow different from the one of the present paper, still leveraging the representational theory of measurement.

\citet{BusinMizzaro2013},  \citet{MaddalenaMizzaro2014} and \citet{AmigoMizzaro2020} proposed a unifying framework for ranking, classification, and clustering measures, which is rooted in the representational theory of measurement as well. They considered scales but as a way of mapping between relevance judgements (assessor scales) and \ac{RSV} (system scales) and of introducing axioms over them rather than a way of studying which are the scales actually used by \ac{IR} evaluation measures and their impact on actual experiments.

As already discussed, \citet{FerranteEtAl2015b} relied on the representational theory of measurement to formally study when evaluation measures are on an ordinal scale while \citet{FerranteEtAl2017,FerranteEtAl2018b} proposed a more general theory of evaluation measures, proving when they are on an interval scale or not. Finally, \citet{FerranteEtAl2019c} conducted a preliminary experimental investigation of the effects of \ac{IR} measures being interval scales or not.

Even if not specifically focused on scales and their relationship to \ac{IR} evaluation measures, there is a bulk of research on studying which constraints define the core properties of evaluation measures: \citet{AmigoEtAl2009,AmigoEtAl2013b,AmigoEtAl2018,AmigoEtAl2019} and \citet{Sebastiani2020} face this issue from a formal and theoretical point of view, applying it to various tasks such as ranking, filtering, diversity and quantification, while \citet{Moffat2013} adopts a more numerical approach.

As it emerges from the above literature review, to the best of our knowledge, no one has dealt yet with the problem of considering the meaningfulness of \ac{IR} experimental results and of transforming \ac{IR} evaluation measures into interval scales.

\section{Transforming IR measures to interval scales}
\label{sec:methodology}

Let ($REL$, $\preceq$) be a totally ordered set of relevance degrees, with a minimum  called the non-relevant relevance degree $\text{nr} = \min(REL)$ and a maximum $\text{rr} = \max(REL)$; in the case of binary relevance, we set $REL=\{0,1\}$ without any loss of generality. Let $N$ be the length of a run, i.e. the number of retrieved documents, we call judged run $\hat r_t \in REL^N$ the vector of relevance degrees associated to each retrieved document, denoting by $\hat r[j]$ the j-th element of the vector.

Any \ac{IR} evaluation measure $M$ naturally defines an order among system runs. Indeed, taken any two runs $\hat{r},\hat{s}\in REL^N$,  we order them as follows
\begin{equation}
\hat r \prec \hat s \; \Leftrightarrow  \; M(\hat{r})< M(\hat{s}) \;.
\label{eq:wto}
\end{equation}

Note that this is a {\em weak total order}, since $M(\hat{r})=M(\hat{s})$ does not imply that $\hat{r}=\hat{s}$, and that it is the order called induced total order by~\citet{FerranteEtAl2018b}. It has the following characteristics, as discussed in the previous sections:
\begin{itemize}
	\item it differs from measure to measure, i.e. each measure may produce a different \ac{RoS};
	\item it typically is not an interval scale, i.e. the produced values are not equi-spaced.
\end{itemize}

The basic idea of our approach is to keep the weak total order~\eqref{eq:wto} produced by the measure $M$ but making sure that all the possible values are equi-spaced. 

The simplest way to achieve this result is to define first the nonlinear transformation $\varphi$ from $[0,1]$ into $\mathbb{N}$ that maps each value $m$ in the image of the measure $M$ into its rank number. Then, we define the \emph{ranked version of the measure}, i.e. the \emph{interval-scaled version} of it, as $M_R=\varphi(M)$. Note that this approach is in line with what suggested by~\citet{Gaito1959} to transform ordinal scales into interval ones.

Most of the measures are not one-to-one mappings and thus the cardinality of their image is strictly smaller than the cardinality of their domain, i.e. $|M(REL^N)|< |REL^N|$. The runs which are assigned the same value by the measure are called \emph{ties}. As pointed out before, this is the reason why the order induced on $REL^N$ by a measure in general is just a {\em weak} total order.

The map $\varphi$ is then defined for any value $m$ in the image $M(REL^N)$ as
\begin{equation}
\varphi(m)=|\{x\in M(REL^N): x\le m\}|
\end{equation}

$M_R=\varphi(M)$ is an interval scale since the ranks are equi-spaced by construction; moreover, it preserves the \ac{RoS} of $M$ and thus it constitutes an interval-scaled  version of it.

Finally, we have to deal with \emph{tied} values in the measure. In statistics there are many ways of breaking ties~\cite{GibbonsChakraborti2011}) and the most common are: average/mid, min, or max rank.  However, each of these alternative strategies would result in a scale where the possible values are no longer equi-spaced. Indeed, suppose you have the following values $M = [0.00 \, 0.25 \, 0.40 \, 0.40 \, 0.70]$; the tied value $0.40$ has ranks $3$ and $4$. If we chose the mid-rank tie breaking strategy, we would obtain $M_R = [1 \, 2 \, 3.5 \, 3.5, 5]$; using min-rank, we would obtain $M_R = [1 \, 2 \, 3 \, 3 \, 5]$; using max-rank, we would obtain $M_R = [1 \, 2 \, 4 \, 4 \, 5]$. In all these cases, the resulting scale would be no more equi-spaced.

Therefore, we simply eliminate the duplicate rank values and assign the same rank at all the tied positions, calling this tie-breaking strategy unique (\texttt{unq}). In the previous example, we would obtain $M_R [1 \, 2 \, 3 \, 3 \, 4]$.

\begin{example}[Mapping \ac{DCG} to an interval scale]

Let us consider the case of \ac{DCG} with $\log_2$ in Figure~\ref{fig:dcg_steps}: there are 16 runs and 12 possible values of \ac{DCG}, being some runs tied. Therefore, we have that: $\varphi(3.131)=\varphi\big(M(1,1,1,1)\big)=12$,  $\varphi(2.631)=\varphi\big(M(1,1,1,0)\big)=11$,  $\varphi(2.500)=\varphi\big(M(1,1,0,1)\big)=10$,  $\varphi(2.131)=\varphi\big(M(1,0,1,1)\big)=\varphi\big(M(0,1,1,1)\big)=9$ and so on until $\varphi(0.500)=\varphi\big(M(0,0,0,1)\big)=2$, and $\varphi(0.000)=\varphi\big(M(0,0,0,0)\big)=1$.
\end{example}

For \ac{RBP} with $p=1/2$ we have that $\varphi(m)=2^N m$ while for \ac{RR} we have that $\varphi(m)=N+1-\frac{1}{m}$. However, in general, the function $\varphi$ does not have any analytical expression, it is {\bf nonlinear}  and it varies from measure to measure.

\subsection{Runs of different length}

When working with runs all with the same length, the proposed approach maps a measure into a proper interval scale, actually the same scale for all the runs, and this allows us to compute aggregates across runs with the same length.

 If we work we runs of different length, the proposed approach maps each length (run) into a proper interval scale but it differs from length to length. For example, in the case of \ac{DCG} with $\log_2$  there are there are $24$ distinct values for $N=5$, 768 for $N=10$, $24,576$ for $N=15$, and so on; all of them correspond to a ranked measure (interval scale) with a different number of steps. As a consequence, even using our approach, we cannot aggregate across runs with different length.
 
However, as already discussed, this is a problem easily manageable in practice. Indeed, for small run lengths or low cut-offs of typical interest, such as the top 10 documents, it is reasonable to assume that the runs have all the same length, since runs are usually able to retrieve enough documents. In the more general case and for bigger run lengths, if a run does not retrieve enough documents, it could be padded with not relevant documents. Therefore, we can consider our approach as generally applicable with respect to this issue.

\subsection{Different topics}

Let us now assume that we have fixed a run length which is the same for all the runs and which allows us to compute aggregates across runs. What happens if we need to compute aggregates across topics?

\subsubsection{Measures not depending on the recall base}

In the case of measures not depending on the recall base, since the length of the run is the same for all the runs across all the topics, our approach maps a measure into the same interval scale for all the runs and all the topics. Therefore, we can safely compute aggregates also across topics.

\subsubsection{Measures depending on the recall base}

In the case of the measures depending on the recall base, as already explained, due to the recall base changing from topic to topic, it does not exist a single (interval) scale which can be used across all the topics. As a consequence our approach could not be directly applied. However, we could use it as a surrogate that brings, at least, some more ``intervalness'' to a measure. 

Indeed, on each single topic, our approach maps a measure depending on the recall base into a proper interval scale, whose steps are equi-spaced. When we deal with two (or more) topics, we would need to find an interval scale where it is possible to match the steps from the (two) scales of each topic into some ``bigger'' set of equi-spaced steps, which can accommodate all of them. However, as shown in Example~\ref{ex:recall-different-scales} and in Section~\ref{subsec:scale-change}, this common super-set of steps does not exist, if not in trivial cases. 

Therefore, as an approximation, we can pretend that the scale for each topic is the overall common scale -- and, as said above, this is exactly what happens in the case of measures not depending on the recall base -- and use it across topics, even if this will actually stretch the steps of each topic.

\begin{example}[Surrogating Recall to an interval scale]

Suppose we are dealing with runs of length $N=2$, i.e. $r_0 = [0 \, 0]$, $r_1 = [0 \, 1]$, $r_2 = [1 \, 0]$, $r_3 = [1 \, 1]$. If the recall base for the first topic $q_1$ is $RB_1 = 2$, these runs are mapped to the following Recall values $\{0, \frac{1}{2}, \frac{1}{2}, 1\}$; if the recall base for the second topic $q_2$ is $RB_2 = 3$, these runs are mapped to the following Recall values $\{0, \frac{1}{3}, \frac{1}{3},  \frac{2}{3}\}$. 

Our transformation approach maps the runs of both topics to $\{1, 2, 2, 3\}$, which is a proper interval scale on each topic separately. However, if we look at the two topics together and we use this mapped scale, we are slightly stretching the steps of the original scales. For example, if we compute the difference between $r_2$ and $r_3$, on this mapped scale it is same, i.e. $1$, for both $q_1$ and $q_2$ while on the original scales it is $\frac{1}{2}$ for $q_1$ and $\frac{1}{3}$ for $q_2$. This means that our transformation also effects ordinal scales when a recall base is involved.
\label{ex:surrogating-recall}
\end{example}

\section{Experimental Setup}
\label{sec:setup}

We consider the following evaluation measures: Precision (P) and Recall (R)~\cite{vanRijsbergen1979},  \ac{AP}~\cite{BuckleyVoorhees2005},  \ac{RBP}~\cite{MoffatZobel2008},  \ac{DCG} and \ac{nDCG}~\cite{JarvelinKekalainen2002} and \ac{RR}~\cite{SinghalEtAl1997}. We calculated RBP by setting $p \in \{0.3, 0.5, 0.8\}$, indicated respectively as RBP\_p03, RBP\_p05, and RBP\_p08; for \ac{DCG} and \ac{nDCG} we use a $log_{2}$ and a $log_{10}$ discounting, indicated respectively as DCG\_b02, nDCG\_b02, DCG\_b10, and nDCG\_b10.

We considered the following datasets:
\begin{itemize}
	\item \textbf{Adhoc track \texttt{T08}}~\cite{VoorheesHarman1999}: 528,155 documents of the TIPSTER disks 4-5 corpus minus congressional record; \texttt{T08} provides 50 topics, each with binary relevance judgments and a pool depth of 100; 129 system runs retrieving 1,000 documents for each topic were submitted to \texttt{T08}.
	\item \textbf{Common Core track \texttt{T26}}~\cite{AllanEtAl2017b}: 1,855,658 documents of the New York Times corpus; \texttt{T26} provides 50 topics, each with multi-graded relevance judgments (not relevant, relevant, highly relevant); relevance judgements were done mixing depth-10 pools with multi-armed bandit approaches~\cite{LosadaEtAl2016,Voorhees2018}; 75 system runs retrieving 10,000 documents for each topic were submitted to \texttt{T26}.
	\item \textbf{Common Core track \texttt{T27}}~\cite{AllanEtAl2018b}: 595,037 documents of the Washington Post corpus; 50 topics, each with multi-graded relevance judgments (not relevant, relevant, highly relevant); relevance judgements were done adding stratified sampling~\cite{CarteretteEtAl2008} and move-to-front~\cite{CormackEtAl1998} approaches to the \texttt{T26} procedure; 72 system runs retrieving 10,000 documents for each topic were submitted to \texttt{T27}.
\end{itemize}

In the case of the \texttt{T26} and \texttt{T27} tracks we mapped their multi-graded relevance judgement to binary ones using a lenient strategy, i.e. whatever above not relevant is considered as relevant\footnote{ The case of multi-graded relevance is left for future work, due to its exponential explosion in the number of possible cases. For example, switching from a binary to a 4-valued scale, for the run length $N=30$ the number of possible runs would grow from $2^{30} \sim 10^9$ to $4^{30}=2^{60} \sim 10^{18}$.}.

For each track we experimented the following run lengths $N \in \{5, 10, 20, 30\}$, i.e. we cut runs at the top-$N$ retrieved documents. In terms of our transformation methodology, this means considering a space of possible runs containing, roughly, $\{32, 10^3,  10^6, 10^9\}$ runs, respectively\footnote{To give an idea of the computational resources required, runs of length $N=30$ mean an occupation of $2^{30}*30*8 = 240$ GByte of memory, just for holding all the possible runs. A length $N=40$ would mean 320 TByte of memory, which is not feasible in practice. 

The code is implemented in Matlab and thus we considered 8 bytes for representing a digit, since this is the size of a double. Even if we considered a more compact representation, in some other language like C, using just 1 bit per digit, it would have meant 40 TByte of memory for runs of length $N=40$.}. We indicated the run length in the identifier of the track, e.g. \texttt{T08\_10} means  \texttt{T08} runs cut down at length $10$.

In significance tests, we used a confidence level $\alpha = 0.05$. 

To ease the reproducibility of the experiments, all the source code needed to run them is available in the following repository: \url{https://bitbucket.org/frrncl/tois2021-fff-code/src/master/}.

\section{Experiments}
\label{sec:experiments}

In Section~\ref{subsec:correlation-rnk} we validate our approach and answer the research question ``How far a measure is from being an interval scale?''. Then, in the next two sections we investigate on the effects of using or not a proper interval scale in \ac{IR} experiments. In particular, in Section~\ref{subsec:correlation-msr} we study how this affects the correlation among evaluation measures, i.e. the main tool we use to determine how close are two measures. In Section~\ref{subsec:significance-test} we analyse how this impacts on the significance tests, both parametric and non-parametric, i.e. the main tool we use to determine when \ac{IR} systems are significantly different or not.

The following sections report, separately, the case of measures not depending on the recall base  -- namely, \ac{RBP}, \ac{DCG}, \ac{RR}, and P  -- and measures depending on the recall base --  namely, \ac{AP}, nDCG, and R. Indeed, as previously explained, fixed a run length, in the former case it is possible to find an overall interval scale, which is the same across all the topics, and apply our transformation approach in an exact way; in the latter case,  an overall interval scale common to all the topics does not exist and our transformation is just the best surrogate that can be figured out.

\subsection{Correlation between measures and their ranked version. How far a measure is from being an interval scale?}
\label{subsec:correlation-rnk}

In this section we study the relationship between each measure and its ranked version, i.e. its mapping towards an interval scale, as explained in Section~\ref{sec:methodology}. This analysis allows us: 1) to validate our approach, verifying that it produces the expected results; 2) to understand how much a measure changes when it is transformed, seeking an explanation for this change; 3) to understand what happens when we apply our transformation approach in a ``surrogate'' way in the case of measures depending on the recall base. 

We compute both the overall and the topic-by-topic Kendall's $\tau$ correlation between each measure $M$ and its ranked version\footnote{To avoid errors due to floating point computations, we rounded averages to 8 decimal digits.}. \citet{Ferro2017} has shown that, even if the absolute correlation values are different, removing or not the lower quartile runs produces the same ranking of measures in terms of correlation; similarly, he has shown that both $\tau$ and AP correlation $\tau_{ap}$~\cite{YilmazEtAl2008} produce the same ranking of measures in terms of correlation. Therefore, we focus only on Kendall's $\tau$ without removing lower quartile systems.

As explained in Section~\ref{subsec:averaging-across-topics}, the \emph{topic-by-topic correlation} is expected to be always $1.0$, since the ranked version of a measure preserves the same order of runs on each topic by construction. As a sanity check, we verified that the topic-by-topic correlation is indeed $1.0$ in all the cases and we do not report it in the following tables for space reasons. On the contrary, the \emph{overall correlation}, i.e. the traditional one, can be different from $1.0$ for the reasons discussed above: the preliminary average operation would be not allowed in the case of an original measure which is not an interval scale while it is allowed in the case of the corresponding ranked measure; and, different recall bases across topics can lead to different scales which should not be averaged together. 

In general, we can consider the overall Kendall's $\tau$ correlation between a measure and its ranked version as a \emph{proxy} providing us with an estimation of how far a measure is from both being a proper interval scale and, in the case of measures depending on the recall base, also being safely averaged across topics. Note that this approach is in line and extends what proposed by~\citet{FerranteEtAl2019c} when they suggested to use the overall Kendall's $\tau$ correlation between a measure and the \acf{SBTO} and \ac{RBTO} interval scales as an estimation of how much a measure is an interval scale.

Table~\ref{tab:smrytau-ranked-unq} summarizes the outcomes of the overall correlation analysis between measures and their ranked version, e.g. we computed the  Kendall's $\tau$ correlation between \ac{DCG} and \ac{DCG}$_R$, the interval-scaled version of \ac{DCG} according to the approach we described in Section~\ref{sec:methodology}. 

From a very high level glance at Table~\ref{tab:smrytau-ranked-unq} we can see that Kendall's $\tau$ correlation changes due to the transformation but not in a too excessive way which suggests that we are not running into any pathological situation.

\begin{table}[tb] 
\centering 
\caption{Kendall's $\tau$ overall correlation analysis between each measure and its respective ranked version, using the \texttt{unq} tie breaking approach.}
\label{tab:smrytau-ranked-unq}
\resizebox{\columnwidth}{!}{
\begin{tabular}{|l||*{6}{r|}r||*{4}{r|}} 
\hline\hline 
\multicolumn{1}{|c||}{\textbf{Track}} & \multicolumn{1}{c|}{\textbf{P}} & \multicolumn{1}{c|}{\textbf{RBP\_p05}} & \multicolumn{1}{c|}{\textbf{RR}} & \multicolumn{1}{c|}{\textbf{RBP\_p03}} & \multicolumn{1}{c|}{\textbf{RBP\_p08}} & \multicolumn{1}{c|}{\textbf{DCG\_b02}} & \multicolumn{1}{c||}{\textbf{DCG\_b10}} & \multicolumn{1}{c|}{\textbf{R}} & \multicolumn{1}{c|}{\textbf{AP}} & \multicolumn{1}{c|}{\textbf{nDCG\_b02}} & \multicolumn{1}{c|}{\textbf{nDCG\_b10}} \\ 
\hline 
\texttt{T08\_05}  & 1.0000  & 1.0000  & 0.9211  & 0.9522  & 0.9605  & 0.9759  & 1.0000  & 0.8145  & 0.8219  & 0.9759  & 1.0000   \\ 
\hline 
\texttt{T08\_10}  & 1.0000  & 1.0000  & 0.8466  & 0.9500  & 0.9527  & 0.9553  & 1.0000  & 0.8030  & 0.8243  & 0.9541  & 0.9965   \\ 
\hline 
\texttt{T08\_20}  & 1.0000  & 1.0000  & 0.7677  & 0.9498  & 0.9498  & 0.9334  & 0.9537  & 0.7943  & 0.8197  & 0.9285  & 0.9452   \\ 
\hline 
\texttt{T08\_30}  & 1.0000  & 1.0000  & 0.7329  & 0.9498  & 0.9508  & 0.9128  & 0.9261  & 0.7948  & 0.8377  & 0.9072  & 0.9147   \\ 
\hline\hline 
\texttt{T26\_05}  & 1.0000  & 1.0000  & 0.9219  & 0.9706  & 0.9661  & 0.9717  & 1.0000  & 0.6974  & 0.6903  & 0.9717  & 1.0000   \\ 
\hline 
\texttt{T26\_10}  & 1.0000  & 1.0000  & 0.8232  & 0.9704  & 0.9610  & 0.9567  & 1.0000  & 0.8207  & 0.7633  & 0.9582  & 0.9982   \\ 
\hline 
\texttt{T26\_20}  & 1.0000  & 1.0000  & 0.7500  & 0.9704  & 0.9560  & 0.9517  & 0.9690  & 0.8848  & 0.8600  & 0.9560  & 0.9661   \\ 
\hline 
\texttt{T26\_30}  & 1.0000  & 1.0000  & 0.6725  & 0.9704  & 0.9582  & 0.9264  & 0.9127  & 0.8901  & 0.8701  & 0.9264  & 0.9141   \\ 
\hline\hline 
\texttt{T27\_05}  & 1.0000  & 1.0000  & 0.9350  & 0.9597  & 0.9536  & 0.9730  & 1.0000  & 0.7540  & 0.8312  & 0.9730  & 1.0000   \\ 
\hline 
\texttt{T27\_10}  & 1.0000  & 1.0000  & 0.8830  & 0.9601  & 0.9476  & 0.9436  & 1.0000  & 0.7860  & 0.8442  & 0.9491  & 0.9912   \\ 
\hline 
\texttt{T27\_20}  & 1.0000  & 1.0000  & 0.8227  & 0.9601  & 0.9288  & 0.9272  & 0.9358  & 0.7929  & 0.8309  & 0.9155  & 0.9068   \\ 
\hline 
\texttt{T27\_30}  & 1.0000  & 1.0000  & 0.7958  & 0.9601  & 0.9303  & 0.9295  & 0.9397  & 0.8191  & 0.8380  & 0.9068  & 0.9139   \\ 
\hline 
\hline 
\end{tabular} 

 }
\end{table}

\subsubsection{Measures not depending on the recall base}

As previously discussed, Precision is already an interval scale -- a different scale for each run length but, fixed a length, the same scale for all the topics, allowing to safely average across them. In this case, our transformation is just a mapping between interval scales, as the transformation between Celsius and Fahrenheit is. We can see as the overall Kendall's $\tau$ correlation is always $1.0$, experimentally confirming the correctness of our transformation approach and that everything is working as expected.

The other case in which we see this happening is RBP\_p05, which we already know to be an interval scale, but different from the one of Precision.

On the other extreme, there is \ac{RR}, which is the farthest away from being an interval scale, among the measures not depending on the recall base. We can observe as the overall Kendall's $\tau$ correlation is in the range $0.67 - 0.93$ and it is systematically lower then the correlation of all the other measures in this group -- namely P, RBP, and DCG. This suggests that transforming \ac{RR} into an interval scale requires a more marked correction or, in other terms, that it experiences a drop in ``intervalness'' in the range $7\% - 33\%$. We can observe also another stable pattern in the case of \ac{RR}: the bigger the length of the run, the lower the correlation between \ac{RR} and its ranked version. This suggests that \ac{RR} departs more and more from the interval scale assumption as the run length increases; we will now see why this is happening. 

\begin{figure}[tb]
    \centering
    \includegraphics[width=\linewidth]{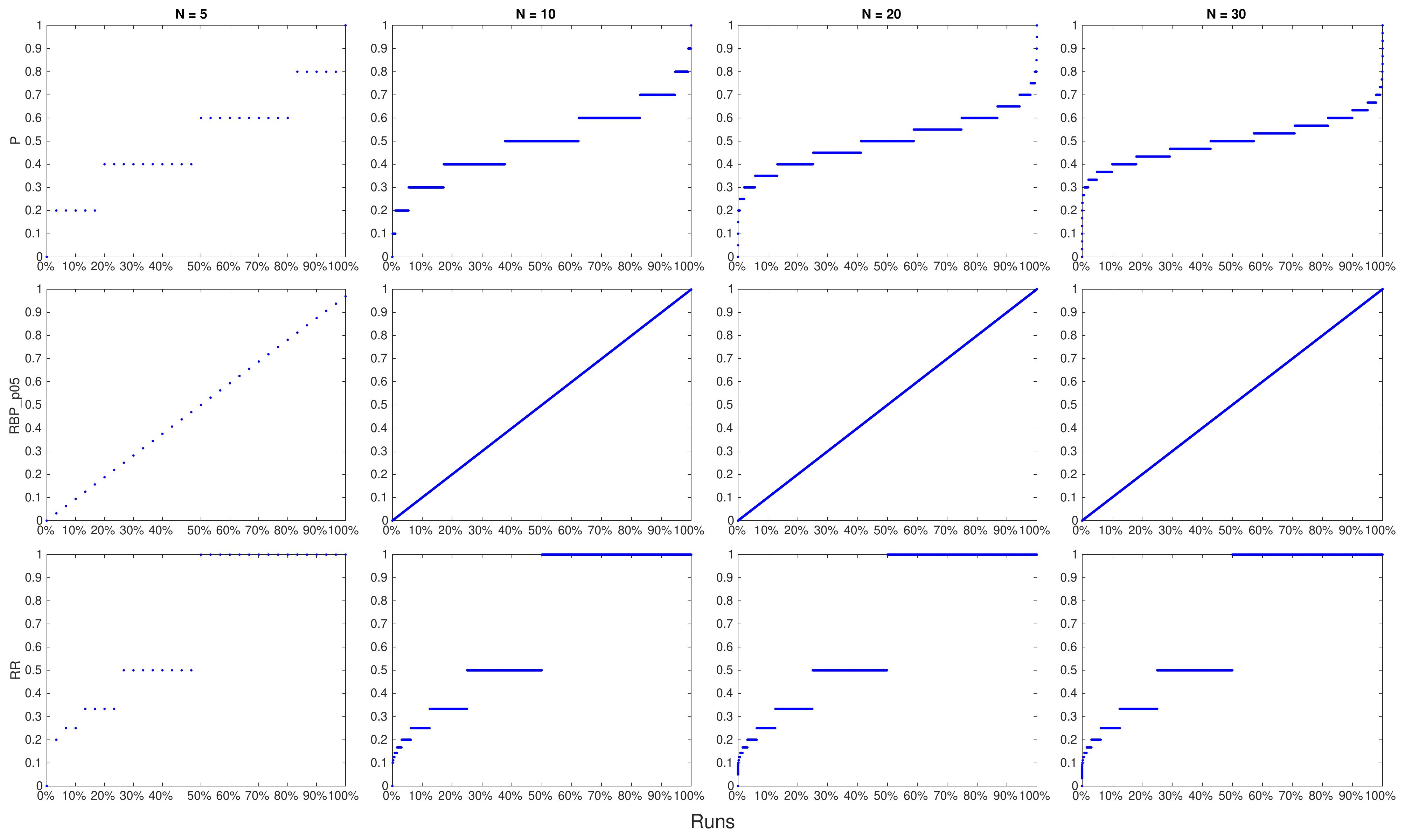}
    \caption{All the possible values of P, \ac{RBP} with $p=0.5$, and \ac{RR} for runs of length $\{5, 10, 20, 30\}$.}
    \label{fig:acr_p_rbp05_rr_steps}
\end{figure}

Figure~\ref{fig:acr_p_rbp05_rr_steps} plots the values of P, \ac{RBP} with $p=0.5$, and \ac{RR} for all the possible runs of a given length. On the X axis there are the runs, increasingly ordered by the value of the measure; this is the order of runs considered by the ranked version of the measure which then just equi-spaces the values and removes ties. The Y axis reports the value of the measure for each run. The labels on the X axis report the fraction of runs up to that point; so, for example, in the case of P and $N=5$, we can understand that $20\%$ of the runs assume the value $P=0.2$, $30\%$ the value $P=0.4$, $30\%$ the value $P=0.6$, and $20\%$ the value $P=0.8$; $1$ run the value $0.0$ and $1$ run the value $1.0$.

We can observe as \ac{RBP} with $p=0.5$ produces distinct equi-spaced values for each of the possible runs or, in other terms, it produces equi-spaced clusters of values containing one single value in each cluster. In the case of P, we can see how the clusters are still equi-spaced but they contain tied values, visible as horizontal segments. Finally, in the case of \ac{RR}, not only the clusters are not equi-spaced -- and this breaks the interval scale assumption -- but they also increase more and more, and only in one region of the range, as the run length increases, making \ac{RR} less and less interval. Indeed, the number of clusters is not equi-spaced but always constant to $5$ in the range $[0.2, 1.0]$, independently from the run length; on the other hand, in the range $[0, 0.2)$ it increases from $1$ to $6$, $16$, and $26$ as the run length increases. 

Moreover, Figure~\ref{fig:acr_p_rbp05_rr_steps} visually shows us why different run lengths correspond to different scales -- interval or not depending on whether clusters are equi-spaced or not. In all the cases, the number of clusters increases as the length of the run increases and this makes the scale to be different. Note that this behaviour is not like getting a more and more accurate scale, which would be a desirable property, but it is rather like saying that the height scale would become denser and denser as you measure taller and taller people. Therefore, as already said, we should avoid to mix values of measures coming from runs with different lengths, e.g. by averaging.

When it comes to \ac{RBP}, we know from~\citet{FerranteEtAl2018b,FerranteEtAl2017} that: RBP\_p05 is an interval scale; RBP\_p03 is an ordinal scale keeping the same ordering as RBP\_p05 but being no more an interval scale; and, RBP\_p08 uses a different ordering from RBP\_p05 and it is not an interval scale. This is also reflected in the overall correlation values. As already noted, and expected, the overall correlation for RBP\_p05 is always $1.0$ while it drops in the range $0.95-0.97$ for RBP\_p03. RBP\_p03 and RBP\_p05 order runs in the same way, which also means that their ranked version is the same. Therefore, the $3\%-5\%$  difference between RBP\_p03 and RBP\_p05 depends only on the lack of equi-spacing of RBP\_p03 and the problems it causes when averaging. This also means that this drop in ``intervalness'' of RBP\_p03 is not the effect of a user model somehow different from the one of RBP\_p05, possibly resulting in a different order of the systems, which is the typical explanation provided in these cases instead. In the case of RBP\_p08 we observe a similar behaviour and the correlation drops in the range $0.93-0.96$ with an ``intervalness'' loss in the range $4\%-7\%$.

\begin{figure}[tb]
    \centering
    \includegraphics[width=\linewidth]{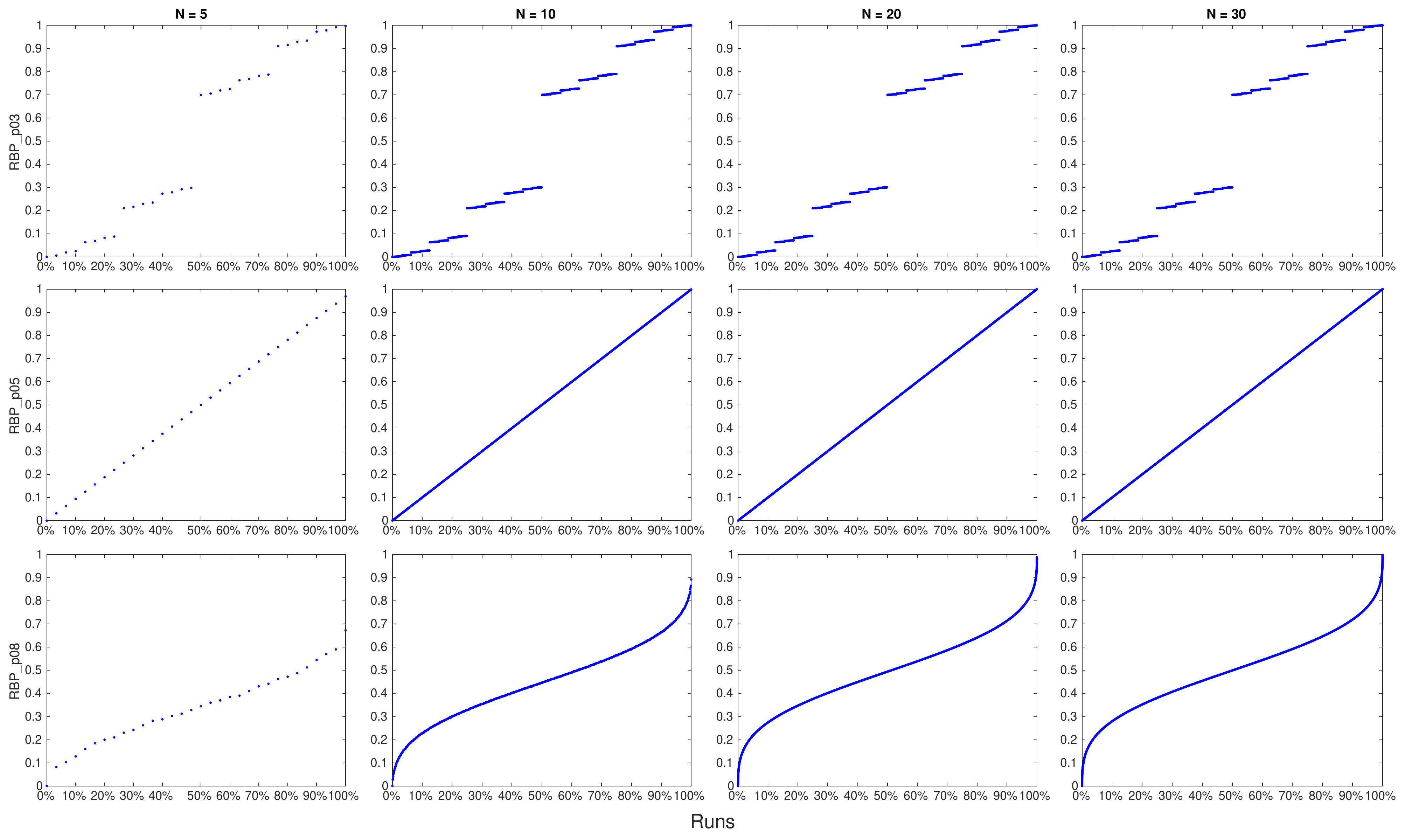}
    \caption{All the possible values of \ac{RBP} with $p \in \{0.3, 0.5, 0.8\}$ for runs of length $\{5, 10, 20, 30\}$.}
    \label{fig:acr_rbp_steps}
\end{figure}

When it comes to increasing run lengths, we can observe that the correlation values of \ac{RBP} oscillate a bit, they tend to get more stable as the run length increases and this happens more for RBP\_p08 than for RBP\_p03. While this still might be partially due to the measure being more or less interval scale depending on the run length, we think that in the case of \ac{RBP} this is mostly motivated by another reason. Indeed, as previously discussed, \ac{RBP} does not use the full range $[0, 1]$ because of the $\frac{1}{1-p}$ overestimation which impacts more as $p$ increases and the length of the run is smaller. Therefore, we think that the increase in range of \ac{RBP} is the motivation of the observed small changes in the correlation values. We can clearly see this behaviour in Figure~\ref{fig:acr_rbp_steps} for RBP\_p08 whose values fall in the full range  $[0, 1]$ only for $N=20$ and $N=30$, while this effect is mostly negligible for RBP\_p03 and RBP\_p05. As a consequence, correlation values tend to get more stable for $N=20$ and $N=30$ in the case of RBP\_p08 while they are quite stable for RBP\_p03 and RBP\_p05, independently from the run length.

\begin{figure}[tb]
    \centering
    \includegraphics[width=\linewidth]{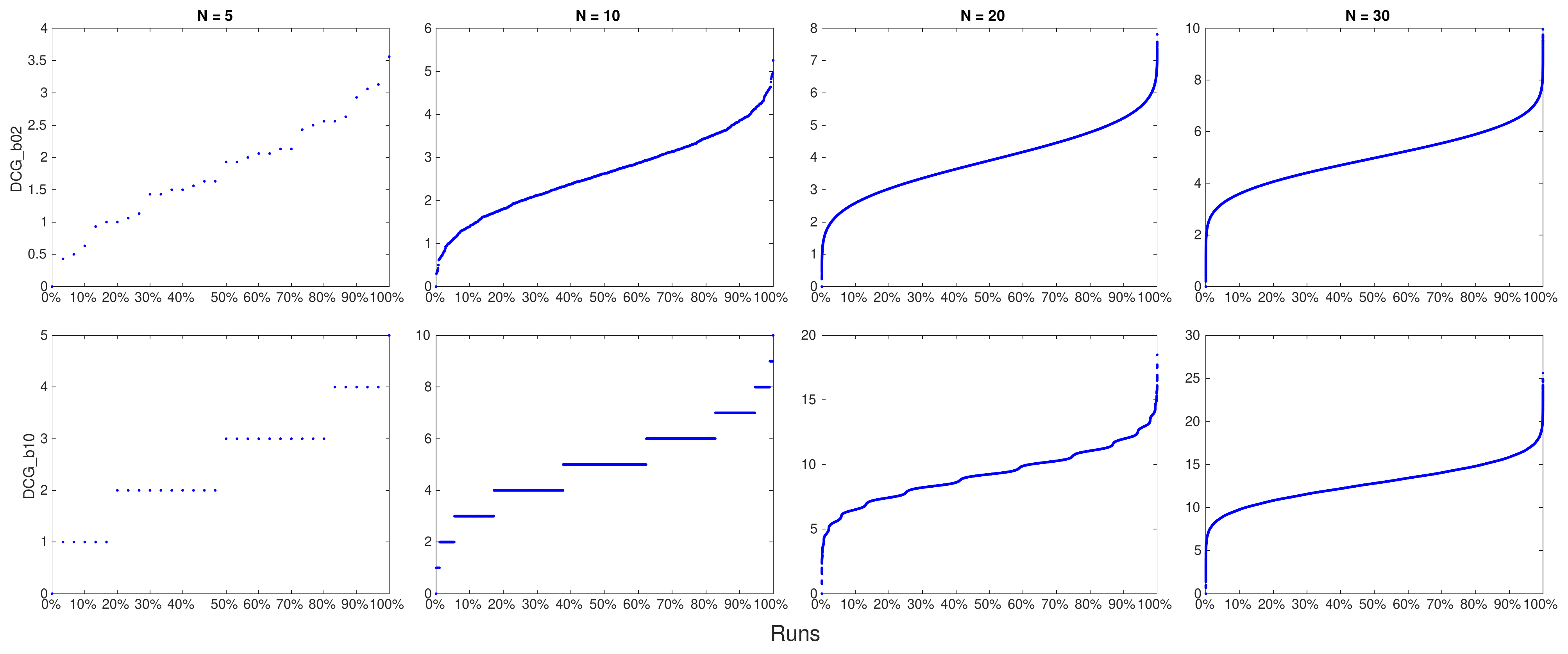}
    \caption{All the possible values of DCG with log base $b=2$ and $b=10$ for runs of length $\{5, 10, 20, 30\}$.}
    \label{fig:acr_dcg_steps}
\end{figure}

Finally, when it comes to \ac{DCG}, we can observe from Table~\ref{tab:smrytau-ranked-unq} that its overall correlation is above $0.9$ with an ``intervalness'' loss in the range $2\% - 9\%$, suggesting it is moderately departing from its ranked version. We can also observe for DCG\_b10 that the correlation for run lengths $N=5$ and $N=10$ is always $1.0$, which may look surprising; this is actually an artefact of the log base 10 which causes the discount to be applied from the 11th rank onwards. Therefore, for run lengths up to the log base, DCG is basically counting the number of relevant retrieved documents, as it is clear from Figure~\ref{fig:acr_dcg_steps}, and this produces the same interval scale as P. However, we should be aware of this somehow unusual behaviour of DCG\_b10 because it changes from being an interval scale for runs up to length 10 to not being it anymore afterwards.

Moreover, as a general trend, \ac{DCG} tends to be less and less an interval scale as the run length increases. If we look at the possible values of \ac{DCG} in Figure~\ref{fig:acr_dcg_steps}, this may sound surprising since \ac{DCG} visually behaves in a very similar way to RBP\_p08, at least after the run length is big enough to compensate for possible effects of the log base itself. However, while RBP\_08 does not have tied values, \ac{DCG} exhibits an increasing number of tied values, clustered unevenly across the range -- this is not visible from the figure, especially for DCB\_b02, due to the small size of the tied clusters but we have verified it on the numerical data underlying the plot. Therefore, the increasing number of uneven tied cluster explains, as in the case of \ac{RR}, why \ac{DCG} is less and less an interval scale.

\subsubsection{Measures depending on the recall base}
\label{subsubsec:measures-recall-base}

Let us go back to Table~\ref{tab:smrytau-ranked-unq} and consider the case of R. We know that Precision and Recall, on each topic separately, are already interval scales and just a transformation of the same interval scale. Therefore, when we map them to their ranked version, it is actually the same interval scale for both of them and it is yet another transformation of their common original interval scale. However, while this means an overall correlation $1.0$ in the case of Precision, it drops in the range $0.7 - 0.9$ in the case of Recall. This $10\% - 30\%$ loss in ``intervalness'' is entirely due to the effect of the recall base and let us understand how careful we should be before averaging across topics.

\ac{AP} follows a somehow similar pattern with overall correlation values in the range $0.69 - 0.86$ with an ``intervalness'' loss in the range $14\% - 31\%$.

On the other hand, \ac{nDCG} exhibit overall correlation values very close to those of \ac{DCG}, all above $0.9$ with an ``intervalness'' loss in the range $2\% - 10\%$. We observe another somehow surprising behaviour of nDCG\_b10: for runs of length $N=5$ the correlation is always $1.0$, indicating that it is an interval scale and, most of all, that there is no effect of the recall base. The fact is that on all the tracks under examination, all the topics have at least $5$ relevant documents, so the recall base is never below $5$; when you trim runs to length $N=5$, the DCG\_b10 of the ideal run, i.e. the factor used to normalize \ac{DCG} in \ac{nDCG}, is constant to $5$ for all the topics and so there is not recall base effect for this reason. On the other hand, there is $1$ topic with less than $10$ relevant documents on both \texttt{T08} and \texttt{T26} and 4 topics on \texttt{T27}. As a consequence, nDCG\_b10 drops slightly below $1.0$ on  \texttt{T08\_10} and \texttt{T26\_10} and a bit more on  \texttt{T27\_10}. This further stresses the need to be careful, or at least aware of, that DCG/nDCG may change behaviour and nature for document cut-offs below the log base. Moreover, this  gives us an idea of how much even very small changes in the recall base can have an impact and how careful we should be when aggregating across topics.

Finally, both \ac{DCG} and \ac{nDCG} are mapped to the same ranked measure, exactly as P and R are mapped to the same ranked measure. However the loss of ``intervalness'' of R is much bigger than the one of \ac{nDCG}. We hypothesise that this is due to how the recall base is accounted for in the measure: in the case of R it is a straight division by the recall base itself while in the case of \ac{nDCG} it is a division by the \ac{DCG} of the ideal run, which is also one of the possible runs considered in the mapping. The latter is a much smoother normalisation than just an integer number representing the total number of relevant documents. The behaviour of \ac{AP}, very close to the one of R, supports this intuition, since also \ac{AP} adopts a straight division by the recall base itself.

\subsubsection{Impact of the tie breaking strategy}

\begin{table}[tb] 
\centering 
\caption{Kendall's $\tau$ overall correlation analysis between each measure and its respective ranked version, using the \texttt{mid}-rank tie breaking approach.}
\label{tab:smrytau-ranked-mid}
\resizebox{\columnwidth}{!}{
\begin{tabular}{|l||*{6}{r|}r||*{4}{r|}} 
\hline\hline 
\multicolumn{1}{|c||}{\textbf{Track}} & \multicolumn{1}{c|}{\textbf{P}} & \multicolumn{1}{c|}{\textbf{RBP\_p05}} & \multicolumn{1}{c|}{\textbf{RR}} & \multicolumn{1}{c|}{\textbf{RBP\_p03}} & \multicolumn{1}{c|}{\textbf{RBP\_p08}} & \multicolumn{1}{c|}{\textbf{DCG\_b02}} & \multicolumn{1}{c||}{\textbf{DCG\_b10}} & \multicolumn{1}{c|}{\textbf{R}} & \multicolumn{1}{c|}{\textbf{AP}} & \multicolumn{1}{c|}{\textbf{nDCG\_b02}} & \multicolumn{1}{c|}{\textbf{nDCG\_b10}} \\ 
\hline 
\texttt{T08\_05}  & 0.9680  & 1.0000  & 0.9694  & 0.9522  & 0.9605  & 0.9622  & 0.9680  & 0.7973  & 0.8208  & 0.9622  & 0.9680   \\ 
\hline 
\texttt{T08\_10}  & 0.9392  & 1.0000  & 0.9713  & 0.9500  & 0.9527  & 0.9462  & 0.9392  & 0.7725  & 0.8213  & 0.9450  & 0.9362   \\ 
\hline 
\texttt{T08\_20}  & 0.9043  & 1.0000  & 0.9736  & 0.9498  & 0.9498  & 0.9276  & 0.9045  & 0.7809  & 0.8161  & 0.9227  & 0.8970   \\ 
\hline 
\texttt{T08\_30}  & 0.8978  & 1.0000  & 0.9743  & 0.9498  & 0.9508  & 0.9094  & 0.8931  & 0.7625  & 0.8401  & 0.9038  & 0.8851   \\ 
\hline\hline 
\texttt{T26\_05}  & 0.9670  & 1.0000  & 0.9794  & 0.9706  & 0.9661  & 0.9614  & 0.9670  & 0.6816  & 0.6907  & 0.9614  & 0.9670   \\ 
\hline 
\texttt{T26\_10}  & 0.9624  & 1.0000  & 0.9782  & 0.9704  & 0.9610  & 0.9502  & 0.9624  & 0.8153  & 0.7540  & 0.9502  & 0.9599   \\ 
\hline 
\texttt{T26\_20}  & 0.9245  & 1.0000  & 0.9854  & 0.9704  & 0.9560  & 0.9488  & 0.9242  & 0.8615  & 0.8557  & 0.9517  & 0.9257   \\ 
\hline 
\texttt{T26\_30}  & 0.8785  & 1.0000  & 0.9832  & 0.9704  & 0.9582  & 0.9228  & 0.8831  & 0.8319  & 0.8665  & 0.9228  & 0.8846   \\ 
\hline\hline 
\texttt{T27\_05}  & 0.9535  & 1.0000  & 0.9782  & 0.9597  & 0.9536  & 0.9607  & 0.9535  & 0.7501  & 0.8294  & 0.9607  & 0.9535   \\ 
\hline 
\texttt{T27\_10}  & 0.9432  & 1.0000  & 0.9827  & 0.9601  & 0.9476  & 0.9272  & 0.9432  & 0.7628  & 0.8521  & 0.9327  & 0.9354   \\ 
\hline 
\texttt{T27\_20}  & 0.9086  & 1.0000  & 0.9843  & 0.9601  & 0.9288  & 0.9225  & 0.9084  & 0.7352  & 0.8356  & 0.9076  & 0.8873   \\ 
\hline 
\texttt{T27\_30}  & 0.9103  & 1.0000  & 0.9835  & 0.9601  & 0.9303  & 0.9264  & 0.9100  & 0.7650  & 0.8395  & 0.9037  & 0.8873   \\ 
\hline 
\hline 
\end{tabular} 
}
\end{table} 

In this section, we perform a further validation of our mapping approach. As explained in Section~\ref{sec:methodology}, evaluation measures often produce tied values and we remove these tied values by assigning them their unique rank position, since this ensures that values are kept equi-spaced. However, as pointed out by~\citet{GibbonsChakraborti2011}, there are many other common ways of breaking ties, one of which if the mid-rank strategy, i.e. keeping the average of the ranks of the tied values.

Table~\ref{tab:smrytau-ranked-mid} shows what happens to our transformation approach when using the \texttt{mid}-rank tie breaking instead of the \texttt{unq} one used in Table~\ref{tab:smrytau-ranked-unq}. Let us consider Precision whose overall correlation values drops from $1.0$ to the range $0.90 - 0.97$. Since Precision is already an interval scale, this drop is entirely due to the fact that the \texttt{mid}-rank tie breaking strategy produces a scale whose values are not equi-spaced anymore and thus a scale which is not interval anymore.  Only RBP\_p05 keeps the overall correlation $1.0$ because it is already an interval scale but it does not have any tied value, so it is insensitive to the tie breaking strategy. As a general trend, we can see that the overall correlation values in Table~\ref{tab:smrytau-ranked-mid}  are lower than those in Table~\ref{tab:smrytau-ranked-unq} due to the loss of ``intervalness'' caused by the tie breaking strategy.

Therefore, we validated that the appropriate way of implementing our transformation approach is by using the \texttt{unq} tie breaking strategy.  Moreover, this further stresses how much the lack of equi-spaced values, lack due to any reason, impacts on our measurement process.

\subsection{Correlation among measures and among their ranked versions. Unveiling the ``true'' correlation among evaluation measures}
\label{subsec:correlation-msr}

Table~\ref{tab:smrytau-msr} summarizes the outcomes of the correlation analysis among measures and among their ranked versions, i.e. on the one side we compute  Kendall's $\tau$ overall correlation among all pairs of measures, on the other side we compute  Kendall's $\tau$ overall correlation among the same pairs of ranked measures. In this way, we can study whether and how the estimated relationship among measures changes when passing to their ranked version or, in other terms, to what extent being an interval scale or not biases our estimations. In particular, the column \texttt{$\Delta\%$} reports the percent increase/decrease of the correlation between the ranked measures (labelled \texttt{RnkMsr}) with respect to the correlation between the original measures (labelled \texttt{Msr}), i.e. how much the correlation between two measures is underestimated/overestimated due to the fact that a measure is not an interval scale. Table~\ref{tab:smrytau-msr} reports results for the \texttt{T08\_30}, \texttt{T26\_30}, and \texttt{T27\_30} tracks; results for the other tracks are similar but not shown here for space reasons.

\begin{table}[p] 
\centering 
\caption{Kendall's $\tau$ overall correlation analysis between each pair of measures (labelled \texttt{Msr}) and between each pair of ranked measures (labelled \texttt{RnkMsr}) on tracks \texttt{T08\_30}, \texttt{T26\_30}, and \texttt{T27\_30}, using the \texttt{unq} tie breaking approach. The $\Delta\%$ column reports the percent increase/decrease of the  \texttt{RnkMsr} correlation with respect to the \texttt{Msr} one.}
\label{tab:smrytau-msr}
\small
\resizebox{0.95\columnwidth}{!}{
\begin{tabular}{|l*{3}{||r|r|r}|} 
\hline\hline 
\multicolumn{1}{|c||}{\textbf{Measure}} & \multicolumn{3}{c||}{\textbf{\texttt{T08\_30}}} & \multicolumn{3}{c||}{\textbf{\texttt{T26\_30}}} & \multicolumn{3}{c|}{\textbf{\texttt{T27\_30}}} \\ 
\cline{2-10}
& \multicolumn{1}{c|}{\textbf{Msr}} & \multicolumn{1}{c|}{\textbf{RnkMsr}} & \multicolumn{1}{c||}{$\mathbf{\Delta\%}$} & \multicolumn{1}{c|}{\textbf{Msr}} & \multicolumn{1}{c|}{\textbf{RnkMsr}} & \multicolumn{1}{c||}{$\mathbf{\Delta\%}$} & \multicolumn{1}{c|}{\textbf{Msr}} & \multicolumn{1}{c|}{\textbf{RnkMsr}} & \multicolumn{1}{c|}{$\mathbf{\Delta\%}$} \\ 
\hline 
P vs RBP\_p05  & 0.7858 & 0.7858 & +0.00\%  & 0.8604 & 0.8604 & +0.00\%  & 0.7963 & 0.7963 & +0.00\%   \\ 
\hline 
P vs RR  & 0.7151 & 0.6322 & -11.60\%  & 0.8126 & 0.6049 & -25.56\%  & 0.7615 & 0.6764 & -11.18\%   \\ 
\hline 
P vs RBP\_p03  & 0.7494 & 0.7858 & +4.86\%  & 0.8503 & 0.8604 & +1.19\%  & 0.7798 & 0.7963 & +2.11\%   \\ 
\hline 
P vs RBP\_p08  & 0.8641 & 0.8447 & -2.24\%  & 0.9009 & 0.8944 & -0.72\%  & 0.8748 & 0.8465 & -3.23\%   \\ 
\hline 
P vs DCG\_b02  & 0.9352 & 0.8962 & -4.17\%  & 0.9623 & 0.9312 & -3.23\%  & 0.9478 & 0.9203 & -2.90\%   \\ 
\hline 
P vs DCG\_b10  & 0.9866 & 0.9243 & -6.32\%  & 0.9861 & 0.9254 & -6.15\%  & 0.9871 & 0.9360 & -5.17\%   \\ 
\hline 
RBP\_p05 vs RBP\_p03  & 0.9498 & 1.0000 & +5.29\%  & 0.9704 & 1.0000 & +3.05\%  & 0.9601 & 1.0000 & +4.16\%   \\ 
\hline 
RBP\_p05 vs RBP\_p08  & 0.9045 & 0.9082 & +0.40\%  & 0.9250 & 0.9351 & +1.09\%  & 0.9068 & 0.8998 & -0.78\%   \\ 
\hline 
RBP\_p05 vs RR  & 0.8840 & 0.6733 & -23.84\%  & 0.9046 & 0.6228 & -31.15\%  & 0.9230 & 0.7590 & -17.77\%   \\ 
\hline 
RBP\_p05 vs DCG\_b02  & 0.8461 & 0.8166 & -3.49\%  & 0.8896 & 0.8593 & -3.41\%  & 0.8489 & 0.8082 & -4.79\%   \\ 
\hline 
RBP\_p05 vs DCG\_b10  & 0.7931 & 0.7686 & -3.09\%  & 0.8687 & 0.7958 & -8.39\%  & 0.8082 & 0.7886 & -2.42\%   \\ 
\hline 
RBP\_p03 vs RBP\_p08  & 0.8606 & 0.9082 & +5.53\%  & 0.9069 & 0.9351 & +3.10\%  & 0.8701 & 0.8998 & +3.42\%   \\ 
\hline 
RBP\_p03 vs RR  & 0.9144 & 0.6733 & -26.37\%  & 0.9154 & 0.6228 & -31.97\%  & 0.9394 & 0.7590 & -19.21\%   \\ 
\hline 
RBP\_p03 vs DCG\_b02  & 0.8054 & 0.8166 & +1.39\%  & 0.8773 & 0.8593 & -2.06\%  & 0.8215 & 0.8082 & -1.62\%   \\ 
\hline 
RBP\_p03 vs DCG\_b10  & 0.7547 & 0.7686 & +1.84\%  & 0.8564 & 0.7958 & -7.08\%  & 0.7918 & 0.7886 & -0.40\%   \\ 
\hline 
RBP\_p08 vs RR  & 0.8133 & 0.6430 & -20.94\%  & 0.8656 & 0.6307 & -27.13\%  & 0.8547 & 0.7213 & -15.61\%   \\ 
\hline 
RBP\_p08 vs DCG\_b02  & 0.9266 & 0.8822 & -4.79\%  & 0.9372 & 0.8939 & -4.62\%  & 0.9233 & 0.8646 & -6.36\%   \\ 
\hline 
RBP\_p08 vs DCG\_b10  & 0.8745 & 0.8333 & -4.71\%  & 0.9149 & 0.8304 & -9.23\%  & 0.8857 & 0.8434 & -4.77\%   \\ 
\hline 
RR vs DCG\_b02  & 0.7668 & 0.5971 & -22.13\%  & 0.8375 & 0.5902 & -29.52\%  & 0.7998 & 0.6545 & -18.17\%   \\ 
\hline 
RR vs DCG\_b10  & 0.7214 & 0.5896 & -18.28\%  & 0.8209 & 0.5475 & -33.30\%  & 0.7700 & 0.6514 & -15.41\%   \\ 
\hline 
DCG\_b02 vs DCG\_b10  & 0.9435 & 0.9273 & -1.72\%  & 0.9747 & 0.9221 & -5.40\%  & 0.9593 & 0.9632 & +0.41\%   \\ 
\hline 
R vs P  & 0.7948 & 1.0000 & +25.82\%  & 0.8901 & 1.0000 & +12.35\%  & 0.8191 & 1.0000 & +22.08\%   \\ 
\hline 
R vs RBP\_p05  & 0.6848 & 0.7858 & +14.76\%  & 0.8102 & 0.8604 & +6.20\%  & 0.8098 & 0.7963 & -1.67\%   \\ 
\hline 
R vs RBP\_p03  & 0.6648 & 0.7858 & +18.20\%  & 0.8052 & 0.8604 & +6.86\%  & 0.8169 & 0.7963 & -2.52\%   \\ 
\hline 
R vs RBP\_p08  & 0.7424 & 0.8447 & +13.78\%  & 0.8391 & 0.8944 & +6.59\%  & 0.8082 & 0.8465 & +4.74\%   \\ 
\hline 
R vs RR  & 0.6611 & 0.6322 & -4.37\%  & 0.7841 & 0.6049 & -22.86\%  & 0.7971 & 0.6764 & -15.14\%   \\ 
\hline 
R vs DCG\_b02  & 0.7756 & 0.8962 & +15.54\%  & 0.8730 & 0.9312 & +6.67\%  & 0.8169 & 0.9203 & +12.67\%   \\ 
\hline 
R vs DCG\_b10  & 0.7957 & 0.9243 & +16.15\%  & 0.8853 & 0.9254 & +4.54\%  & 0.8184 & 0.9360 & +14.37\%   \\ 
\hline 
R vs AP  & 0.8859 & 0.8932 & +0.83\%  & 0.8709 & 0.9319 & +7.01\%  & 0.9078 & 0.9211 & +1.46\%   \\ 
\hline 
R vs nDCG\_b02  & 0.7899 & 0.8962 & +13.45\%  & 0.8788 & 0.9312 & +5.97\%  & 0.8427 & 0.9203 & +9.21\%   \\ 
\hline 
R vs nDCG\_b10  & 0.8338 & 0.9243 & +10.85\%  & 0.8911 & 0.9254 & +3.86\%  & 0.8490 & 0.9360 & +10.25\%   \\ 
\hline 
AP vs P  & 0.8244 & 0.8932 & +8.35\%  & 0.8583 & 0.9319 & +8.58\%  & 0.8481 & 0.9211 & +8.61\%   \\ 
\hline 
AP vs RBP\_p05  & 0.7524 & 0.7936 & +5.48\%  & 0.8326 & 0.8413 & +1.04\%  & 0.8614 & 0.7980 & -7.36\%   \\ 
\hline 
AP vs RBP\_p03  & 0.7271 & 0.7936 & +9.14\%  & 0.8160 & 0.8413 & +3.09\%  & 0.8575 & 0.7980 & -6.94\%   \\ 
\hline 
AP vs RBP\_p08  & 0.8081 & 0.8587 & +6.27\%  & 0.8672 & 0.8730 & +0.67\%  & 0.8685 & 0.8466 & -2.52\%   \\ 
\hline 
AP vs RR  & 0.7030 & 0.5869 & -16.52\%  & 0.7761 & 0.5707 & -26.47\%  & 0.8249 & 0.6545 & -20.66\%   \\ 
\hline 
AP vs DCG\_b02  & 0.8316 & 0.9673 & +16.32\%  & 0.8651 & 0.9632 & +11.34\%  & 0.8748 & 0.9757 & +11.54\%   \\ 
\hline 
AP vs DCG\_b10  & 0.8280 & 0.9411 & +13.67\%  & 0.8672 & 0.9444 & +8.90\%  & 0.8575 & 0.9750 & +13.69\%   \\ 
\hline 
AP vs nDCG\_b02  & 0.8454 & 0.9673 & +14.42\%  & 0.8709 & 0.9632 & +10.60\%  & 0.8912 & 0.9757 & +9.49\%   \\ 
\hline 
AP vs nDCG\_b10  & 0.8563 & 0.9411 & +9.90\%  & 0.8687 & 0.9444 & +8.72\%  & 0.8849 & 0.9750 & +10.17\%   \\ 
\hline 
nDCG\_b02 vs P  & 0.9330 & 0.8962 & -3.95\%  & 0.9579 & 0.9312 & -2.79\%  & 0.9305 & 0.9203 & -1.10\%   \\ 
\hline 
nDCG\_b02 vs RBP\_p05  & 0.8449 & 0.8166 & -3.36\%  & 0.8925 & 0.8593 & -3.72\%  & 0.8638 & 0.8082 & -6.43\%   \\ 
\hline 
nDCG\_b02 vs RBP\_p03  & 0.8051 & 0.8166 & +1.42\%  & 0.8773 & 0.8593 & -2.06\%  & 0.8395 & 0.8082 & -3.73\%   \\ 
\hline 
nDCG\_b02 vs RBP\_p08  & 0.9283 & 0.8822 & -4.96\%  & 0.9401 & 0.8939 & -4.91\%  & 0.9319 & 0.8646 & -7.22\%   \\ 
\hline 
nDCG\_b02 vs RR  & 0.7684 & 0.5971 & -22.30\%  & 0.8389 & 0.5902 & -29.65\%  & 0.8179 & 0.6545 & -19.98\%   \\ 
\hline 
nDCG\_b02 vs DCG\_b02  & 0.9804 & 1.0000 & +2.00\%  & 0.9913 & 1.0000 & +0.87\%  & 0.9695 & 1.0000 & +3.15\%   \\ 
\hline 
nDCG\_b02 vs DCG\_b10  & 0.9418 & 0.9273 & -1.54\%  & 0.9719 & 0.9221 & -5.12\%  & 0.9429 & 0.9632 & +2.16\%   \\ 
\hline 
nDCG\_b02 vs nDCG\_b10  & 0.9368 & 0.9273 & -1.01\%  & 0.9690 & 0.9221 & -4.84\%  & 0.9577 & 0.9632 & +0.57\%   \\ 
\hline 
nDCG\_b10 vs P  & 0.9568 & 0.9243 & -3.40\%  & 0.9832 & 0.9254 & -5.88\%  & 0.9525 & 0.9360 & -1.73\%   \\ 
\hline 
nDCG\_b10 vs RBP\_p05  & 0.7841 & 0.7686 & -1.98\%  & 0.8644 & 0.7958 & -7.93\%  & 0.8278 & 0.7886 & -4.73\%   \\ 
\hline 
nDCG\_b10 vs RBP\_p03  & 0.7472 & 0.7686 & +2.86\%  & 0.8535 & 0.7958 & -6.76\%  & 0.8129 & 0.7886 & -2.99\%   \\ 
\hline 
nDCG\_b10 vs RBP\_p08  & 0.8684 & 0.8333 & -4.05\%  & 0.9120 & 0.8304 & -8.94\%  & 0.8928 & 0.8434 & -5.52\%   \\ 
\hline 
nDCG\_b10 vs RR  & 0.7212 & 0.5896 & -18.25\%  & 0.8194 & 0.5475 & -33.18\%  & 0.7881 & 0.6514 & -17.35\%   \\ 
\hline 
nDCG\_b10 vs DCG\_b02  & 0.9288 & 0.9273 & -0.16\%  & 0.9719 & 0.9221 & -5.12\%  & 0.9507 & 0.9632 & +1.32\%   \\ 
\hline 
nDCG\_b10 vs DCG\_b10  & 0.9610 & 1.0000 & +4.06\%  & 0.9899 & 1.0000 & +1.02\%  & 0.9616 & 1.0000 & +3.99\%   \\ 
\hline 
\hline 
\end{tabular} 
}
\end{table}

We can observe from Table~\ref{tab:smrytau-msr},  as very coarse and general trends, that correlation is overestimated (\texttt{$\Delta\%$} column negative) in the range $[-33.30\%, \, -0.16\%]$, i.e. two evaluation measures are less close to each other than what we would be induced to think; conversely, correlation is underestimated (\texttt{$\Delta\%$} column positive) in the range $[0.4\%, \, 25.82\%]$, i.e. two evaluation measures are closer to each other than what we would be induced to think. This observation opens up a relevant question for IR experiments: are \ac{IR} measures really that different? Do we need all of them? Are we really scoring runs according to different user viewpoints or are these differences just an artefact of violating the scale assumptions? How much of what reported in the literature is due just to this scale violation bias?

In the following sections we discuss a few examples from Table~\ref{tab:smrytau-msr} of how the correlation may change.

\subsubsection{Measures not depending on the recall base}

Let us start from the correlation between Precision and \ac{RBP} with $p=0.5$. We already know that they are interval scales and, therefore, their ranked version is just another mapping of their respective interval scales -- and this is why in Table~\ref{tab:smrytau-ranked-unq} their overall correlation is $1.0$. We can observe from Table~\ref{tab:smrytau-msr} that on \texttt{T08\_30} the correlation between RBP\_p05 and P is $0.7858$ and, as expected, the correlation between their ranked versions is the same, since the interval scale behind the original measures and their ranked version is the same. The same happens for the other tracks,  i.e. \texttt{T26\_30} and  \texttt{T27\_30}. 

Note that ``the correlation between Precision and \ac{RBP} with $p=0.5$ on \texttt{T08\_30} is $0.7858$'' is an example of a \emph{meaningful statement} in \ac{IR}, since it is invariant to a permissible transformation of the interval scales of these two measures and it does not change its truth value.

The correlation between \ac{RBP} with $p=0.3$ and \ac{RBP} with $p=0.5$ is $0.9498$ while the correlation between their ranked versions is $1.0$. As we discussed in Section~\ref{subsec:correlation-rnk}, RBP\_p03 and RBP\_p05 order runs in the same way and so their correlation should be $1.0$. Therefore, this $5.3\%$ underestimation of the similarity between them is just due to RBP\_p03 not being an interval scale. Note that this case is somehow particularly severe since it induces us to attribute this $5.3\%$ change to other reasons; typical explanations for such changes you may find in studies about correlation among evaluation measures are: ``the user model behind RBP\_p03 slightly differs from the RBP\_p05 since it represents a more impatient or less motivated user'' or ``due to the smaller value of $p$, RBP\_p03 is a slightly more top-heavy measure''; unfortunately, none of these explanations would be correct since this $5.3\%$ change is just due to  fact that the values of RBP\_p03 are not equi-spaced, still ordering runs in exactly the same way as RBP\_p05.

For the sake of completeness, we can observe that $0.9498$ is the same correlation value reported in Table~\ref{tab:smrytau-ranked-unq} between RBP\_p03 and its ranked version. This is indeed correct since both RBP\_p03 and RBP\_p05 are mapped to the same ranked interval scale, which is just another mapping of the interval scale of RBP\_p05; therefore, the correlation between RBP\_p03 and its ranked version is the same as the correlation between RBP\_p03 and RBP\_p05.

Another interesting case is \ac{RR}: its correlation with respect to P, \ac{RBP}, and \ac{DCG} is way over-estimated -- in the range $12\% - 26\%$ more on \texttt{T08\_30}, $26\% - 33\%$ more on \texttt{T26\_30}, and $11\% - 19\%$ more on \texttt{T27\_30}, --  mistakenly suggesting us that this measure is closer to the others much more than what it actually is. As it emerges from the previous discussion, \ac{RR} is one of the measures which departs more from being an interval scale and which also has the highest number of tied values. Therefore, the computation of averages on \ac{RR} and on its ranked version leads to sensibly different \acf{RoS}, as it is clearly shown in Table~\ref{tab:smrytau-ranked-unq} when comparing \ac{RR} to its ranked version. As a consequence, the correlation of the ranked version of \ac{RR} with the other measures changes more than in the other cases. 

\subsubsection{Measures depending on the recall base}

Before proceeding, a word of caution has to be made remembering that in the case of measures depending on the recall base our approach is just a surrogate, which improves the ``intervalness'' of a measure but stretches the steps of the scale. Therefore, all the increases/decreases in correlations should be taken as tendency to overestimation/underestimation rather than exact quantification of it.

Let us consider Precision and Recall: we know that on each topic they are the same interval scale and this is reflected in Table~\ref{tab:smrytau-msr} in the correlation between their ranked version being $1.0$. On the other hand, the correlation between the original measures tends to be underestimated by $26\%$ on \texttt{T08\_30}, $12\%$ on \texttt{T26\_30}, and $22\%$ on \texttt{T27\_30}. Apart from suggesting that these two measures should be considered closer to each other than what they usually are, this wide range of underestimation further stresses how much just the recall base can affect the averaging across topics and how careful we should be with such averages -- not to say that we should avoid them at all.

Coherently with what discussed in Section~\ref{subsubsec:measures-recall-base}, we can observe that the correlation between \ac{DCG} and \ac{nDCG} tends to be underestimated by just $1\% - 4\%$, suggesting that they are practically equivalent. Therefore, even if usually \ac{nDCG} is preferred over \ac{DCG} because it is bounded and normalised, it could be actually better to use \ac{DCG} instead, since it avoids issues with the recall base and it can be easily turned into a proper interval scale by using our transformation approach.

Let us now discuss \ac{AP} with respect to Precision and Recall: the correlation with R is higher than the one with P and this is usually attributed to \ac{AP} embedding the recall base in the same way as  R does. However, when we turn to the ranked measures, we can see how the correlation between AP and R and AP and P is the same (for all the reasons already explained) and, especially, how this tends to be underestimated in the range $1\% - 9\%$, suggesting that \ac{AP} is slightly closer to these two measures than what is usually thought.

Finally, let us consider \ac{AP} with respect to \ac{DCG}: their correlation tends to be underestimated in the range $9\% - 16\%$ and the correlation between their ranked versions is actually quite -- high, between $0.94$ and $0.97$. This suggests that, even if these two measures have two quite different formulations and the user model of \ac{DCG} is considered much more realistic than the somehow artificial one of \ac{AP}, when they are turned into their interval scale version, they are much closer than expected and that part of their difference could have been  just due to their lack of ``intervalness''.

\subsection{Significance Testing Analysis. What systems are actually different, or not?}
\label{subsec:significance-test}

In this section, we analyse how the results of statistical significance tests change when using a measure or its ranked version. In other terms, we study how much statistical significance tests are impacted by using or not a proper interval scale. Indeed, as discussed in Section~\ref{subsec:significance-test}, there are significance tests which assume to work with just an ordinal scale and others which assume to work with an interval scale and they should be somehow affected by using a measure which matches or not their assumptions. By impacted, we mean that we can observe some change in which systems are considered significantly different or not.

Moreover, as discussed in the previous sections, the recall base makes working across topics problematic at best and statistical significance tests typically perform some aggregation across topics. Therefore, they may be further affected by the recall base.

As described in Section~\ref{subsec:significance-test}, we consider the following tests: Sign (ordinal scale assumption), Wilcoxon Rank Sum (ordinal scale assumption), Wilcoxon Signed Rank (interval scale assumption), Student's t (interval scale assumption), ANOVA (interval scale assumption), Kruskal-Wallis (ordinal scale assumption), and Friedman (ordinal scale assumption). For the \ac{ANOVA} case we consider two alternatives:
\begin{itemize}
 	\item \textbf{One-way ANOVA for the System Effect}: $y_{ij} = \mu_{\cdot\cdot} + \alpha_j + \varepsilon_{ij}$ checks for the effect of $\alpha_j, \, j = 1, \ldots, q$ different systems. It can be considered as an extension of the Student's t test to the comparison of multiple systems at the same time and it is the parametric counterpart of the Kruskal-Wallis Test.
	
 \item \textbf{Two-way ANOVA for the Topic and System Effects}: a more accurate model $y_{ij} = \mu_{\cdot\cdot} + \tau_i + \alpha_j + \varepsilon_{ij}$ which accounts also for the effect of  $\tau_i, \, i = 1, \ldots, p$ topics, thus improving the estimation of the system effect as well. Note that this is the ANOVA model adopted by~\citet{Tague-SutcliffeBlustein1994} and~\citet{BanksEtAl1999} when analysing \ac{TREC} data. It is the parametric counterpart of the Friedman Test.
\end{itemize}
In the case of the ANOVA, Kruskal-Wallis, and Friedman tests we performed a Tukey \ac{HSD} adjustment for multiple comparisons~\cite{HochbergTamhane1987,Tukey1949}.

Table~\ref{tab:smrysigcnts-norb} (measures not depending on the recall base) and Table~\ref{tab:smrysigcnts-rb} (measures depending on the recall base) show the results of the analyses in the case of the \texttt{T08\_30}, \texttt{T26\_30}, and \texttt{T27\_30} tracks. Results for the other tracks are similar but not shown here for space reasons. For each test, the tables report:
\begin{itemize}
        \item \textbf{Sig}: the total number of significantly different system pair using the original measure;
	\item \textbf{S2NS}:  number of pairs changed from significantly to not significantly different when passing from the original measure to its ranked version; within parentheses we report their ratio with respect to \texttt{Sig};
	\item \textbf{NS2S}: number of pairs changed from not significantly to significantly different when passing from the original measure to its ranked version; within parentheses we report their ratio with respect to \texttt{Sig};
	\item \textbf{$\Delta\%$}: $\frac{\text{S2NS} + \text{NS2S}}{\text{Sig}}$ the ratio of the total number of pairs that changed significance when passing from the original measure to its ranked version.
\end{itemize}
In Tables~\ref{tab:smrysigcnts-norb} and~\ref{tab:smrysigcnts-rb} rows corresponding to significance tests based on an ordinal scale assumption are highlighted in grey. 

In an ideal situation an oracle would have told us which pairs of systems are significantly different and which are not and this would have allowed us to exactly determine which pairs of systems were correctly detected by each measure and test. Unfortunately, this a priori knowledge is not available in practice. On the other hand, we are comparing a measure to its ranked version and we know that changes in the decision about what is significantly different and what is not are a consequence of the steps of the scale being rearranged from not equi-spaced and un-evenly distributed across the their range to equi-spaced and evenly distributed across their range. Therefore, we can interpret the \texttt{S2NS} count as a tendency to \emph{false positives}, since it accounts for significantly different systems which are not significant when you remove the effect of uneven steps in the scale; in other terms, \texttt{S2NS} can be interpreted as a tendency of the ranked measure (the interval scale) to reduce \emph{Type I errors}. Symmetrically, we can interpret the  \texttt{NS2S} count as a tendency to \emph{false negatives}, since it accounts for not significantly different systems which are significant when you remove the effect of uneven steps in the scale; in other terms, \texttt{NS2S} can be interpreted as a tendency of the ranked measure (the interval scale) to reduce \emph{Type II errors}. Note that we are not claiming that an interval scale detects/removes false positives/negatives in any absolute sense; we are rather saying that, starting from whatever unknown level of false positive/negatives, we can interpret the \texttt{S2NS} and \texttt{NS2S} counts as a relative tendency to reduce false positives/negatives.

As a side note not regarding measures being interval scales or not, in Tables~\ref{tab:smrysigcnts-norb} and~\ref{tab:smrysigcnts-rb} we can observe that, as expected, parametric significance tests are more powerful than non-parametric ones, since they discover more significantly different pairs. Moreover, we can also observe as the Sign, Wilcoxon Rank Sum, Wilcoxon Signed Rank, and Student's t tests find many more significantly different system pairs than the ANOVA, Kruskal-Wallis, and Friedman tests. This increase is not due to more powerful tests but rather to the increase in Type I errors due to the lack of adjustment in multiple comparisons for the former tests. This further stresses the need for always adjusting for multiple comparisons, as also pointed out by \citet{Fuhr2017,Sakai2020}.

We can observe from Tables~\ref{tab:smrysigcnts-norb} and~\ref{tab:smrysigcnts-rb},  as very coarse and general trends, that the less close an evaluation measure is to be an interval scale, the stronger are the changes in statistical significance tests based on an interval scale assumption while those based on an ordinal scale assumption are not affected. On the other hand, the presence of the recall base generally affects both tests based on the ordinal scale assumption and those based on the interval scale assumption, being the latter more affected. 

In particular, we have found that:
\begin{itemize}
	\item  for measures not depending on the recall base and significance test assuming an interval scale, we have an overall average increase in the \texttt{S2NS} count around $13\%$ and in the \texttt{NS2S} around $5\%$. This suggests that the major impact is on reducing Type I errors still improving Type II errors and making the test more powerful;
	\item for measures depending on the recall base and significance test assuming an ordinal scale, we have an overall average increase in the \texttt{S2NS} count around $2\%$ and in the \texttt{NS2S} around $4\%$. This suggests that there is a small reduction in Type I errors and some improvement in Type II errors, making the test a bit more powerful;
	\item for measures depending on the recall base and significance test assuming an interval scale, we have an overall average increase in the \texttt{S2NS} count around $10\%$ and in the \texttt{NS2S} around $45\%$. This suggests that there is a sizeable reduction in Type I errors and a quite substantial improvement in Type II errors, making the test much more powerful.
\end{itemize}
In general, these results indicate that adopting a proper interval scale tends to reduce the Type I errors and, when the situation get more complicated because of the effect of the recall base across topics, it also brings substantially more power to the test.

Overall, if we consider the grand mean across all the tracks, measures, and significance test, we observe an overall change \texttt{$\Delta\%$} around $25\% \pm 11\%$ in the decision about what is significantly different and what is not. Even without wishing to interpret it in term of Type I or Type II errors, this figure let us understand how big is the impact of using an interval scale or not, as well as the effect of the recall base.

As in the case of the correlation analysis, these observations open some questions about IR experimentation: since violating the scale assumptions has an impact on the number of significant/not-significant detected pairs and on Type I and  Type II errors, when we compare systems and algorithms, how much of the observed differences is just due to the scale violation bias? How many false positives/negatives are we observing? How much have the findings in the literature been affected by these phenomena?

\subsubsection{Measures not depending on the recall base}

\begin{table}[tbp] 
\centering 
\caption{Measures not depending on the recall base: changes in significance test analyses between using a measure and its ranked version on tracks \texttt{T08\_30}, \texttt{T26\_30}, and \texttt{T27\_30}, using the \texttt{unq} tie breaking approach.}
\label{tab:smrysigcnts-norb}
\resizebox{\columnwidth}{!}{
\begin{tabular}{|l*{3}{||r|r|r|r}|} 
\hline 
& \multicolumn{4}{c||}{\textbf{\texttt{T08\_30} -- 8256 system pairs}} & \multicolumn{4}{c||}{\textbf{\texttt{T26\_30} -- 2775 system pairs}} & \multicolumn{4}{c|}{\textbf{\texttt{T27\_30} -- 2556 system pairs}} \\ 
\cline{2-13} 
\multicolumn{1}{|c||}{\textbf{P}} & \multicolumn{1}{c|}{\textbf{Sig}}  & \multicolumn{1}{c|}{\textbf{S2NS (\%)}} & \multicolumn{1}{c|}{\textbf{NS2S (\%)}} & \multicolumn{1}{c||}{$\mathbf{\Delta\%}$}   & \multicolumn{1}{c|}{\textbf{Sig}}  & \multicolumn{1}{c|}{\textbf{S2NS (\%)}} & \multicolumn{1}{c|}{\textbf{NS2S (\%)}} & \multicolumn{1}{c||}{$\mathbf{\Delta\%}$}  & \multicolumn{1}{c|}{\textbf{Sig}}  & \multicolumn{1}{c|}{\textbf{S2NS (\%)}} & \multicolumn{1}{c|}{\textbf{NS2S (\%)}} & \multicolumn{1}{c|}{$\mathbf{\Delta\%}$} \\ 
\hline 
\rowcolor{Gray}  
Sign Test  & 5153 & 0 (0.00\%) & 0 (0.00\%) & 0.00\%  & 1848 & 0 (0.00\%) & 0 (0.00\%) & 0.00\%  & 1746 & 0 (0.00\%) & 0 (0.00\%) & 0.00\%  \\ 
\hline 
\rowcolor{Gray}  
Wilcoxon Rank Sum Test  & 4276 & 0 (0.00\%) & 0 (0.00\%) & 0.00\%  & 1453 & 0 (0.00\%) & 0 (0.00\%) & 0.00\%  & 1473 & 0 (0.00\%) & 0 (0.00\%) & 0.00\%  \\ 
\hline 
Wilcoxon Signed Rank Test  & 5644 & 0 (0.00\%) & 0 (0.00\%) & 0.00\%  & 2017 & 0 (0.00\%) & 0 (0.00\%) & 0.00\%  & 1857 & 0 (0.00\%) & 0 (0.00\%) & 0.00\%  \\ 
\hline 
Student's t Test  & 5633 & 0 (0.00\%) & 0 (0.00\%) & 0.00\%  & 2007 & 0 (0.00\%) & 0 (0.00\%) & 0.00\%  & 1830 & 0 (0.00\%) & 0 (0.00\%) & 0.00\%  \\ 
\hline 
One-way ANOVA  & 1923 & 0 (0.00\%) & 0 (0.00\%) & 0.00\%  & 454 & 0 (0.00\%) & 0 (0.00\%) & 0.00\%  & 724 & 0 (0.00\%) & 0 (0.00\%) & 0.00\%  \\ 
\hline 
\rowcolor{Gray}  
Kruskal-Wallis Test  & 1740 & 0 (0.00\%) & 0 (0.00\%) & 0.00\%  & 391 & 0 (0.00\%) & 0 (0.00\%) & 0.00\%  & 692 & 0 (0.00\%) & 0 (0.00\%) & 0.00\%  \\ 
\hline 
Two-way ANOVA  & 3362 & 0 (0.00\%) & 0 (0.00\%) & 0.00\%  & 1155 & 0 (0.00\%) & 0 (0.00\%) & 0.00\%  & 1214 & 0 (0.00\%) & 0 (0.00\%) & 0.00\%  \\ 
\hline 
\rowcolor{Gray}  
Friedman Test  & 2595 & 0 (0.00\%) & 0 (0.00\%) & 0.00\%  & 883 & 0 (0.00\%) & 0 (0.00\%) & 0.00\%  & 925 & 0 (0.00\%) & 0 (0.00\%) & 0.00\%  \\ 
\hline \hline 
\multicolumn{1}{|c||}{\textbf{RBP\_p05}} & \multicolumn{1}{c|}{\textbf{Sig}}  & \multicolumn{1}{c|}{\textbf{S2NS (\%)}} & \multicolumn{1}{c|}{\textbf{NS2S (\%)}} & \multicolumn{1}{c||}{$\mathbf{\Delta\%}$}   & \multicolumn{1}{c|}{\textbf{Sig}}  & \multicolumn{1}{c|}{\textbf{S2NS (\%)}} & \multicolumn{1}{c|}{\textbf{NS2S (\%)}} & \multicolumn{1}{c||}{$\mathbf{\Delta\%}$}  & \multicolumn{1}{c|}{\textbf{Sig}}  & \multicolumn{1}{c|}{\textbf{S2NS (\%)}} & \multicolumn{1}{c|}{\textbf{NS2S (\%)}} & \multicolumn{1}{c|}{$\mathbf{\Delta\%}$} \\ 
\hline 
\rowcolor{Gray}  
Sign Test  & 4724 & 0 (0.00\%) & 0 (0.00\%) & 0.00\%  & 1414 & 0 (0.00\%) & 0 (0.00\%) & 0.00\%  & 1613 & 0 (0.00\%) & 0 (0.00\%) & 0.00\%  \\ 
\hline 
\rowcolor{Gray}  
Wilcoxon Rank Sum Test  & 4302 & 0 (0.00\%) & 0 (0.00\%) & 0.00\%  & 1328 & 0 (0.00\%) & 0 (0.00\%) & 0.00\%  & 1545 & 0 (0.00\%) & 0 (0.00\%) & 0.00\%  \\ 
\hline 
Wilcoxon Signed Rank Test  & 5108 & 0 (0.00\%) & 0 (0.00\%) & 0.00\%  & 1683 & 0 (0.00\%) & 0 (0.00\%) & 0.00\%  & 1679 & 0 (0.00\%) & 0 (0.00\%) & 0.00\%  \\ 
\hline 
Student's t Test  & 5091 & 0 (0.00\%) & 0 (0.00\%) & 0.00\%  & 1711 & 0 (0.00\%) & 0 (0.00\%) & 0.00\%  & 1635 & 0 (0.00\%) & 0 (0.00\%) & 0.00\%  \\ 
\hline 
One-way ANOVA  & 1945 & 0 (0.00\%) & 0 (0.00\%) & 0.00\%  & 377 & 0 (0.00\%) & 0 (0.00\%) & 0.00\%  & 660 & 0 (0.00\%) & 0 (0.00\%) & 0.00\%  \\ 
\hline 
\rowcolor{Gray}  
Kruskal-Wallis Test  & 1678 & 0 (0.00\%) & 0 (0.00\%) & 0.00\%  & 326 & 0 (0.00\%) & 0 (0.00\%) & 0.00\%  & 680 & 0 (0.00\%) & 0 (0.00\%) & 0.00\%  \\ 
\hline 
Two-way ANOVA  & 2861 & 0 (0.00\%) & 0 (0.00\%) & 0.00\%  & 733 & 0 (0.00\%) & 0 (0.00\%) & 0.00\%  & 927 & 0 (0.00\%) & 0 (0.00\%) & 0.00\%  \\ 
\hline 
\rowcolor{Gray}  
Friedman Test  & 2228 & 0 (0.00\%) & 0 (0.00\%) & 0.00\%  & 588 & 0 (0.00\%) & 0 (0.00\%) & 0.00\%  & 806 & 0 (0.00\%) & 0 (0.00\%) & 0.00\%  \\ 
\hline \hline 
\multicolumn{1}{|c||}{\textbf{RBP\_p03}} & \multicolumn{1}{c|}{\textbf{Sig}}  & \multicolumn{1}{c|}{\textbf{S2NS (\%)}} & \multicolumn{1}{c|}{\textbf{NS2S (\%)}} & \multicolumn{1}{c||}{$\mathbf{\Delta\%}$}   & \multicolumn{1}{c|}{\textbf{Sig}}  & \multicolumn{1}{c|}{\textbf{S2NS (\%)}} & \multicolumn{1}{c|}{\textbf{NS2S (\%)}} & \multicolumn{1}{c||}{$\mathbf{\Delta\%}$}  & \multicolumn{1}{c|}{\textbf{Sig}}  & \multicolumn{1}{c|}{\textbf{S2NS (\%)}} & \multicolumn{1}{c|}{\textbf{NS2S (\%)}} & \multicolumn{1}{c|}{$\mathbf{\Delta\%}$} \\ 
\hline 
\rowcolor{Gray}  
Sign Test  & 4724 & 0 (0.00\%) & 0 (0.00\%) & 0.00\%  & 1414 & 0 (0.00\%) & 0 (0.00\%) & 0.00\%  & 1613 & 0 (0.00\%) & 0 (0.00\%) & 0.00\%  \\ 
\hline 
\rowcolor{Gray}  
Wilcoxon Rank Sum Test  & 4302 & 0 (0.00\%) & 0 (0.00\%) & 0.00\%  & 1328 & 0 (0.00\%) & 0 (0.00\%) & 0.00\%  & 1545 & 0 (0.00\%) & 0 (0.00\%) & 0.00\%  \\ 
\hline 
Wilcoxon Signed Rank Test  & 5027 & 29 (0.58\%) & 110 (2.19\%) & 2.77\%  & 1638 & 10 (0.61\%) & 55 (3.36\%) & 3.97\%  & 1651 & 3 (0.18\%) & 31 (1.88\%) & 2.06\%  \\ 
\hline 
Student's t Test  & 4801 & 70 (1.46\%) & 360 (7.50\%) & 8.96\%  & 1551 & 19 (1.23\%) & 179 (11.54\%) & 12.77\%  & 1486 & 7 (0.47\%) & 156 (10.50\%) & 10.97\%  \\ 
\hline 
One-way ANOVA  & 1730 & 18 (1.04\%) & 233 (13.47\%) & 14.51\%  & 317 & 3 (0.95\%) & 63 (19.87\%) & 20.82\%  & 606 & 4 (0.66\%) & 58 (9.57\%) & 10.23\%  \\ 
\hline 
\rowcolor{Gray}  
Kruskal-Wallis Test  & 1678 & 0 (0.00\%) & 0 (0.00\%) & 0.00\%  & 326 & 0 (0.00\%) & 0 (0.00\%) & 0.00\%  & 680 & 0 (0.00\%) & 0 (0.00\%) & 0.00\%  \\ 
\hline 
Two-way ANOVA  & 2432 & 8 (0.33\%) & 437 (17.97\%) & 18.30\%  & 573 & 2 (0.35\%) & 162 (28.27\%) & 28.62\%  & 807 & 0 (0.00\%) & 120 (14.87\%) & 14.87\%  \\ 
\hline 
\rowcolor{Gray}  
Friedman Test  & 2228 & 0 (0.00\%) & 0 (0.00\%) & 0.00\%  & 588 & 0 (0.00\%) & 0 (0.00\%) & 0.00\%  & 806 & 0 (0.00\%) & 0 (0.00\%) & 0.00\%  \\ 
\hline \hline 
\multicolumn{1}{|c||}{\textbf{RBP\_p08}} & \multicolumn{1}{c|}{\textbf{Sig}}  & \multicolumn{1}{c|}{\textbf{S2NS (\%)}} & \multicolumn{1}{c|}{\textbf{NS2S (\%)}} & \multicolumn{1}{c||}{$\mathbf{\Delta\%}$}   & \multicolumn{1}{c|}{\textbf{Sig}}  & \multicolumn{1}{c|}{\textbf{S2NS (\%)}} & \multicolumn{1}{c|}{\textbf{NS2S (\%)}} & \multicolumn{1}{c||}{$\mathbf{\Delta\%}$}  & \multicolumn{1}{c|}{\textbf{Sig}}  & \multicolumn{1}{c|}{\textbf{S2NS (\%)}} & \multicolumn{1}{c|}{\textbf{NS2S (\%)}} & \multicolumn{1}{c|}{$\mathbf{\Delta\%}$} \\ 
\hline 
\rowcolor{Gray}  
Sign Test  & 5105 & 0 (0.00\%) & 0 (0.00\%) & 0.00\%  & 1615 & 0 (0.00\%) & 0 (0.00\%) & 0.00\%  & 1678 & 0 (0.00\%) & 0 (0.00\%) & 0.00\%  \\ 
\hline 
\rowcolor{Gray}  
Wilcoxon Rank Sum Test  & 4375 & 0 (0.00\%) & 0 (0.00\%) & 0.00\%  & 1378 & 0 (0.00\%) & 0 (0.00\%) & 0.00\%  & 1487 & 0 (0.00\%) & 0 (0.00\%) & 0.00\%  \\ 
\hline 
Wilcoxon Signed Rank Test  & 5508 & 165 (3.00\%) & 114 (2.07\%) & 5.07\%  & 1879 & 70 (3.73\%) & 30 (1.60\%) & 5.32\%  & 1846 & 110 (5.96\%) & 3 (0.16\%) & 6.12\%  \\ 
\hline 
Student's t Test  & 5425 & 221 (4.07\%) & 123 (2.27\%) & 6.34\%  & 1924 & 144 (7.48\%) & 60 (3.12\%) & 10.60\%  & 1853 & 236 (12.74\%) & 14 (0.76\%) & 13.49\%  \\ 
\hline 
One-way ANOVA  & 2183 & 200 (9.16\%) & 44 (2.02\%) & 11.18\%  & 469 & 59 (12.58\%) & 15 (3.20\%) & 15.78\%  & 712 & 98 (13.76\%) & 1 (0.14\%) & 13.90\%  \\ 
\hline 
\rowcolor{Gray}  
Kruskal-Wallis Test  & 1781 & 0 (0.00\%) & 0 (0.00\%) & 0.00\%  & 384 & 0 (0.00\%) & 0 (0.00\%) & 0.00\%  & 691 & 0 (0.00\%) & 0 (0.00\%) & 0.00\%  \\ 
\hline 
Two-way ANOVA  & 3474 & 257 (7.40\%) & 20 (0.58\%) & 7.97\%  & 950 & 104 (10.95\%) & 21 (2.21\%) & 13.16\%  & 1105 & 187 (16.92\%) & 6 (0.54\%) & 17.47\%  \\ 
\hline 
\rowcolor{Gray}  
Friedman Test  & 2589 & 0 (0.00\%) & 0 (0.00\%) & 0.00\%  & 741 & 0 (0.00\%) & 0 (0.00\%) & 0.00\%  & 851 & 0 (0.00\%) & 0 (0.00\%) & 0.00\%  \\ 
\hline \hline 
\multicolumn{1}{|c||}{\textbf{DCG\_b02}} & \multicolumn{1}{c|}{\textbf{Sig}}  & \multicolumn{1}{c|}{\textbf{S2NS (\%)}} & \multicolumn{1}{c|}{\textbf{NS2S (\%)}} & \multicolumn{1}{c||}{$\mathbf{\Delta\%}$}   & \multicolumn{1}{c|}{\textbf{Sig}}  & \multicolumn{1}{c|}{\textbf{S2NS (\%)}} & \multicolumn{1}{c|}{\textbf{NS2S (\%)}} & \multicolumn{1}{c||}{$\mathbf{\Delta\%}$}  & \multicolumn{1}{c|}{\textbf{Sig}}  & \multicolumn{1}{c|}{\textbf{S2NS (\%)}} & \multicolumn{1}{c|}{\textbf{NS2S (\%)}} & \multicolumn{1}{c|}{$\mathbf{\Delta\%}$} \\ 
\hline 
\rowcolor{Gray}  
Sign Test  & 5227 & 0 (0.00\%) & 0 (0.00\%) & 0.00\%  & 1781 & 0 (0.00\%) & 0 (0.00\%) & 0.00\%  & 1738 & 0 (0.00\%) & 0 (0.00\%) & 0.00\%  \\ 
\hline 
\rowcolor{Gray}  
Wilcoxon Rank Sum Test  & 4425 & 0 (0.00\%) & 0 (0.00\%) & 0.00\%  & 1457 & 0 (0.00\%) & 0 (0.00\%) & 0.00\%  & 1499 & 0 (0.00\%) & 0 (0.00\%) & 0.00\%  \\ 
\hline 
Wilcoxon Signed Rank Test  & 5743 & 260 (4.53\%) & 147 (2.56\%) & 7.09\%  & 2020 & 127 (6.29\%) & 40 (1.98\%) & 8.27\%  & 1887 & 91 (4.82\%) & 21 (1.11\%) & 5.94\%  \\ 
\hline 
Student's t Test  & 5664 & 408 (7.20\%) & 161 (2.84\%) & 10.05\%  & 2026 & 176 (8.69\%) & 47 (2.32\%) & 11.01\%  & 1881 & 155 (8.24\%) & 15 (0.80\%) & 9.04\%  \\ 
\hline 
One-way ANOVA  & 2125 & 633 (29.79\%) & 59 (2.78\%) & 32.56\%  & 495 & 113 (22.83\%) & 3 (0.61\%) & 23.43\%  & 734 & 153 (20.84\%) & 2 (0.27\%) & 21.12\%  \\ 
\hline 
\rowcolor{Gray}  
Kruskal-Wallis Test  & 1812 & 0 (0.00\%) & 0 (0.00\%) & 0.00\%  & 400 & 0 (0.00\%) & 0 (0.00\%) & 0.00\%  & 695 & 0 (0.00\%) & 0 (0.00\%) & 0.00\%  \\ 
\hline 
Two-way ANOVA  & 3609 & 601 (16.65\%) & 25 (0.69\%) & 17.35\%  & 1129 & 179 (15.85\%) & 36 (3.19\%) & 19.04\%  & 1201 & 328 (27.31\%) & 4 (0.33\%) & 27.64\%  \\ 
\hline 
\rowcolor{Gray}  
Friedman Test  & 2706 & 0 (0.00\%) & 0 (0.00\%) & 0.00\%  & 876 & 0 (0.00\%) & 0 (0.00\%) & 0.00\%  & 912 & 0 (0.00\%) & 0 (0.00\%) & 0.00\%  \\ 
\hline \hline 
\multicolumn{1}{|c||}{\textbf{DCG\_b10}} & \multicolumn{1}{c|}{\textbf{Sig}}  & \multicolumn{1}{c|}{\textbf{S2NS (\%)}} & \multicolumn{1}{c|}{\textbf{NS2S (\%)}} & \multicolumn{1}{c||}{$\mathbf{\Delta\%}$}   & \multicolumn{1}{c|}{\textbf{Sig}}  & \multicolumn{1}{c|}{\textbf{S2NS (\%)}} & \multicolumn{1}{c|}{\textbf{NS2S (\%)}} & \multicolumn{1}{c||}{$\mathbf{\Delta\%}$}  & \multicolumn{1}{c|}{\textbf{Sig}}  & \multicolumn{1}{c|}{\textbf{S2NS (\%)}} & \multicolumn{1}{c|}{\textbf{NS2S (\%)}} & \multicolumn{1}{c|}{$\mathbf{\Delta\%}$} \\ 
\hline 
\rowcolor{Gray}  
Sign Test  & 5095 & 0 (0.00\%) & 0 (0.00\%) & 0.00\%  & 1813 & 0 (0.00\%) & 0 (0.00\%) & 0.00\%  & 1755 & 0 (0.00\%) & 0 (0.00\%) & 0.00\%  \\ 
\hline 
\rowcolor{Gray}  
Wilcoxon Rank Sum Test  & 4307 & 0 (0.00\%) & 0 (0.00\%) & 0.00\%  & 1458 & 0 (0.00\%) & 0 (0.00\%) & 0.00\%  & 1485 & 0 (0.00\%) & 0 (0.00\%) & 0.00\%  \\ 
\hline 
Wilcoxon Signed Rank Test  & 5663 & 260 (4.59\%) & 127 (2.24\%) & 6.83\%  & 2019 & 105 (5.20\%) & 32 (1.58\%) & 6.79\%  & 1861 & 60 (3.22\%) & 35 (1.88\%) & 5.10\%  \\ 
\hline 
Student's t Test  & 5636 & 370 (6.56\%) & 132 (2.34\%) & 8.91\%  & 2021 & 169 (8.36\%) & 50 (2.47\%) & 10.84\%  & 1843 & 114 (6.19\%) & 36 (1.95\%) & 8.14\%  \\ 
\hline 
One-way ANOVA  & 1958 & 747 (38.15\%) & 37 (1.89\%) & 40.04\%  & 468 & 86 (18.38\%) & 12 (2.56\%) & 20.94\%  & 727 & 187 (25.72\%) & 0 (0.00\%) & 25.72\%  \\ 
\hline 
\rowcolor{Gray}  
Kruskal-Wallis Test  & 1750 & 0 (0.00\%) & 0 (0.00\%) & 0.00\%  & 390 & 0 (0.00\%) & 0 (0.00\%) & 0.00\%  & 687 & 0 (0.00\%) & 0 (0.00\%) & 0.00\%  \\ 
\hline 
Two-way ANOVA  & 3438 & 623 (18.12\%) & 41 (1.19\%) & 19.31\%  & 1139 & 130 (11.41\%) & 36 (3.16\%) & 14.57\%  & 1216 & 256 (21.05\%) & 4 (0.33\%) & 21.38\%  \\ 
\hline 
\rowcolor{Gray}  
Friedman Test  & 2637 & 0 (0.00\%) & 0 (0.00\%) & 0.00\%  & 882 & 0 (0.00\%) & 0 (0.00\%) & 0.00\%  & 925 & 0 (0.00\%) & 0 (0.00\%) & 0.00\%  \\ 
\hline \hline 
\multicolumn{1}{|c||}{\textbf{RR}} & \multicolumn{1}{c|}{\textbf{Sig}}  & \multicolumn{1}{c|}{\textbf{S2NS (\%)}} & \multicolumn{1}{c|}{\textbf{NS2S (\%)}} & \multicolumn{1}{c||}{$\mathbf{\Delta\%}$}   & \multicolumn{1}{c|}{\textbf{Sig}}  & \multicolumn{1}{c|}{\textbf{S2NS (\%)}} & \multicolumn{1}{c|}{\textbf{NS2S (\%)}} & \multicolumn{1}{c||}{$\mathbf{\Delta\%}$}  & \multicolumn{1}{c|}{\textbf{Sig}}  & \multicolumn{1}{c|}{\textbf{S2NS (\%)}} & \multicolumn{1}{c|}{\textbf{NS2S (\%)}} & \multicolumn{1}{c|}{$\mathbf{\Delta\%}$} \\ 
\hline 
\rowcolor{Gray}  
Sign Test  & 4074 & 0 (0.00\%) & 0 (0.00\%) & 0.00\%  & 1305 & 0 (0.00\%) & 0 (0.00\%) & 0.00\%  & 1309 & 0 (0.00\%) & 0 (0.00\%) & 0.00\%  \\ 
\hline 
\rowcolor{Gray}  
Wilcoxon Rank Sum Test  & 3951 & 0 (0.00\%) & 0 (0.00\%) & 0.00\%  & 1198 & 0 (0.00\%) & 0 (0.00\%) & 0.00\%  & 1280 & 0 (0.00\%) & 0 (0.00\%) & 0.00\%  \\ 
\hline 
Wilcoxon Signed Rank Test  & 4578 & 479 (10.46\%) & 362 (7.91\%) & 18.37\%  & 1497 & 201 (13.43\%) & 180 (12.02\%) & 25.45\%  & 1462 & 94 (6.43\%) & 111 (7.59\%) & 14.02\%  \\ 
\hline 
Student's t Test  & 4582 & 903 (19.71\%) & 454 (9.91\%) & 29.62\%  & 1517 & 455 (29.99\%) & 180 (11.87\%) & 41.86\%  & 1454 & 202 (13.89\%) & 110 (7.57\%) & 21.46\%  \\ 
\hline 
One-way ANOVA  & 1691 & 288 (17.03\%) & 162 (9.58\%) & 26.61\%  & 285 & 203 (71.23\%) & 2 (0.70\%) & 71.93\%  & 696 & 92 (13.22\%) & 148 (21.26\%) & 34.48\%  \\ 
\hline 
\rowcolor{Gray}  
Kruskal-Wallis Test  & 1500 & 0 (0.00\%) & 0 (0.00\%) & 0.00\%  & 218 & 0 (0.00\%) & 0 (0.00\%) & 0.00\%  & 633 & 0 (0.00\%) & 0 (0.00\%) & 0.00\%  \\ 
\hline 
Two-way ANOVA  & 2233 & 511 (22.88\%) & 131 (5.87\%) & 28.75\%  & 527 & 349 (66.22\%) & 23 (4.36\%) & 70.59\%  & 868 & 145 (16.71\%) & 92 (10.60\%) & 27.30\%  \\ 
\hline 
\rowcolor{Gray}  
Friedman Test  & 1813 & 0 (0.00\%) & 0 (0.00\%) & 0.00\%  & 428 & 0 (0.00\%) & 0 (0.00\%) & 0.00\%  & 739 & 0 (0.00\%) & 0 (0.00\%) & 0.00\%  \\ 
\hline \hline 
\end{tabular} 
 }
\end{table}

Let us start from Precision and RBP\_p05 in Table~\ref{tab:smrysigcnts-norb}. As we already know, both of them are interval scales and, as expected, we do not observe any changes in using them or their ranked version. 

As in the case of the correlation, we can take them as an example of \emph{meaningful statements} in \ac{IR}, since a statement like ``There are 1,923 significantly different system pairs for Precision according to one-way ANOVA on \texttt{T08\_30}'' does not change its truth value for a permissible transformation of the scale.

As said, RBP\_p03 orders systems in the same way as RBP\_p05 but it is no more an interval scale. Coherently with this, we can see how the significance tests assuming just an ordinal case detect the same number of significantly different pairs for both RBP\_p03 and RBP\_p05. On the other hand, significance tests assuming an interval scale are affected by this difference between RBP\_p03 and RBP\_p05, causing an overall change \texttt{$\Delta\%$} in the range $2\% - 29\%$. In particular, we observe a marked increase in the number of significantly different pairs (\texttt{NS2S} up to $28\%$),  i.e. the reduction in the number of false negatives, and a very marginal decrease in the number of not significantly different ones (\texttt{S2NS} around $1\%$),  i.e. the reduction in the number of false positives. In the case of RBP\_p08 we can note a much more marked increase in the number of not significantly different pairs (\texttt{S2NS} up to $17\%$), i.e. a reduction in the number of false positives; on the other hand, the increase in the number of significantly different pairs (\texttt{NS2S} around $2\%$), i.e. the reduction in the number of false negatives, is more marginal. 

Why do we observe such a different behaviour between RBP\_p03 and RBP\_p08? If we look at Figure~\ref{fig:acr_rbp_steps}, we can see that RBP\_p03 condenses values at the top and the bottom of the range of possible values, in spans with a very small range of values but containing the same number of runs. As a consequence, when the ranked version of RBP\_p03 equi-spaces these values, runs that before were very close, and possibly not significantly different (\texttt{NS}) in the ranked version become more distant, and possibly significantly different (\texttt{S}); and this can explain why the \texttt{NS2S} case is more prominent for RBP\_p03. On the other hand, RBP\_p08 uses all the possible range of values but very few runs, roughly $20\%$ packed at the bottom and at the top, cover almost $50\%$ of the range of values while the remaining $80\%$ of the runs, in the middle part, cover the other $50\%$ of the range. Therefore, when we pass to the ranked version of the measure, very few runs which were very distant, and possibly  significantly different (\texttt{S}), become closer, and possibly not significantly different (\texttt{NS}); viceversa, many runs which were very close, and possibly not significantly different (\texttt{NS}), may become a little bit more distant, and possibly (but not necessarily) significantly different (\texttt{S}). As a consequence, the effect on \texttt{S2NS} is more prominent than the one on \texttt{NS2S}.

In the case of \ac{DCG} we observe a behaviour similar to the one of RBP\_p08, being the increase on \texttt{S2NS} and the reduction in the false positives even more prominent. If we look at Figure~\ref{fig:acr_dcg_steps}, we can see how \ac{DCG} is sharper than RBP\_p08 at the top and bottom of the range -- less than $10\%$ of the runs account for almost $50\%$ of the range of values -- making even fewer runs falling more apart. 

Finally, \ac{RR} exhibits both effects: a very remarkable increase in \texttt{S2NS}, i.e. a reduction in the false positives, and a sizeable increase in \texttt{NS2S}, i.e. a reduction in the false negatives, causing an overall change \texttt{$\Delta\%$} up to a $72\%$. If we consider Figure~\ref{fig:acr_p_rbp05_rr_steps}, we can see how most of the runs, over $90\%$, are concentrated in just 4 possible values which are quite distant, possibly making them significantly different (\texttt{S}); when we move to the ranked version, these 4 values become much closer, possibly making the runs not significantly different (\texttt{NS}); and this explains the big \texttt{S2NS} effect. Vice versa, few runs, less than $10\%$, account for just the $20\%$ of the range of values in the lower quartile; when we move to the ranked version,  these values become more distant, possibly making the runs significantly different (\texttt{S}); since this change may affect a smaller number of runs, this explains why \texttt{NS2S} tends to be more moderate with respect to \texttt{S2NS}.

\subsubsection{Measures depending on the recall base}

\begin{table}[tbp] 
\centering 
\caption{Measures depending on the recall base: changes in significance test analyses between using a measure and its ranked version on tracks \texttt{T08\_30}, \texttt{T26\_30}, and \texttt{T27\_30}, using the \texttt{unq} tie breaking approach.}
\label{tab:smrysigcnts-rb}
\resizebox{\columnwidth}{!}{
\begin{tabular}{|l*{3}{||r|r|r|r}|} 
\hline 
& \multicolumn{4}{c||}{\textbf{\texttt{T08\_30} -- 8256 system pairs}} & \multicolumn{4}{c||}{\textbf{\texttt{T26\_30} -- 2775 system pairs}} & \multicolumn{4}{c|}{\textbf{\texttt{T27\_30} -- 2556 system pairs}} \\ 
\cline{2-13} 
\multicolumn{1}{|c||}{\textbf{R}} & \multicolumn{1}{c|}{\textbf{Sig}}  & \multicolumn{1}{c|}{\textbf{S2NS (\%)}} & \multicolumn{1}{c|}{\textbf{NS2S (\%)}} & \multicolumn{1}{c||}{$\mathbf{\Delta\%}$}   & \multicolumn{1}{c|}{\textbf{Sig}}  & \multicolumn{1}{c|}{\textbf{S2NS (\%)}} & \multicolumn{1}{c|}{\textbf{NS2S (\%)}} & \multicolumn{1}{c||}{$\mathbf{\Delta\%}$}  & \multicolumn{1}{c|}{\textbf{Sig}}  & \multicolumn{1}{c|}{\textbf{S2NS (\%)}} & \multicolumn{1}{c|}{\textbf{NS2S (\%)}} & \multicolumn{1}{c|}{$\mathbf{\Delta\%}$} \\ 
\hline 
\rowcolor{Gray}  
Sign Test  & 5153 & 0 (0.00\%) & 0 (0.00\%) & 0.00\%  & 1848 & 0 (0.00\%) & 0 (0.00\%) & 0.00\%  & 1746 & 0 (0.00\%) & 0 (0.00\%) & 0.00\%  \\ 
\hline 
\rowcolor{Gray}  
Wilcoxon Rank Sum Test  & 3494 & 14 (0.40\%) & 796 (22.78\%) & 23.18\%  & 1239 & 25 (2.02\%) & 239 (19.29\%) & 21.31\%  & 1254 & 17 (1.36\%) & 236 (18.82\%) & 20.18\%  \\ 
\hline 
Wilcoxon Signed Rank Test  & 5434 & 185 (3.40\%) & 395 (7.27\%) & 10.67\%  & 1892 & 35 (1.85\%) & 160 (8.46\%) & 10.31\%  & 1799 & 27 (1.50\%) & 85 (4.72\%) & 6.23\%  \\ 
\hline 
Student's t Test  & 5073 & 303 (5.97\%) & 863 (17.01\%) & 22.98\%  & 1723 & 62 (3.60\%) & 346 (20.08\%) & 23.68\%  & 1554 & 34 (2.19\%) & 310 (19.95\%) & 22.14\%  \\ 
\hline 
One-way ANOVA  & 409 & 0 (0.00\%) & 1514 (370.17\%) & 370.17\%  & 69 & 0 (0.00\%) & 385 (557.97\%) & 557.97\%  & 574 & 3 (0.52\%) & 153 (26.66\%) & 27.18\%  \\ 
\hline 
\rowcolor{Gray}  
Kruskal-Wallis Test  & 1417 & 28 (1.98\%) & 351 (24.77\%) & 26.75\%  & 259 & 0 (0.00\%) & 132 (50.97\%) & 50.97\%  & 653 & 16 (2.45\%) & 55 (8.42\%) & 10.87\%  \\ 
\hline 
Two-way ANOVA  & 2440 & 47 (1.93\%) & 969 (39.71\%) & 41.64\%  & 683 & 3 (0.44\%) & 475 (69.55\%) & 69.99\%  & 923 & 15 (1.63\%) & 306 (33.15\%) & 34.78\%  \\ 
\hline 
\rowcolor{Gray}  
Friedman Test  & 2595 & 0 (0.00\%) & 0 (0.00\%) & 0.00\%  & 883 & 0 (0.00\%) & 0 (0.00\%) & 0.00\%  & 925 & 0 (0.00\%) & 0 (0.00\%) & 0.00\%  \\ 
\hline \hline 
\multicolumn{1}{|c||}{\textbf{AP}} & \multicolumn{1}{c|}{\textbf{Sig}}  & \multicolumn{1}{c|}{\textbf{S2NS (\%)}} & \multicolumn{1}{c|}{\textbf{NS2S (\%)}} & \multicolumn{1}{c||}{$\mathbf{\Delta\%}$}   & \multicolumn{1}{c|}{\textbf{Sig}}  & \multicolumn{1}{c|}{\textbf{S2NS (\%)}} & \multicolumn{1}{c|}{\textbf{NS2S (\%)}} & \multicolumn{1}{c||}{$\mathbf{\Delta\%}$}  & \multicolumn{1}{c|}{\textbf{Sig}}  & \multicolumn{1}{c|}{\textbf{S2NS (\%)}} & \multicolumn{1}{c|}{\textbf{NS2S (\%)}} & \multicolumn{1}{c|}{$\mathbf{\Delta\%}$} \\ 
\hline 
\rowcolor{Gray}  
Sign Test  & 5233 & 0 (0.00\%) & 0 (0.00\%) & 0.00\%  & 1801 & 0 (0.00\%) & 0 (0.00\%) & 0.00\%  & 1727 & 0 (0.00\%) & 0 (0.00\%) & 0.00\%  \\ 
\hline 
\rowcolor{Gray}  
Wilcoxon Rank Sum Test  & 4067 & 25 (0.61\%) & 385 (9.47\%) & 10.08\%  & 1455 & 84 (5.77\%) & 85 (5.84\%) & 11.62\%  & 1446 & 48 (3.32\%) & 96 (6.64\%) & 9.96\%  \\ 
\hline 
Wilcoxon Signed Rank Test  & 5659 & 242 (4.28\%) & 169 (2.99\%) & 7.26\%  & 1896 & 91 (4.80\%) & 128 (6.75\%) & 11.55\%  & 1853 & 97 (5.23\%) & 38 (2.05\%) & 7.29\%  \\ 
\hline 
Student's t Test  & 5156 & 339 (6.57\%) & 559 (10.84\%) & 17.42\%  & 1683 & 145 (8.62\%) & 336 (19.96\%) & 28.58\%  & 1759 & 118 (6.71\%) & 109 (6.20\%) & 12.91\%  \\ 
\hline 
One-way ANOVA  & 391 & 3 (0.77\%) & 955 (244.25\%) & 245.01\%  & 59 & 0 (0.00\%) & 295 (500.00\%) & 500.00\%  & 529 & 60 (11.34\%) & 65 (12.29\%) & 23.63\%  \\ 
\hline 
\rowcolor{Gray}  
Kruskal-Wallis Test  & 1593 & 19 (1.19\%) & 225 (14.12\%) & 15.32\%  & 380 & 20 (5.26\%) & 37 (9.74\%) & 15.00\%  & 700 & 21 (3.00\%) & 11 (1.57\%) & 4.57\%  \\ 
\hline 
Two-way ANOVA  & 2508 & 171 (6.82\%) & 560 (22.33\%) & 29.15\%  & 636 & 31 (4.87\%) & 360 (56.60\%) & 61.48\%  & 913 & 105 (11.50\%) & 73 (8.00\%) & 19.50\%  \\ 
\hline 
\rowcolor{Gray}  
Friedman Test  & 2706 & 0 (0.00\%) & 0 (0.00\%) & 0.00\%  & 883 & 0 (0.00\%) & 0 (0.00\%) & 0.00\%  & 915 & 0 (0.00\%) & 0 (0.00\%) & 0.00\%  \\ 
\hline \hline 
\multicolumn{1}{|c||}{\textbf{nDCG\_b02}} & \multicolumn{1}{c|}{\textbf{Sig}}  & \multicolumn{1}{c|}{\textbf{S2NS (\%)}} & \multicolumn{1}{c|}{\textbf{NS2S (\%)}} & \multicolumn{1}{c||}{$\mathbf{\Delta\%}$}   & \multicolumn{1}{c|}{\textbf{Sig}}  & \multicolumn{1}{c|}{\textbf{S2NS (\%)}} & \multicolumn{1}{c|}{\textbf{NS2S (\%)}} & \multicolumn{1}{c||}{$\mathbf{\Delta\%}$}  & \multicolumn{1}{c|}{\textbf{Sig}}  & \multicolumn{1}{c|}{\textbf{S2NS (\%)}} & \multicolumn{1}{c|}{\textbf{NS2S (\%)}} & \multicolumn{1}{c|}{$\mathbf{\Delta\%}$} \\ 
\hline 
\rowcolor{Gray}  
Sign Test  & 5227 & 0 (0.00\%) & 0 (0.00\%) & 0.00\%  & 1781 & 0 (0.00\%) & 0 (0.00\%) & 0.00\%  & 1738 & 0 (0.00\%) & 0 (0.00\%) & 0.00\%  \\ 
\hline 
\rowcolor{Gray}  
Wilcoxon Rank Sum Test  & 4420 & 49 (1.11\%) & 54 (1.22\%) & 2.33\%  & 1457 & 14 (0.96\%) & 14 (0.96\%) & 1.92\%  & 1530 & 52 (3.40\%) & 21 (1.37\%) & 4.77\%  \\ 
\hline 
Wilcoxon Signed Rank Test  & 5726 & 245 (4.28\%) & 149 (2.60\%) & 6.88\%  & 2004 & 115 (5.74\%) & 44 (2.20\%) & 7.93\%  & 1892 & 96 (5.07\%) & 21 (1.11\%) & 6.18\%  \\ 
\hline 
Student's t Test  & 5623 & 384 (6.83\%) & 178 (3.17\%) & 9.99\%  & 2013 & 174 (8.64\%) & 58 (2.88\%) & 11.53\%  & 1883 & 163 (8.66\%) & 21 (1.12\%) & 9.77\%  \\ 
\hline 
One-way ANOVA  & 2159 & 667 (30.89\%) & 59 (2.73\%) & 33.63\%  & 498 & 116 (23.29\%) & 3 (0.60\%) & 23.90\%  & 820 & 237 (28.90\%) & 0 (0.00\%) & 28.90\%  \\ 
\hline 
\rowcolor{Gray}  
Kruskal-Wallis Test  & 1827 & 34 (1.86\%) & 19 (1.04\%) & 2.90\%  & 418 & 20 (4.78\%) & 2 (0.48\%) & 5.26\%  & 718 & 25 (3.48\%) & 2 (0.28\%) & 3.76\%  \\ 
\hline 
Two-way ANOVA  & 3642 & 631 (17.33\%) & 22 (0.60\%) & 17.93\%  & 1120 & 172 (15.36\%) & 38 (3.39\%) & 18.75\%  & 1201 & 330 (27.48\%) & 6 (0.50\%) & 27.98\%  \\ 
\hline 
\rowcolor{Gray}  
Friedman Test  & 2706 & 0 (0.00\%) & 0 (0.00\%) & 0.00\%  & 876 & 0 (0.00\%) & 0 (0.00\%) & 0.00\%  & 912 & 0 (0.00\%) & 0 (0.00\%) & 0.00\%  \\ 
\hline \hline 
\multicolumn{1}{|c||}{\textbf{nDCG\_b10}} & \multicolumn{1}{c|}{\textbf{Sig}}  & \multicolumn{1}{c|}{\textbf{S2NS (\%)}} & \multicolumn{1}{c|}{\textbf{NS2S (\%)}} & \multicolumn{1}{c||}{$\mathbf{\Delta\%}$}   & \multicolumn{1}{c|}{\textbf{Sig}}  & \multicolumn{1}{c|}{\textbf{S2NS (\%)}} & \multicolumn{1}{c|}{\textbf{NS2S (\%)}} & \multicolumn{1}{c||}{$\mathbf{\Delta\%}$}  & \multicolumn{1}{c|}{\textbf{Sig}}  & \multicolumn{1}{c|}{\textbf{S2NS (\%)}} & \multicolumn{1}{c|}{\textbf{NS2S (\%)}} & \multicolumn{1}{c|}{$\mathbf{\Delta\%}$} \\ 
\hline 
\rowcolor{Gray}  
Sign Test  & 5095 & 0 (0.00\%) & 0 (0.00\%) & 0.00\%  & 1813 & 0 (0.00\%) & 0 (0.00\%) & 0.00\%  & 1755 & 0 (0.00\%) & 0 (0.00\%) & 0.00\%  \\ 
\hline 
\rowcolor{Gray}  
Wilcoxon Rank Sum Test  & 4312 & 52 (1.21\%) & 47 (1.09\%) & 2.30\%  & 1453 & 15 (1.03\%) & 20 (1.38\%) & 2.41\%  & 1532 & 62 (4.05\%) & 15 (0.98\%) & 5.03\%  \\ 
\hline 
Wilcoxon Signed Rank Test  & 5632 & 250 (4.44\%) & 148 (2.63\%) & 7.07\%  & 2021 & 111 (5.49\%) & 36 (1.78\%) & 7.27\%  & 1851 & 65 (3.51\%) & 50 (2.70\%) & 6.21\%  \\ 
\hline 
Student's t Test  & 5579 & 370 (6.63\%) & 189 (3.39\%) & 10.02\%  & 2012 & 162 (8.05\%) & 52 (2.58\%) & 10.64\%  & 1816 & 122 (6.72\%) & 71 (3.91\%) & 10.63\%  \\ 
\hline 
One-way ANOVA  & 2012 & 814 (40.46\%) & 50 (2.49\%) & 42.94\%  & 480 & 97 (20.21\%) & 11 (2.29\%) & 22.50\%  & 830 & 290 (34.94\%) & 0 (0.00\%) & 34.94\%  \\ 
\hline 
\rowcolor{Gray}  
Kruskal-Wallis Test  & 1782 & 50 (2.81\%) & 18 (1.01\%) & 3.82\%  & 397 & 7 (1.76\%) & 0 (0.00\%) & 1.76\%  & 728 & 44 (6.04\%) & 3 (0.41\%) & 6.46\%  \\ 
\hline 
Two-way ANOVA  & 3478 & 670 (19.26\%) & 48 (1.38\%) & 20.64\%  & 1125 & 121 (10.76\%) & 41 (3.64\%) & 14.40\%  & 1203 & 281 (23.36\%) & 42 (3.49\%) & 26.85\%  \\ 
\hline 
\rowcolor{Gray}  
Friedman Test  & 2637 & 0 (0.00\%) & 0 (0.00\%) & 0.00\%  & 882 & 0 (0.00\%) & 0 (0.00\%) & 0.00\%  & 925 & 0 (0.00\%) & 0 (0.00\%) & 0.00\%  \\ 
\hline \hline 
\end{tabular} 
 }
\end{table}

As in the case of the correlation among measures, a word of caution has to be made remembering that in the case of measures depending on the recall base our approach is just a surrogate, which improves the ``intervalness'' of a measure but stretches the steps of the scale. Therefore, all the changes in the significantly different system pairs should be taken as tendency rather than exact quantification.

From a glance at Table~\ref{tab:smrysigcnts-rb}  we can note as, in this case, also the significance tests assuming just an ordinal scale, with the exception of the Sign and Friedman tests, are affected by the transformation to an interval scale for an overall change \texttt{$\Delta\%$}  up to a $51\%$. This further confirms that aggregating across topics when the recall base changes can cause variations which go well beyond the loss of ``intervalness''.

As another general trend, we can see that significance tests based on an interval scale assumption are generally more affected, since they experience both the violation of their assumptions and the effect of the recall base.

If we consider Recall, we can see how the most prominent effect is the underestimation of significant differences with a very big increase in the number of significantly different pairs (\texttt{NS2S}), i.e. a reduction in the number of false positives, up to a striking $558\%$ for the one-way ANOVA. Considering that the interval scale behind Recall is the same as the one behind Precision, these figures tell us how big is the loss of power for Recall, mostly due to the impact of aggregating across topics with different recall bases.

In the case of \ac{nDCG} we can observe a behaviour quite similar to one of \ac{DCG}, with an overall change \texttt{$\Delta\%$} just a bit bigger than the one of \ac{DCG}. Considering that both \ac{DCG} and \ac{nDCG} share the same interval scale, this further suggests to use \ac{DCG} to avoid the further bias due to the recall base.

\section{Conclusions and Future Work}
\label{sec:conclusions}

We have  addressed the problem that \ac{IR} measures typically are not interval scales. This issue has severe consequences: you should neither compute means, variances, and confidence intervals nor perform statistical significance tests which assume an interval scale. We have provided a detailed discussion on the motivations and needs behind the interval scales, both in the general field of the representational theory of measurement and in the \ac{IR} context in particular, presenting viewpoints and opinions both supporting and opposing these two ``prohibitions''. It is a matter of fact that these two ``prohibitions'' have been constantly overlooked in the \ac{IR} community. However, when applying improper methods, the results should not be called valid (according to general scientific standards), especially before the impact of these violations has been thoroughly investigated, as has so far been the case in \ac{IR}.

The main  motivation for \ac{IR} measures not being interval scales is that their values are not equi-spaced. Therefore, we have proposed a straightforward yet powerful way to turn any measure into an interval scale by considering how all the possible runs are ranked by the measure and keeping the unique ranks, i.e.\ after removing tied values, as values of the mapped measures. These ranks are equi-spaced by construction and preserve the same order of runs of the original measure. In this way, we obtain an interval scale able to represent the order of runs produced by the original measure.

We have also shown that the situation in IR is worsened by the fact that mixing runs of different length and different recall bases for different topics actually means mixing different scales, being them interval or not. Therefore, computing aggregations across runs and topics in such a way can lead to invalid results. While the run length issue can be mitigated by ensuring that all the runs have the same length, the recall base one is more problematic, since you cannot force a single recall base for all the topics. Therefore, this discourages the use of measures depending on the recall base.

Overall, this discussion led us to raise the fundamental question that \ac{IR} should be more concerned with being able to rely on \emph{meaningful statements}, i.e.\ statements whose truth values do not change when you perform legitimate transformations of the underlying scale, since they ensure for more valid and generalisable inferences.

Relying on several \ac{TREC} collections, we have conducted a thorough experimentation on several (popular) state-of-the-art evaluation measures in order to assess the differences between using an evaluation measure and its interval-scaled version. 

The correlation analysis has shown that the relationship between evaluation measures and their interval-scaled version matches the expected theoretical properties and that not using an interval scale somehow inflates the differences among evaluation measures. Notably, \ac{RR} represents an exception since its departure from being an interval scale makes it look to be more similar to other measures than what it actually is. 

Most importantly, the correlation analysis provides us with a rough estimator of how much interval scale an evaluation measure is and it represents the first attempt to quantify how much evaluation measures depart from their assumptions.

The analysis on many different types of statistical significance tests has clearly shown the impact of passing from an evaluation measure to its  interval-scaled version. In particular, for measures not depending on the recall base, the transformation provides benefits in terms of reduced Type I error and some increase in power of the test. While for measures depending on the recall base, it produces sizeable improvements in terms  of Type II error and power of the test, still delivering substantial enhancements in terms of Type I error. Even apart from any interpretation in terms of Type I and Type II errors, we observed an overall mean change around $25\%$ in the decision about which systems are significantly different and which are not.

Our results on both the correlation analysis and the statistical significance tests open the question about which claims and findings in the IR literature would be impacted by these difference or, in other terms, which statements made in \ac{IR} so far would be actually \emph{meaningful}.

The main limitation of the proposed approach is practical, since first you need to generate all the possible runs of $REL^N$ and then you have compute the evaluation measures on all these  runs. For increasing values of $N$, and even more in the case of multi-graded relevance, this becomes practically infeasible. Therefore, our future work will concern approximating this generation process in order to make it possible to deal with runs of whatever length.

\acrodef{3G}[3G]{Third Generation Mobile System}
\acrodef{5S}[5S]{Streams, Structures, Spaces, Scenarios, Societies}
\acrodef{AA}[AA]{Active Agreements}
\acrodef{AAAI}[AAAI]{Association for the Advancement of Artificial Intelligence}
\acrodef{AAL}[AAL]{Annotation Abstraction Layer}
\acrodef{AAM}[AAM]{Automatic Annotation Manager}
\acrodef{AAP}[AAP]{Average Average Precision}
\acrodef{ACLIA}[ACLIA]{Advanced Cross-Lingual Information Access}
\acrodef{ACM}[ACM]{Association for Computing Machinery}
\acrodef{AD}[AD]{Active Disagreements}
\acrodef{ADSL}[ADSL]{Asymmetric Digital Subscriber Line}
\acrodef{ADUI}[ADUI]{ADministrator User Interface}
\acrodef{AIP}[AIP]{Archival Information Package}
\acrodef{AJAX}[AJAX]{Asynchronous JavaScript Technology and \acs{XML}}
\acrodef{ALU}[ALU]{Aritmetic-Logic Unit}
\acrodef{AMUSID}[AMUSID]{Adaptive MUSeological IDentity-service}
\acrodef{ANOVA}[ANOVA]{ANalysis Of VAriance}
\acrodef{ANSI}[ANSI]{American National Standards Institute}
\acrodef{AP}[AP]{Average Precision}
\acrodef{APC}[APC]{AP Correlation}
\acrodef{API}[API]{Application Program Interface}
\acrodef{AR}[AR]{Address Register}
\acrodef{AS}[AS]{Annotation Service}
\acrodef{ASAP}[ASAP]{Adaptable Software Architecture Performance}
\acrodef{ASI}[ASI]{Annotation Service Integrator}
\acrodef{ASL}[ASL]{Achieved Significance Level}
\acrodef{ASM}[ASM]{Annotation Storing Manager}
\acrodef{ASR}[ASR]{Automatic Speech Recognition}
\acrodef{ASUI}[ASUI]{ASsessor User Interface}
\acrodef{ATIM}[ATIM]{Annotation Textual Indexing Manager}
\acrodef{AUC}[AUC]{Area Under the ROC Curve}
\acrodef{AUI}[AUI]{Administrative User Interface}
\acrodef{AWARE}[AWARE]{Assessor-driven Weighted Averages for Retrieval Evaluation}
\acrodef{BANKS-I}[BANKS-I]{Browsing ANd Keyword Searching I}
\acrodef{BANKS-II}[BANKS-II]{Browsing ANd Keyword Searching II}
\acrodef{BH}[BH]{Benjamini-Hochberg}
\acrodef{bpref}[bpref]{Binary Preference}
\acrodef{BNF}[BNF]{Backus and Naur Form}
\acrodef{BPM}[BPM]{Bejeweled Player Model}
\acrodef{BRICKS}[BRICKS]{Building Resources for Integrated Cultural Knowledge Services}
\acrodef{CAN}[CAN]{Content Addressable Netword}
\acrodef{CAS}[CAS]{Content-And-Structure}
\acrodef{CBSD}[CBSD]{Component-Based Software Developlement}
\acrodef{CBSE}[CBSE]{Component-Based Software Engineering}
\acrodef{CB-SPE}[CB-SPE]{Component-Based \acs{SPE}}
\acrodef{CD}[CD]{Collaboration Diagram}
\acrodef{CD}[CD]{Compact Disk}
\acrodef{CDF}[CDF]{Cumulative Density Function}
\acrodef{CENL}[CENL]{Conference of European National Librarians}
\acrodef{CIDOC CRM}[CIDOC CRM]{CIDOC Conceptual Reference Model}
\acrodef{CIR}[CIR]{Current Instruction Register}
\acrodef{CIRCO}[CIRCO]{Coordinated Information Retrieval Components Orchestration}
\acrodef{CG}[CG]{Cumulated Gain}
\acrodef{CL}[CL]{Curriculum Learning}
\acrodef{CL-ESA}[CL-ESA]{Cross-Lingual Explicit Semantic Analysis}
\acrodef{CLAIRE}[CLAIRE]{Combinatorial visuaL Analytics system for Information Retrieval Evaluation}
\acrodef{CLEF1}[CLEF]{Cross-Language Evaluation Forum}
\acrodef{CLEF}[CLEF]{Conference and Labs of the Evaluation Forum}
\acrodef{CLIR}[CLIR]{Cross Language Information Retrieval}
\acrodef{CM}[CM]{Continuation Methods}
\acrodef{CMS}[CMS]{Content Management System}
\acrodef{CMT}[CMT]{Campaign Management Tool}
\acrodef{CNR}[CNR]{Italian National Council of Research}
\acrodef{CO}[CO]{Content-Only}
\acrodef{COD}[COD]{Code On Demand}
\acrodef{CODATA}[CODATA]{Committee on Data for Science and Technology}
\acrodef{COLLATE}[COLLATE]{Collaboratory for Annotation Indexing and Retrieval of Digitized Historical Archive Material}
\acrodef{CP}[CP]{Characteristic Pattern}
\acrodef{CPE}[CPE]{Control Processor Element}
\acrodef{CPU}[CPU]{Central Processing Unit}
\acrodef{CQL}[CQL]{Contextual Query Language}
\acrodef{CRP}[CRP]{Cumulated Relative Position}
\acrodef{CRUD}[CRUD]{Create--Read--Update--Delete}
\acrodef{CS}[CS]{Characteristic Structure}
\acrodef{CSM}[CSM]{Campaign Storing Manager}
\acrodef{CSS}[CSS]{Cascading Style Sheets}
\acrodef{CTR}[CTR]{Click-Through Rate}
\acrodef{CU}[CU]{Control Unit}
\acrodef{CUI}[CUI]{Client User Interface}
\acrodef{CV}[CV]{Cross-Validation}
\acrodef{DAFFODIL}[DAFFODIL]{Distributed Agents for User-Friendly Access of Digital Libraries}
\acrodef{DAO}[DAO]{Data Access Object}
\acrodef{DARE}[DARE]{Drawing Adequate REpresentations}
\acrodef{DARPA}[DARPA]{Defense Advanced Research Projects Agency}
\acrodef{DAS}[DAS]{Distributed Annotation System}
\acrodef{DB}[DB]{DataBase}
\acrodef{DBMS}[DBMS]{DataBase Management System}
\acrodef{DC}[DC]{Dublin Core}
\acrodef{DCG}[DCG]{Discounted Cumulative Gain}
\acrodef{DCMI}[DCMI]{Dublin Core Metadata Initiative}
\acrodef{DCV}[DCV]{Document Cut--off Value}
\acrodef{DD}[DD]{Deployment Diagram}
\acrodef{DDC}[DDC]{Dewey Decimal Classification}
\acrodef{DDS}[DDS]{Direct Data Structure}
\acrodef{DF}[DF]{Degrees of Freedom}
\acrodef{DFI}[DFI]{Divergence From Independence}
\acrodef{DFR}[DFR]{Divergence From Randomness}
\acrodef{DHT}[DHT]{Distributed Hash Table}
\acrodef{DI}[DI]{Digital Image}
\acrodef{DIKW}[DIKW]{Data, Information, Knowledge, Wisdom}
\acrodef{DIL}[DIL]{\acs{DIRECT} Integration Layer}
\acrodef{DiLAS}[DiLAS]{Digital Library Annotation Service}
\acrodef{DIRECT}[DIRECT]{Distributed Information Retrieval Evaluation Campaign Tool}
\acrodef{DKMS}[DKMS]{Data and Knowledge Management System}
\acrodef{DL}[DL]{Digital Library}
\acrodefplural{DL}[DL]{Digital Libraries}
\acrodef{DLMS}[DLMS]{Digital Library Management System}
\acrodef{DLOG}[DL]{Description Logics}
\acrodef{DLS}[DLS]{Digital Library System}
\acrodef{DLSS}[DLSS]{Digital Library Service System}
\acrodef{DM}[DM]{Data Mining}
\acrodef{DO}[DO]{Digital Object}
\acrodef{DOI}[DOI]{Digital Object Identifier}
\acrodef{DOM}[DOM]{Document Object Model}
\acrodef{DoMDL}[DoMDL]{Document Model for Digital Libraries}
\acrodef{DP}[DP]{Discriminative Power}
\acrodef{DPBF}[DPBF]{Dynamic Programming Best-First}
\acrodef{DR}[DR]{Data Register}
\acrodef{DRIVER}[DRIVER]{Digital Repository Infrastructure Vision for European Research}
\acrodef{DTD}[DTD]{Document Type Definition}
\acrodef{DVD}[DVD]{Digital Versatile Disk}
\acrodef{EAC-CPF}[EAC-CPF]{Encoded Archival Context for Corporate Bodies, Persons, and Families}
\acrodef{EAD}[EAD]{Encoded Archival Description}
\acrodef{EAN}[EAN]{International Article Number}
\acrodef{EBU}[EBU]{Expected Browsing Utility}
\acrodef{ECD}[ECD]{Enhanced Contenty Delivery}
\acrodef{ECDL}[ECDL]{European Conference on Research and Advanced Technology for Digital Libraries}
\acrodef{EDM}[EDM]{Europeana Data Model}
\acrodef{EG}[EG]{Execution Graph}
\acrodef{ELDA}[ELDA]{Evaluation and Language resources Distribution Agency}
\acrodef{ELRA}[ELRA]{European Language Resources Association}
\acrodef{EM}[EM]{Expectation Maximization}
\acrodef{EMMA}[EMMA]{Extensible MultiModal Annotation}
\acrodef{EPROM}[EPROM]{Erasable Programmable \acs{ROM}}
\acrodef{EQNM}[EQNM]{Extended Queueing Network Model}
\acrodef{ER}[ER]{Entity--Relationship}
\acrodef{ERR}[ERR]{Expected Reciprocal Rank}
\acrodef{ERS}[ERS]{Empirical Relational System}
\acrodef{ESA}[ESA]{Explicit Semantic Analysis}
\acrodef{ESL}[ESL]{Expected Search Length}
\acrodef{ETL}[ETL]{Extract-Transform-Load}
\acrodef{FAST}[FAST]{Flexible Annotation Service Tool}
\acrodef{FDR}[FDR]{False Discovery Rate}
\acrodef{FIFO}[FIFO]{First-In / First-Out}
\acrodef{FIRE}[FIRE]{Forum for Information Retrieval Evaluation}
\acrodef{FN}[FN]{False Negative}
\acrodef{FNR}[FNR]{False Negative Rate}
\acrodef{FOAF}[FOAF]{Friend of a Friend}
\acrodef{FORESEE}[FORESEE]{FOod REcommentation sErvER}
\acrodef{FP}[FP]{False Positive}
\acrodef{FPR}[FPR]{False Positive Rate}
\acrodef{FWER}[FWER]{Family-wise Error Rate}
\acrodef{GIF}[GIF]{Graphics Interchange Format}
\acrodef{GIR}[GIR]{Geografic Information Retrieval}
\acrodef{GAP}[GAP]{Graded Average Precision}
\acrodef{GLM}[GLM]{General Linear Model}
\acrodef{GLMM}[GLMM]{General Linear Mixed Model}
\acrodef{GMAP}[GMAP]{Geometric Mean Average Precision}
\acrodef{GoP}[GoP]{Grid of Points}
\acrodef{GPRS}[GPRS]{General Packet Radio Service}
\acrodef{gP}[gP]{Generalized Precision}
\acrodef{gR}[gR]{Generalized Recall}
\acrodef{gRBP}[gRBP]{Graded Rank-Biased Precision}
\acrodef{GT}[GT]{Generalizability Theory}
\acrodef{GTIN}[GTIN]{Global Trade Item Number}
\acrodef{GUI}[GUI]{Graphical User Interface}
\acrodef{GW}[GW]{Gateway}
\acrodef{HCI}[HCI]{Human Computer Interaction}
\acrodef{HDS}[HDS]{Hybrid Data Structure}
\acrodef{HIR}[HIR]{Hypertext Information Retrieval}
\acrodef{HIT}[HIT]{Human Intelligent Task}
\acrodef{HITS}[HITS]{Hyperlink-Induced Topic Search}
\acrodef{HMM}[HMM]{Hidden Markov Model}
\acrodef{HTML}[HTML]{HyperText Markup Language}
\acrodef{HTTP}[HTTP]{HyperText Transfer Protocol}
\acrodef{HSD}[HSD]{Honestly Significant Difference}
\acrodef{ICA}[ICA]{International Council on Archives}
\acrodef{ICSU}[ICSU]{International Council for Science}
\acrodef{IDF}[IDF]{Inverse Document Frequency}
\acrodef{IDS}[IDS]{Inverse Data Structure}
\acrodef{IEEE}[IEEE]{Institute of Electrical and Electronics Engineers}
\acrodef{IEI}[IEI]{Istituto della Enciclopedia Italiana fondata da Giovanni Treccani}
\acrodef{IETF}[IETF]{Internet Engineering Task Force}
\acrodef{IIR}[IIR]{Interactive Information Retrieval}
\acrodef{IMS}[IMS]{Information Management System}
\acrodef{IMSPD}[IMS]{Information Management Systems Research Group}
\acrodef{indAP}[indAP]{Induced Average Precision}
\acrodef{infAP}[infAP]{Inferred Average Precision}
\acrodef{INEX}[INEX]{INitiative for the Evaluation of \acs{XML} Retrieval}
\acrodef{INS-M}[INS-M]{Inverse Set Data Model}
\acrodef{INTR}[INTR]{Interrupt Register}
\acrodef{IP}[IP]{Internet Protocol}
\acrodef{IPSA}[IPSA]{Imaginum Patavinae Scientiae Archivum}
\acrodef{IR}[IR]{Information Retrieval}
\acrodef{IRON}[IRON]{Information Retrieval ON}
\acrodef{IRON2}[IRON$^2$]{Information Retrieval On aNNotations}
\acrodef{IRON-SAT}[IRON-SAT]{\acs{IRON} - Statistical Analysis Tool}
\acrodef{IRS}[IRS]{Information Retrieval System}
\acrodef{ISAD(G)}[ISAD(G)]{International Standard for Archival Description (General)}
\acrodef{ISBN}[ISBN]{International Standard Book Number}
\acrodef{ISIS}[ISIS]{Interactive SImilarity Search}
\acrodef{ISJ}[ISJ]{Interactive Searching and Judging}
\acrodef{ISO}[ISO]{International Organization for Standardization}
\acrodef{ITU}[ITU]{International Telecommunication Union }
\acrodef{ITU-T}[ITU-T]{Telecommunication Standardization Sector of \acs{ITU}}
\acrodef{IV}[IV]{Information Visualization}
\acrodef{JAN}[JAN]{Japanese Article Number}
\acrodef{JDBC}[JDBC]{Java DataBase Connectivity}
\acrodef{JMB}[JMB]{Java--Matlab Bridge}
\acrodef{JPEG}[JPEG]{Joint Photographic Experts Group}
\acrodef{JSON}[JSON]{JavaScript Object Notation}
\acrodef{JSP}[JSP]{Java Server Pages}
\acrodef{JTE}[JTE]{Java-Treceval Engine}
\acrodef{KDE}[KDE]{Kernel Density Estimation}
\acrodef{KLD}[KLD]{Kullback-Leibler Divergence}
\acrodef{KLAPER}[KLAPER]{Kernel LAnguage for PErformance and Reliability analysis}
\acrodef{LAM}[LAM]{Libraries, Archives, and Museums}
\acrodef{LAM2}[LAM]{Logistic Average Misclassification}
\acrodef{LAN}[LAN]{Local Area Network}
\acrodef{LD}[LD]{Linked Data}
\acrodef{LEAF}[LEAF]{Linking and Exploring Authority Files}
\acrodef{LIDO}[LIDO]{Lightweight Information Describing Objects}
\acrodef{LIFO}[LIFO]{Last-In / First-Out}
\acrodef{LM}[LM]{Language Model}
\acrodef{LMT}[LMT]{Log Management Tool}
\acrodef{LOD}[LOD]{Linked Open Data}
\acrodef{LODE}[LODE]{Linking Open Descriptions of Events}
\acrodef{LpO}[LpO]{Leave-$p$-Out}
\acrodef{LRM}[LRM]{Local Relational Model}
\acrodef{LRU}[LRU]{Last Recently Used}
\acrodef{LS}[LS]{Lexical Signature}
\acrodef{LSM}[LSM]{Log Storing Manager}
\acrodef{LtR}[LtR]{Learning to Rank}
\acrodef{LUG}[LUG]{Lexical Unit Generator}
\acrodef{MA}[MA]{Mobile Agent}
\acrodef{MA}[MA]{Moving Average}
\acrodef{MACS}[MACS]{Multilingual ACcess to Subjects}
\acrodef{MADCOW}[MADCOW]{Multimedia Annotation of Digital Content Over the Web}
\acrodef{MAD}[MAD]{Mean Assessed Documents}
\acrodef{MADP}[MADP]{Mean Assessed Documents Precision}
\acrodef{MADS}[MADS]{Metadata Authority Description Standard}
\acrodef{MAP}[MAP]{Mean Average Precision}
\acrodef{MARC}[MARC]{Machine Readable Cataloging}
\acrodef{MATTERS}[MATTERS]{MATlab Toolkit for Evaluation of information Retrieval Systems}
\acrodef{MDA}[MDA]{Model Driven Architecture}
\acrodef{MDD}[MDD]{Model-Driven Development}
\acrodef{METS}[METS]{Metadata Encoding and Transmission Standard}
\acrodef{MIDI}[MIDI]{Musical Instrument Digital Interface}
\acrodef{MIME}[MIME]{Multipurpose Internet Mail Extensions}
\acrodef{ML}[ML]{Machine Learning}
\acrodef{MLE}[MLE]{Maximum Likelihood Estimation}
\acrodef{MLIA}[MLIA]{MultiLingual Information Access}
\acrodef{MM}[MM]{Machinery Model}
\acrodef{MMU}[MMU]{Memory Management Unit}
\acrodef{MODS}[MODS]{Metadata Object Description Schema}
\acrodef{MOF}[MOF]{Meta-Object Facility}
\acrodef{MP}[MP]{Markov Precision}
\acrodef{MPEG}[MPEG]{Motion Picture Experts Group}
\acrodef{MRD}[MRD]{Machine Readable Dictionary}
\acrodef{MRF}[MRF]{Markov Random Field}
\acrodef{MRR}[MRR]{Mean Reciprocal Rank}
\acrodef{MS}[MS]{Mean Squares}
\acrodef{MSAC}[MSAC]{Multilingual Subject Access to Catalogues}
\acrodef{MSE}[MSE]{Mean Square Error}
\acrodef{MT}[MT]{Machine Translation}
\acrodef{MV}[MV]{Majority Vote}
\acrodef{MVC}[MVC]{Model-View-Controller}
\acrodef{NACSIS}[NACSIS]{NAtional Center for Science Information Systems}
\acrodef{NAP}[NAP]{Network processors Applications Profile}
\acrodef{NCP}[NCP]{Normalized Cumulative Precision}
\acrodef{nCG}[nCG]{Normalized Cumulated Gain}
\acrodef{nCRP}[nCRP]{Normalized Cumulated Relative Position}
\acrodef{nDCG}[nDCG]{Normalized Discounted Cumulative Gain}
\acrodef{nMCG}[nMCG]{Normalized Markov Cumulated Gain}
\acrodef{NESTOR}[NESTOR]{NEsted SeTs for Object hieRarchies}
\acrodef{NEXI}[NEXI]{Narrowed Extended XPath I}
\acrodef{NII}[NII]{National Institute of Informatics}
\acrodef{NISO}[NISO]{National Information Standards Organization}
\acrodef{NIST}[NIST]{National Institute of Standards and Technology}
\acrodef{NLP}[NLP]{Natural Language Processing}
\acrodef{NN}[NN]{Neural Network}
\acrodef{NP}[NP]{Network Processor}
\acrodef{NR}[NR]{Normalized Recall}
\acrodef{NRS}[NRS]{Numerical Relational System}
\acrodef{NS-M}[NS-M]{Nested Set Model}
\acrodef{NTCIR}[NTCIR]{NII Testbeds and Community for Information access Research}
\acrodef{OAI}[OAI]{Open Archives Initiative}
\acrodef{OAI-ORE}[OAI-ORE]{Open Archives Initiative Object Reuse and Exchange}
\acrodef{OAI-PMH}[OAI-PMH]{Open Archives Initiative Protocol for Metadata Harvesting}
\acrodef{OAIS}[OAIS]{Open Archival Information System}
\acrodef{OC}[OC]{Operation Code}
\acrodef{OCLC}[OCLC]{Online Computer Library Center}
\acrodef{OMG}[OMG]{Object Management Group}
\acrodef{OO}[OO]{Object Oriented}
\acrodef{OODB}[OODB]{Object-Oriented \acs{DB}}
\acrodef{OODBMS}[OODBMS]{Object-Oriented \acs{DBMS}}
\acrodef{OPAC}[OPAC]{Online Public Access Catalog}
\acrodef{OQL}[OQL]{Object Query Language}
\acrodef{ORP}[ORP]{Open Relevance Project}
\acrodef{OSIRIS}[OSIRIS]{Open Service Infrastructure for Reliable and Integrated process Support}
\acrodef{P}[P]{Precision}
\acrodef{P2P}[P2P]{Peer-To-Peer}
\acrodef{PA}[PA]{Passive Agreements}
\acrodef{PAMT}[PAMT]{Pool-Assessment Management Tool}
\acrodef{PASM}[PASM]{Pool-Assessment Storing Manager}
\acrodef{PC}[PC]{Program Counter}
\acrodef{PCP}[PCP]{Pre-Commercial Procurement}
\acrodef{PCR}[PCR]{Peripherical Command Register}
\acrodef{PD}[PD]{Passive Disagreements}
\acrodef{PDA}[PDA]{Personal Digital Assistant}
\acrodef{PDF}[PDF]{Probability Density Function}
\acrodef{PDR}[PDR]{Peripherical Data Register}
\acrodef{PIR}[PIR]{Personalized Information Retrieval}
\acrodef{POI}[POI]{\acs{PURL}-based Object Identifier}
\acrodef{PoS}[PoS]{Part of Speech}
\acrodef{PAA}[PAA]{Proportion of Active Agreements}
\acrodef{PPA}[PPA]{Proportion of Passive Agreements}
\acrodef{PPE}[PPE]{Programmable Processing Engine}
\acrodef{PREFORMA}[PREFORMA]{PREservation FORMAts for culture information/e-archives}
\acrodef{PRIMAD}[PRIMAD]{Platform, Research goal, Implementation, Method, Actor, and Data}
\acrodef{PRIMAmob-UML}[PRIMAmob-UML]{mobile \acs{PRIMA-UML}}
\acrodef{PRIMA-UML}[PRIMA-UML]{PeRformance IncreMental vAlidation in \acs{UML}}
\acrodef{PROM}[PROM]{Programmable \acs{ROM}}
\acrodef{PROMISE}[PROMISE]{Participative Research labOratory  for Multimedia and Multilingual Information Systems Evaluation}
\acrodef{pSQL}[pSQL]{propagate \acs{SQL}}
\acrodef{PUI}[PUI]{Participant User Interface}
\acrodef{PURL}[PURL]{Persistent \acs{URL}}
\acrodef{QA}[QA]{Question Answering}
\acrodef{QE}[QE]{Query Expansion}
\acrodef{QoS-UML}[QoS-UML]{\acs{UML} Profile for QoS and Fault Tolerance}
\acrodef{QPA}[QPA]{Query Performance Analyzer}
\acrodef{QPP}[QPP]{Query Performance Prediction}
\acrodef{R}[R]{Recall}
\acrodef{RAM}[RAM]{Random Access Memory}
\acrodef{RAMM}[RAM]{Random Access Machine}
\acrodef{RBO}[RBO]{Rank-Biased Overlap}
\acrodef{RBP}[RBP]{Rank-Biased Precision}
\acrodef{RBTO}[RBTO]{Rank-Based Total Order}
\acrodef{RDBMS}[RDBMS]{Relational \acs{DBMS}}
\acrodef{RDF}[RDF]{Resource Description Framework}
\acrodef{REST}[REST]{REpresentational State Transfer}
\acrodef{REV}[REV]{Remote Evaluation}
\acrodef{RF}[RF]{Relevance Feedback}
\acrodef{RFC}[RFC]{Request for Comments}
\acrodef{RIA}[RIA]{Reliable Information Access}
\acrodef{RMSE}[RMSE]{Root Mean Square Error}
\acrodef{RMT}[RMT]{Run Management Tool}
\acrodef{ROM}[ROM]{Read Only Memory}
\acrodef{ROMIP}[ROMIP]{Russian Information Retrieval Evaluation Seminar}
\acrodef{RoMP}[RoMP]{Rankings of Measure Pairs}
\acrodef{RoS}[RoS]{Rankings of Systems}
\acrodef{RP}[RP]{Relative Position}
\acrodef{RR}[RR]{Reciprocal Rank}
\acrodef{RSM}[RSM]{Run Storing Manager}
\acrodef{RST}[RST]{Rhetorical Structure Theory}
\acrodef{RSV}[RSV]{Retrieval Status Value}
\acrodef{RT-UML}[RT-UML]{\acs{UML} Profile for Schedulability, Performance and Time}
\acrodef{SA}[SA]{Software Architecture}
\acrodef{SAL}[SAL]{Storing Abstraction Layer}
\acrodef{SAMT}[SAMT]{Statistical Analysis Management Tool}
\acrodef{SAN}[SAN]{Sistema Archivistico Nazionale}
\acrodef{SASM}[SASM]{Statistical Analysis Storing Manager}
\acrodef{SBTO}[SBTO]{Set-Based Total Order}
\acrodef{SD}[SD]{Sequence Diagram}
\acrodef{SE}[SE]{Search Engine}
\acrodef{SEBD}[SEBD]{Convegno Nazionale su Sistemi Evoluti per Basi di Dati}
\acrodef{SEM}[SEM]{Standard Error of the Mean}
\acrodef{SERP}[SERP]{Search Engine Result Page}
\acrodef{SFT}[SFT]{Satisfaction--Frustration--Total}
\acrodef{SIL}[SIL]{Service Integration Layer}
\acrodef{SIP}[SIP]{Submission Information Package}
\acrodef{SKOS}[SKOS]{Simple Knowledge Organization System}
\acrodef{SM}[SM]{Software Model}
\acrodef{SME}[SME]{Statistics--Metrics-Experiments}
\acrodef{SMART}[SMART]{System for the Mechanical Analysis and Retrieval of Text}
\acrodef{SoA}[SoA]{Service-oriented Architectures}
\acrodef{SOA}[SOA]{Strength of Association}
\acrodef{SOAP}[SOAP]{Simple Object Access Protocol}
\acrodef{SOM}[SOM]{Self-Organizing Map}
\acrodef{SPARQL}[SPARQL]{Simple Protocol and RDF Query Language}
\acrodef{SPE}[SPE]{Software Performance Engineering}
\acrodef{SPINA}[SPINA]{Superimposed Peer Infrastructure for iNformation Access}
\acrodef{SPLIT}[SPLIT]{Stemming Program for Language Independent Tasks}
\acrodef{SPOOL}[SPOOL]{Simultaneous Peripheral Operations On Line}
\acrodef{SQL}[SQL]{Structured Query Language}
\acrodef{SR}[SR]{Sliding Ratio}
\acrodef{sRBP}[sRBP]{Session Rank Biased Precision}
\acrodef{SRU}[SRU]{Search/Retrieve via \acs{URL}}
\acrodef{SS}[SS]{Sum of Squares}
\acrodef{SSD}[s.s.d.]{statistically significantly different}
\acrodef{SSTF}[SSTF]{Shortest Seek Time First}
\acrodef{STAR}[STAR]{Steiner-Tree Approximation in Relationship graphs}
\acrodef{STON}[STON]{STemming ON}
\acrodef{SVM}[SVM]{Support Vector Machine}
\acrodef{TAC}[TAC]{Text Analysis Conference}
\acrodef{TBG}[TBG]{Time-Biased Gain}
\acrodef{TCP}[TCP]{Transmission Control Protocol}
\acrodef{TEL}[TEL]{The European Library}
\acrodef{TERRIER}[TERRIER]{TERabyte RetrIEveR}
\acrodef{TF}[TF]{Term Frequency}
\acrodef{TFR}[TFR]{True False Rate}
\acrodef{TLD}[TLD]{Top Level Domain}
\acrodef{TME}[TME]{Topics--Metrics-Experiments}
\acrodef{TN}[TN]{True Negative}
\acrodef{TO}[TO]{Transfer Object}
\acrodef{TP}[TP]{True Positve}
\acrodef{TPR}[TPR]{True Positive Rate}
\acrodef{TRAT}[TRAT]{Text Relevance Assessing Task}
\acrodef{TREC}[TREC]{Text REtrieval Conference}
\acrodef{TRECVID}[TRECVID]{TREC Video Retrieval Evaluation}
\acrodef{TTL}[TTL]{Time-To-Live}
\acrodef{UCD}[UCD]{Use Case Diagram}
\acrodef{UDC}[UDC]{Universal Decimal Classification}
\acrodef{uGAP}[uGAP]{User-oriented Graded Average Precision}
\acrodef{UI}[UI]{User Interface}
\acrodef{UML}[UML]{Unified Modeling Language}
\acrodef{UMT}[UMT]{User Management Tool}
\acrodef{UMTS}[UMTS]{Universal Mobile Telecommunication System}
\acrodef{UoM}[UoM]{Utility-oriented Measurement}
\acrodef{UPC}[UPC]{Universal Product Code}
\acrodef{URI}[URI]{Uniform Resource Identifier}
\acrodef{URL}[URL]{Uniform Resource Locator}
\acrodef{URN}[URN]{Uniform Resource Name}
\acrodef{USM}[USM]{User Storing Manager}
\acrodef{VA}[VA]{Visual Analytics}
\acrodef{VAIRE}[VAIR\"{E}]{Visual Analytics for Information Retrieval Evaluation}
\acrodef{VATE}[VATE$^2$]{Visual Analytics Tool for Experimental Evaluation}
\acrodef{VIRTUE}[VIRTUE]{Visual Information Retrieval Tool for Upfront Evaluation}
\acrodef{VD}[VD]{Virtual Document}
\acrodef{VDM}[VDM]{Visual Data Mining}
\acrodef{VIAF}[VIAF]{Virtual International Authority File}
\acrodef{VIM}[VIM]{International Vocabulary of Metrology}
\acrodef{VL}[VL]{Visual Language}
\acrodef{VoIP}[VoIP]{Voice over IP}
\acrodef{VS}[VS]{Visual Sentence}
\acrodef{W3C}[W3C]{World Wide Web Consortium}
\acrodef{WAN}[WAN]{Wide Area Network}
\acrodef{WHO}[WHO]{World Health Organization}
\acrodef{WLAN}[WLAN]{Wireless \acs{LAN}}
\acrodef{WP}[WP]{Work Package}
\acrodef{WS}[WS]{Web Services}
\acrodef{WSD}[WSD]{Word Sense Disambiguation}
\acrodef{WSDL}[WSDL]{Web Services Description Language}
\acrodef{WWW}[WWW]{World Wide Web}
\acrodef{XMI}[XMI]{\acs{XML} Metadata Interchange}
\acrodef{XML}[XML]{eXtensible Markup Language}
\acrodef{XPath}[XPath]{XML Path Language}
\acrodef{XSL}[XSL]{eXtensible Stylesheet Language}
\acrodef{XSL-FO}[XSL-FO]{\acs{XSL} Formatting Objects}
\acrodef{XSLT}[XSLT]{\acs{XSL} Transformations}
\acrodef{YAGO}[YAGO]{Yet Another Great Ontology}
\acrodef{YASS}[YASS]{Yet Another Suffix Stripper}

\bibliographystyle{plainnat}
\bibliography{bibliografia,zz-proceedings}

\end{document}